\documentclass{article} 
\pdfoutput=1
 
\usepackage[utf8]{inputenc}
\usepackage[colorlinks = true,
            linkcolor = blue,
            urlcolor  = blue,
            citecolor = blue,
            anchorcolor = blue]{hyperref}
\usepackage{amsmath}
\usepackage{amsthm}
\usepackage{amssymb}
\usepackage{xcolor}
\colorlet{darkgreen}{green!50!black}

\usepackage{algorithm,caption}
\usepackage{algorithmic}
\usepackage{cleveref}
\usepackage{graphicx}

\usepackage{natbib}
\usepackage{framed}
\usepackage{tcolorbox}
\usepackage{times}

\usepackage{geometry}
\usepackage{tikz}
\usetikzlibrary{matrix,arrows,shapes,trees,positioning, calc}
\usetikzlibrary{decorations.pathreplacing, decorations.pathmorphing}

\makeatletter
\theoremstyle{plain}
\newtheorem{theorem}{\protect\theoremname}
\newtheorem*{theorem*}{\protect\theoremname}
\newtheorem*{proposition*}{\protect\theoremname}
\newtheorem{definition}{\protect\definitionname}
\theoremstyle{definition}

\newtheorem{example}[definition]{\protect\examplename}
\theoremstyle{plain}
\newtheorem{lemma}[definition]{\protect\lemmaname}
\newtheorem{corollary}[definition]{\protect\corollaryname}
\newtheorem*{corollary*}{\protect\corollaryname}

\newtheorem{proposition}[definition]{\protect\propname}

\newtheorem*{question*}{\protect\questionname}
\newtheorem{conjecture}[definition]{\protect\conjecturename}
\newtheorem*{conjecture*}{\protect\conjecturename}
\newtheorem*{assumption*}{\protect\assumptionname}

\makeatother

\newtheorem{innercustomthm}{Theorem}
\newenvironment{customthm}[1]
  {\renewcommand\theinnercustomthm{#1}\innercustomthm}
  {\endinnercustomthm}
\newtheorem{innercustomlem}{Lemma}
\newenvironment{customlem}[1]
  {\renewcommand\theinnercustomlem{#1}\innercustomlem}
  {\endinnercustomlem}

\newtheorem{innercustomconjecture}{Conjecture}
\newenvironment{customconjecture}[1]
  {\renewcommand\theinnercustomconjecture{#1}\innercustomconjecture}
  {\endinnercustomconjecture}

\providecommand{\questionname}{Question}
\providecommand{\conjecturename}{Conjecture}
\providecommand{\assumptionname}{Assumption}
\providecommand{\observationname}{Observation}
\providecommand{\corollaryname}{Corollary}
\providecommand{\definitionname}{Definition}
\providecommand{\lemmaname}{Lemma}

\providecommand{\theoremname}{Theorem}
\providecommand{\exercisename}{Exercise}
\providecommand{\examplename}{Example}
\providecommand{\remarkname}{Remark}
\providecommand{\factname}{Fact}
\providecommand{\propname}{Proposition}

\newcommand{\C}{\mathcal{C}}
\newcommand{\A}{\mathcal{A}}
\newcommand{\R}{\mathbb{R}}
\newcommand{\U}{\mathcal{U}}

\newcommand{\D}{\mathcal{D}}

\newcommand{\inp}{\mathrm{in}}
\newcommand{\out}{\mathrm{out}}
\newcommand{\E}{\mathcal{E}}
\newcommand{\Ex}{\mathop{\mathbb{E}}}
\newcommand{\TV}{\mathtt{TV}}
\newcommand{\VC}{\mathtt{VC}}

\newcommand{\F}{\mathcal{F}}
\newcommand{\G}{\mathcal{G}}
\newcommand{\T}{\mathcal{T}}
\renewcommand{\H}{\mathcal{H}}
\renewcommand{\S}{\mathbb{S}}
\newcommand{\ZZ}{\mathbb{Z}}

\newcommand{\N}{\mathbb{N}}

\renewcommand{\SS}{\mathbb{S}}

\newcommand{\PP}{\mathcal{P}}
\newcommand{\LL}{\mathcal{L}}
\renewcommand{\Pr}{\mathbb{P}}
\newcommand{\List}{\mathtt{LR}}
\newcommand{\Q}{\mathbb{Q}}
\newcommand{\PPP}{\mathbb{P}}
\newcommand{\CC}{\mathbb{C}}
\newcommand{\B}{\mathbb{B}}
\newcommand{\DD}{\mathbb{D}}

\newcommand{\an}{\mathrm{a}}
\newcommand{\ant}{\mathrm{ant}}

\newcommand{\simp}{\mathrm{simp}}

\newcommand{\extr}{\mathrm{extr}}
\newcommand{\diam}{\textrm{diam}}

\newcommand{\str}{\mathtt{shatter}}
\newcommand{\St}{\mathrm{St}}
\newcommand{\Lk}{\mathrm{Lk}}

\newcommand{\ol}[1]{\overline{#1}}
\newcommand{\sd}{\mathtt{sd}}
\newcommand{\disamb}{\mathtt{disamb}}
\newcommand{\OPT}{\mathtt{OPT}}
\newcommand{\outp}{\mathtt{out}}
\newcommand{\SCO}{\mathtt{SCO}}
\newcommand{\sign}{\mathsf{sign}}
\newcommand{\sr}{\mathtt{sign}\mathrm{-}\mathtt{rank}}

\DeclareMathOperator{\Img}{\mathrm{Im}}
\DeclareMathOperator{\supp}{\mathrm{supp}}

\title{Spherical dimension}

\author{Bogdan Chornomaz\footnote{Department of Mathematics, Technion. Supported by the European Union (ERC, GENERALIZATION, 101039692).} 
\and Shay Moran\footnote{Departments of Mathematics, Computer Science, and Data and Decision Sciences, Technion and Google Research.
Robert J.\ Shillman Fellow; supported by ISF grant 1225/20, by BSF grant 2018385, by an Azrieli Faculty Fellowship, by Israel PBC-VATAT, by the Technion Center for Machine Learning and Intelligent Systems (MLIS), and by the European Union (ERC, GENERALIZATION, 101039692). Views and opinions expressed are however those of the author(s) only and do not necessarily reflect those of the European Union or the European Research Council Executive Agency. Neither the European Union nor the granting authority can be held responsible for them.
}
\and Tom Waknine\footnote{Department of Mathematics, Technion}
}

\begin{document}
\maketitle

\begin{abstract}
 We introduce and study the \emph{spherical dimension}, a natural topological relaxation of the VC dimension that unifies several results in learning theory where topology plays a key role in the proofs. The spherical dimension is defined by extending the set of realizable datasets (used to define the VC dimension) to the continuous space of realizable distributions. In this space, a shattered set of size d (in the VC sense) is completed into a continuous object, specifically a d-dimensional sphere of realizable distributions. The spherical dimension is then defined as the dimension of the largest sphere in this space. Thus, the spherical dimension is at least the VC dimension. 

The spherical dimension serves as a common foundation for leveraging the Borsuk-Ulam theorem and related topological tools. We demonstrate the utility of the spherical dimension in diverse applications, including disambiguations of partial concept classes, reductions from classification to stochastic convex optimization, stability and replicability, and sample compression schemes. Perhaps surprisingly, we show that the open question posed by Alon, Hanneke, Holzman, and Moran (FOCS 2021) of whether there exist non-trivial disambiguations for halfspaces with margin is equivalent to the basic open question of whether the VC and spherical dimensions are finite together.
\end{abstract}


\section{Introduction}\label{sec-intro}
The VC dimension is one of the most celebrated parameters in the theory of PAC learning, offering a simple yet powerful combinatorial framework for understanding the complexity of concept classes. Formally, the VC dimension of a binary-labeled concept class \(\mathcal{H}\) is the largest integer \(n\) such that there exist points \(x_1, \ldots, x_n \in X\) where, for any labeling \(y_1, \ldots, y_n \in \{\pm 1\}\), the dataset \((x_1, y_1), \ldots, (x_n, y_n)\) is realizable by \(\mathcal{H}\).

In this work, we study a continuous relaxation of the VC dimension. To define it, we first extend the discrete space of realizable datasets to the continuous space of realizable distributions. In this space, a shattered set corresponds to an embedding of an \(\ell_1\) sphere (see \Cref{fig-DeltaH} below). This observation motivates the definition of the spherical dimension as the maximum dimension of a sphere within the space of realizable distributions. 

\begin{figure}[!hbt]
	\centering
		\begin{tikzpicture}
	[pt/.style={inner sep = 1.0pt, circle, draw, fill=black},
	bpt/.style={inner sep = 1.5pt, circle, draw=black, fill=black}]
		\begin{scope}[yscale=0.6, xscale=0.7]	
		\coordinate(3p) at (5, 0) {};
		\coordinate(1m) at (3,0) {};	
		\coordinate(2m) at (5,2) {};	
		\coordinate(1p) at (7,0) {};	
		\coordinate(2p) at (5,-2) {};
		
		
		\draw [thick, black] (1m)--(2m)--(1p)--(2p)--(1m); 
		
		\node[left] at (1m) {$(1, -)$};
		\node[left] at (2m.north) {$(2, -)$};
		\node[right] at (1p) {$(1, +)$};
		\node[right] at (2p.south) {$(2, +)$};
		\node[bpt] at (1p){};
		\node[bpt] at (2p){};
		\node[bpt] at (1m){};
		\node[bpt] at (2m){};
		
		
		\path (1m) to node(pD)[pt, pos=0.33]{} (2p);	
		\node(D) at (0.5, -1.5) {$\mu: \substack{(1, -)\text{  w.p. } 2/3 \\(2, +)\text{  w.p. } 1/3}$};
		\draw[gray, -stealth] (D.east)--(pD);
		\end{scope}
	\end{tikzpicture}
	\caption{Space of realizable distributions of a four-concept class shattering two points. 
	}
	\label{fig-DeltaH}       
\end{figure}
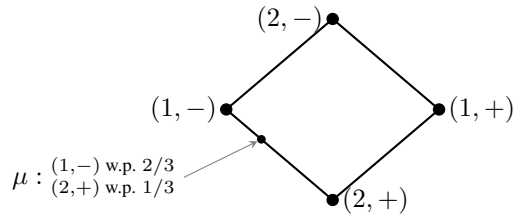

We will later demonstrate that this topological relaxation of the VC dimension has several disparate applications in learning theory, ranging from sample compression schemes to reductions from classification to optimization. But first, we present the formal definition. Let \(\mathcal{H}\) be a class of functions from domain \(X\) to label space \(Y = \{\pm 1\}\), and let \(\Delta = \Delta(\mathcal{H})\) be the space of all possible distributions on \(X \times Y\), realizable by \(\mathcal{H}\). We equip $\Delta$ with a \emph{total variation distance} $\TV$ between the distributions, turning it into a metric space. We then define a natural ``flipping the labels'' map $\rho \mapsto -\rho$ on the space of all distributions over \(X \times Y\), that is, for any event $E\subseteq X \times Y$, $\Pr_{-\rho}(E) = \Pr_\rho(-E)$, where $-E = \{(x, -y)~|~(x,y)\in E\}$. While \(\Delta\) is not necessarily closed under this map, we are concerned with a part of it that is. We thus define \(\Delta^{\ant} = \Delta \cap -\Delta\), on which this map becomes a continuous (with respect to $\TV$) fixed-point free involution\footnote{Recall that an involution is a function $f$ such that $f(f(x))=x$ for all $x$. A fixed-point free involution is an involution for which $f(x)\neq x$ for all $x$.}. Then, our main definition can be stated as:
\begin{definition}[Spherical dimension]\label{def-sd} We say that a finite class $\H$ \emph{admits an $n$-sphere} if there is a continuous map $f \colon \S^n \rightarrow \Delta^\ant_{\H}$ such that $f$ respects the antipodalities on $\S^n$ and $\Delta^\ant_{\H}$, that is, $f(-v) = -f(v)$ for all $v\in \S^n$. The \emph{spherical dimension} $\sd(\H)$ of $\H$ is then defined as a maximal integer $n$ such that $\H$ admits an $n$-sphere. 
\end{definition}
\noindent The restriction to finite classes here is for technical reasons. In~\Cref{sec-sd} we give the full version, as well as certain gradations of this definition. All the statements in the introduction, although formally correct, will be given ignoring the nuances of this gradation, and stated in full detail in their proofs. Let us now discuss the utility of the spherical dimension.

\subsection*{How we can use the spherical dimension} 
As of now, we know of four, rather diverse, applications of the spherical dimension in learning: (i) global stability and list replicability, (ii) reductions to stochastic convex optimization (SCO), (iii) embedding into maximum/extremal classes, (iv) and disambiguations of linear classifiers with margin. Here we outline these areas and applications of spherical dimension in them and push the technical discussion to Sections \ref{sec-lr}--\ref{sec-disamb} (in the same order). 

We note that, typically, $\sd$ is used to lower-bound the parameter of interest via an application of the Borsuk-Ulam (BU) theorem, or one of its relatives. The bounds that this application yields are usually linear, sometimes with a factor of $1/2$, and in many cases they are known to be sharp. We also note that while in the last two applications (list replicability and SCO) the connection to spherical dimension is rather straightforward (modulo published results and with some amount of technical work), the connection in the other two cases is, to our knowledge, either entirely new (disambiguations), or constitute substantial progress in a previously outlined direction (extremal classes).

\subsubsection*{Disambiguations of linear classifiers with margin}

In~\cite{PCS22} the authors initiated a systematic study of PAC learnability for partial concept classes\footnote{Although the first, to our knowledge, paper where this learning setting was considered is \cite{long01}.}. Their motivation for studying this setup was that it naturally captures some practical data-dependent assumptions, not covered by the standard PAC learning model. For example that the data is located on a low-dimensional subspace, or is separable with a margin (such as in the classical Perceptron algorithm~\citep{perceptron58}). 

While the authors did establish that, just as for the PAC learnability, the learnability of partial concept classes is equivalent to the finite VC dimension, they also note a range of stark differences specific to this setup. For example, while in the classical setting the learning can always be done by an \emph{empirical risk minimization} (ERM) algorithm, it is provably not so for partial concept classes. Moreover, there are easily learnable partial classes for which the class of hypotheses returned by the learner necessarily has an infinite VC dimension. 

The central notion in~\cite{PCS22} is a \emph{disambiguation} of a partial class $H$: this is a class~$H'$ of total functions such that each partial function in $H$ is extended by some function in~$H'$. 
A natural question to ask here is whether for a given partial class there is a disambiguation of comparable VC dimension. However, the authors proved that it is, in a strong sense, false: There is a partial class of $\VC=1$ such that any of its disambiguation has an infinite VC dimension. Effectively, the existence of such a class is the reason for the failure of ERM, along with some other impossibility results, for partial classes. This construction, however, relies on a recent breakthrough in graph theory and communication complexity (\cite{goos15,bendavid16,balodis23}) and provides little insight into the structure of such a class. They conjectured, however (Open Question~21), that the partial class of linear classifiers with margin, the very same class as used in connection to the Perceptron model, can give such a construction: for large margin, its VC dimension is trivially $1$ (independently of the euclidean dimension!), and for all ``natural'' disambiguations the VC grows with the dimension of the underlying space.

In this paper, we prove that modulo some technical details, their OQ~$21$: whether the class of linear classifiers with margin, for some nontrivial margin, can be disambiguated with a class of bounded VC dimension, is equivalent to asking if $\sd$ is bounded by a function of VC dimension (that is, if $\sd$ and $\VC$ go to infinity together). Moreover, our connection is not just qualitative, but quantitative, and, in fact, very precise:
\begin{theorem}[Rough equivalence between disambiguations and spherical dimension]\label{th-sd-disamb}~
For integers $a$, $b$, and~$n$, the following are equivalent:
\begin{itemize}
	\item There is a class with $\VC \leq a$, $\VC^{*\an}\leq b$, and $\sd_\simp \geq n$;
	\item There is a disambiguation of a class of linear classifiers of an $n$-dimensional unit sphere with some nontrivial margin, for which $\VC \leq a$ and $\VC^{*\an}\leq b$.
\end{itemize}
\end{theorem}


\subsubsection*{Embeddings into extremal classes}

When one has to study a complicated object, a very natural and common sense idea is to transform it, without losing too much precision, into a nice structured object. In mathematics, it is manifested, for example, in approximating smooth functions with polynomials, considering a convex hull of a set instead of the set itself, you name it.

In learning theory, especially from the perspective of one of the most long-standing open problems, the \emph{sample compression schemes conjecture}~\cite{floyd:95, Warmuth:03}, good candidates for such ``nice structured objects'' are \emph{extremal classes}\footnote{Extremal, or shattering-extremal, classes are those that meet Pajor's inequality with equality; they generalize a perhaps more known family of \emph{maximum} classes that reach equality in Sauer-Shelah-Perles bound, see \Cref{sec-extremal} for technical discussion.}. An unavoidably incomplete list of works along the lines of resolving sample compression conjecture by embedding a low-VC class into an extremal class of comparable VC dimension includes~\cite{Ben-David:98,Kuzmin:07,Rubinstein:09,Rubinstein:12,rubinstein15,moran2016labeled,chepoi:20,Chepoi:21,chepoi:22,Chalopin:22,Rubinstein:22,Chalopin:23}. While the aimed at ``comparable VC dimension'' would usually be linearly bounded, anything subexponential would also be good, at least in the sense of beating the best known bound for the sample compression in~\cite{MY16}. 

While for certain natural families of classes such embedding can be done (for example \cite{Rubinstein:22} constructed such embeddings of linearly bounded VC for intersection-closed classes), in recent work~\citep{chase2024dual} the authors proved that certain classes cannot be embedded into extremal without at least an exponential blowup in the VC dimension. Their main structural result, leading to this conclusion, was that extremal classes satisfy the inequality~$\VC^* \leq 2\VC + 1$ (contrasting with the general exponential upper bound on $\VC^*$ in terms of $\VC$, \cite{assouad83}). The proof of this bound in \cite{chase2024dual} was topological, utilizing the properties of \emph{cubical complexes} of extremal classes, as well as a generalization of the topological Radon theorem. Here we prove that this bound, again, for extremal classes, can be strengthened in terms of spherical dimension to $\sd \leq 2\VC-1$. Furthermore, this bound is a consequence of a strong topological similarity between the space of realizable distributions and the cubical complex of an extremal class, which, structurally, is our main result in this direction. \Cref{fig-EMBEDDING4} below illustrates this similarity. However, substantiating the meaning of this picture requires much deliberation and we will only do it in \Cref{sec-extremal-proofs}.

\begin{figure}[hbt]
\centering
\begin{tikzpicture}
[
pt/.style={inner sep = 0.9pt, circle, draw, fill=black},
ptb/.style={inner sep = 0.9pt, circle, draw, fill=black},
ptl/.style={inner sep = 1.5pt, circle, draw, fill=black},
ptbl/.style={inner sep = 1.5pt, circle, draw=blue, fill=blue}
]

\begin{scope}[yscale=0.7, xscale = 1]  
    \coordinate (+**) at (0,0);
    \coordinate (*+*) at (4,0);
    \coordinate (**+) at (2,3);
    \coordinate (-**) at (6,3);
    \coordinate (*-*) at (4,6);
    \coordinate (**-) at (8,6);
        
    \draw[fill=gray, fill opacity=0.1] (+**)--(**+)--(*+*)--cycle;
    \draw[fill=gray, fill opacity=0.1] (*-*)--(**-)--(-**)--cycle;
    \draw[fill=gray, fill opacity=0.1] (**+)--(*-*)--(-**)--(*+*)--cycle;

    \draw[thick] (+**) to node[ptb, midway] (+*+){} (**+)
    	to node[ptbl, midway] (*++){} (*+*)
    	to node[ptb, midway] (++*){} (+**);
	\node[ptbl](+++) at (2, 1.1) {};
    \draw[thick] (-**) to node[ptb, midway] (-*-){} (**-)
    	to node[ptb, midway] (*--){} (*-*)
    	to node[ptbl, midway] (--*){} (-**);
	\node[ptbl](---) at (6, 4.9) {};

	\draw(**+) to node[pt, midway](*-+){} (*-*);
	\draw(**+) to node[ptbl, midway](-*+){} (-**);
	\draw(*+*) to node[pt, midway](-+*){} (-**);

	\node[ptbl](--+) at (4, 4.1) {};
	\node[ptbl](-++) at (4, 1.9) {};

	\node[ptb] at (+**) {};
	\node[ptb] at (**+) {};
	\node[ptb] at (*-*) {};
	\node[ptb] at (**-) {};
	\node[ptb] at (-**) {};
	\node[ptb] at (*+*) {};

    \node [left] at (+**) {\scriptsize$+**$};
    \node [left] at (**+) {\scriptsize$**+$};
    \node [left] at (*-*) {\scriptsize$*-*$};
    \node [right] at (**-) {\scriptsize$**-$};
    \node [right] at (-**) {\scriptsize$-**$};
    \node [right] at (*+*) {\scriptsize$*+*$};
    
    \node [below] at (++*) {\scriptsize$++*$};
    \node [below] at (+++) {\scriptsize$+++$};
    \node [above] at (*--) {\scriptsize$*--$};
    \node [above] at (---) {\scriptsize$---$};
    \node [left] at (+*+) {\scriptsize$+*+$};
    \node [left] at (*-+) {\scriptsize$*-+$};
    \node [right] at (-*-) {\scriptsize$-*-$};
    \node [right] at (-+*) {\scriptsize$-+*$};

    \node [right] at (-++) {\scriptsize$-++$};
    \node [left] at (--+) {\scriptsize$--+$};
    \node [above right] at (-*+) {\scriptsize$-*+$};
    \node [below right] at (*++) {\scriptsize$*++$};
    \node [above left] at (--*) {\scriptsize$--*$};
    
    \draw[ultra thick, blue, double] (+++)--(*++) (--*)--(---);
    \draw[ultra thick, blue, double] (*++)--(-++)--(-*+)--(--+)--(--*);
    
    \draw[dotted] (+++)--(*+*) (+++)--(++*) (+++)--(+**) (+++)--(+*+) (+++)--(**+)
    	(---)--(*-*) (---)--(*--) (---)--(**-) (---)--(-*-) (---)--(-**)
    	(-++)--(**+) (-++)--(-**) (-++)--(-+*) (-++)--(*+*)
    	(--+)--(**+) (--+)--(*-+) (--+)--(*-*) (--+)--(-**);
\end{scope}

\end{tikzpicture}
\caption{Cubical complex (in blue) of a threshold on $3$ points
 $\E = \{---, --+, -++, +++\}$, embedded into its space of realizable distributions.} 
\label{fig-EMBEDDING4}
\end{figure}
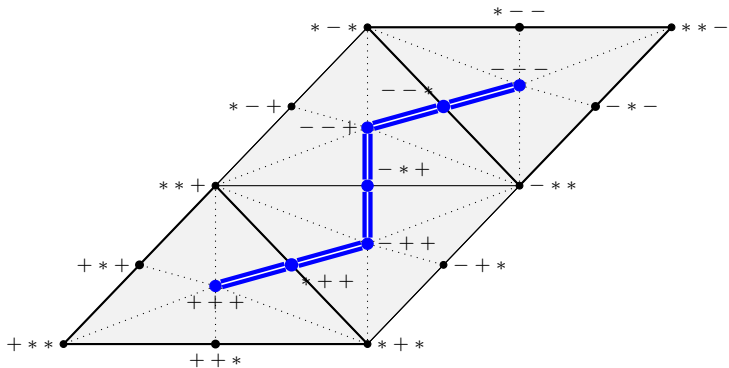

Curiously enough, this result, together with a well-known fact that every $\VC=1$ class can be extended to an extremal $\VC=1$ class, when applied to the setup of disambiguations, proves that any class that disambiguates classifiers with a margin on a $2$-dimensional sphere should have VC dimension at least $2$.

We finally note that, apart from the above connection to the sample compression schemes conjecture, extremal classes found to be useful in other problems in learning theory, such as proper optimal learners~\citep{bousquet:20}, and also in other areas of mathematics such as extremal combinatorics~\citep{Bollobas:89, Bollobas:95}, discrete geometry~\citep{lawrence:83} and functional analysis~\citep{Pajor:1985} (on many occasions, these classes were independently rediscovered, see~\cite{Moran:12}).

\subsubsection*{Global stability and list replicability}
The study of replicability in PAC learning was initiated in the work of \cite{impagliazzo2022reproducibility} to answer, on the theory side, the growing demand for reproducibility in science~\citep{baker2016}. Their definition of replicability, however, assumed shared internal randomness between the runs of the learner. As observed in~\cite{dixon2023list}, without this assumption replicability becomes essentially non-achievable and, to overcome this obstruction, they introduced \emph{list replicability}, on which we concentrate in this paper. Concurrently, \cite{Chase23rep} proved an equivalence (by reciprocity) of list replicability with \emph{global stability}, introduced in \cite{BLM20}, where it was an essential ingredient in the proof of establishing the differentially-private (DP) learnability of online-learnable classes (together with the earlier results in \cite{ALMM19}, this established an equivalence between DP and online learnability). We also note that, apart from an independent interest and the above connection to DP, the notions of stability and replicability also have deep ties to pseudo-determinism~\citep{GG11}. We refer to~\cite{KKVZ24} for a further review of the field, with a focus on computational efficiency.

It was then observed in~\cite{chase2024local} (with a precursor in~\cite{Chase23rep}) that for finite classes the list replicability number $\List$ has a characterization in terms of overlap of certain closed covers of the space of realizable distributions. Further down this road, they showed that $\List$ can be linearly lower-bounded by both VC and dual VC (VC$^*$) dimensions, even for weak learners. In this paper, we show that both these bounds are subsumed by the bound $\sd \leq 2\List - 3$, which, similarly, holds even for weak learners. We note, however, that we consider~\cite{chase2024local} as a crucial stepping stone for our present setup.

We further note that the same topological approach was used in \cite{chase2024local} to show that, surprisingly, \emph{agnostic} replicable learning is impossible (unless in trivial cases). A recent work~\citep{BGHH25} also elaborates on some relaxations of the notion of replicability in the agnostic setting.

\subsubsection*{Reductions to stochastic convex optimization}
While, from the statistical perspective, PAC learnability is fully characterized by the VC dimension, things get much more complicated once we also consider a computational aspect of learning (see, for example, Chapter~8 in~\cite{shais2014}). One reasonable way of dealing with it is by \emph{reducing} the learning problem of interest to the one that can be efficiently solved~\citep{PittW90}. This venue was recently explored in~\cite{reductions}, where the authors focused on reductions to learning with halfspaces~\citep{ForsterSS01, Ben-DavidES02, LinialS08, KamathMS20} or stochastic convex optimization (SCO) problems~\citep{Shalev-ShwartzSSS09}, with a special focus on the, broadly understood, dimensional complexity of such reductions.	

In particular, in~\cite{reductions} the authors proved that the reduction of a class to a $d$-dimensional SCO problem requires $d$ to be at least the VC dimension of this class and, somewhat counterintuitive to the natural interpretation of VC as the ``number of parameters'', for certain classes requires $d$ to be at least exponential in VC. Their approach combines the Borsuk-Ulam theorem with convexity considerations and thus fits naturally into our framework. In particular, here we prove that their result can be refined in terms of lower-bounding an SCO dimension of a class (defined as a minimal $d$ for which the above reduction is possible) with $\sd$. We further note that the SCO dimension can, in turn, be seen as a refinement of \emph{dimension complexity}(see, e.g.~\citep{KamathMS20}), also known, especially from the communication complexity side, as \emph{sign-rank}~\citep{paturi86, Forster01, AMY16, HatamiHM22}, resulting in the bound $\sd-1 \leq \SCO\leq\sr$.

We finally note that the study of reductions in~\cite{reductions} is broader than the particular thread that we concentrate on here and also extends to probabilistic and approximate reductions which, to our knowledge, do not fall under our current perspective.

\subsection*{What do we know about the spherical dimension} 
As it was already argued in~\cite{chase2024local}, the spherical dimension is lower-bounded by both VC and VC$^*$ dimensions. Here we refine the VC$^*$ bound to the one in terms of a certain refinement VC$^{*\an}$ of VC$^*$, which is more natural from the topological perspective, see \Cref{sec-prelim}:
\begin{lemma}\label{lem-vc-lb}
	For a class $\H$, $\sd(\H)\geq \VC(\H) - 1$ and $\sd(\H)\geq \VC^{*\an}(\H) - 2$.
\end{lemma}
We also know how to construct classes of spherical dimension asymptotically larger than both VC and VC$^*$, using \emph{products of classes}:
\begin{lemma}\label{lem-vcvc*-lb}
	For classes $\H_1$ and $\H_2$, $\VC(\H_1 \times \H_2) = \VC(\H_1) + \VC(\H_2)$ and $\sd(\H_1 \times \H_2) \geq \sd(\H_1) + \sd(\H_2) + 1$. In particular, for an $m$'th direct power $\U^m_n$ of the universal class $\U_n$, we have $\VC(\U^m_n) = m\lfloor\log n\rfloor$, $n\leq \VC^*(\U^m_n) \leq n + \lfloor\log m\rfloor$, and $m(n-1) - 1 \leq \sd(\U^m_n) \leq mn$.

	In particular, for the class $\U_n^n$, $\VC(\U^n_n) \sim \VC^*(\U^n_n) \sim n$ (up to a log factor) and $\sd(\U^n_n) \gtrsim n^2$. 
\end{lemma}
In \Cref{sec-proofs-basic}, we formulate Lemmas~\ref{lem-vc-lb} and~\ref{lem-vcvc*-lb} in an extended form, explicating some values for the classes $\C_n$ of \emph{binary hypercubes} and \emph{universal classes} $\U_n$ (minimal classes witnessing VC and VC$^*$). It is also notable that $\sd$ enables to provide a fine-tuned classification of low-VC classes:
\begin{lemma}\label{lem-trivial}
	Let $\H$ be a nonempty class over a finite domain. Then one of the following mutually exclusive cases holds:
	\begin{itemize}
		\item $\H$ is a class with one hypothesis ($\sd = -1$, $\VC=0$);
		\item $\H$  has at least two hypotheses and, up to a bit-flip, is a subclass of thresholds ($\sd = 0$, $\VC=1$);
		\item $\H$ is a $\VC=1$ class that is not, up to a bit-flip, a subclass of thresholds ($\sd = 1$, $\VC=1$);
		\item $\VC(\H)\geq 2$ and $\sd(\H)\geq 1$.
	\end{itemize}
\end{lemma}

At the same time, we have very little understanding of how the spherical dimension can grow with VC and VC$^*$. The following three conjectures (from the easiest to the hardest) summarize what we expect to see and are based mostly on the examples in Lemma~\ref{lem-vc-lb} and Lemma~\ref{lem-vcvc*-lb}. However, we have no means in sight of proving even the first most basic one (or disproving the hardest).
\begin{conjecture}\label{conj-sd-vc-1}
	VC and spherical dimension go to infinity together. That is, there is increasing $f\colon \N\rightarrow \N$ such that for any class $\H$, $\sd(\H)\leq f(\VC(\H))$.
\end{conjecture}

\begin{conjecture}\label{conj-sd-vc-2}
	Spherical dimension is at most exponential in both $\VC$ and $\VC^*$.
\end{conjecture}

\begin{conjecture}\label{conj-sd-vc-3}
	Up to log factors, the spherical dimension is upper-bounded by $\VC\cdot\VC^*$.
\end{conjecture}
We find it intriguing that the only known to us upper-bound of the form $\VC\cdot\VC^*$ is from the famous construction of sample compression schemes~\citep{MY16}; just as in Conjecture~\ref{conj-sd-vc-3}, this bound is up to $\log$-factors. This parallel is further reinforced by the results in \Cref{sec-extremal}, where we use the spherical dimension to improve $\VC^*\leq 2\VC + 1$ bound for extremal classes from~\cite{chase2024dual}, which, in turn, was used to prove that a certain attack on the sample compression schemes conjecture cannot succeed.

Our last theorem is the best we were able to achieve in the direction of proving Conjecture~\ref{conj-sd-vc-1}. 
\begin{theorem}\label{th-sd-vc-ub}
	If $\VC(\H)\leq 1$ then $\sd(\H) \leq 1$. Alternatively, if $\sd(\H) \geq 2$ then $\VC(\H)\geq 2$.
\end{theorem}
Despite the modest statement, the proof relies on a highly nontrivial relation of spherical dimension and extremal classes, see~\Cref{sec-extremal}, specifically~\Cref{t-sd-for-extremal}. Note also while Theorem~\ref{th-sd-vc-ub} formally follows from the classification in Lemma~\ref{lem-trivial}, the actual inference goes the other way round: Theorem~\ref{th-sd-vc-ub} is a hard result, which enables the classification in Lemma~\ref{lem-trivial}.

\begin{figure}[!hbt]
	\centering
	\begin{tikzpicture} 
[
pt/.style={inner sep = 0.0pt, circle, draw, fill=black},
point/.style={inner sep = 1.2pt, circle,draw,fill=black},
spoint/.style={inner sep = 1.2pt, circle,draw,fill=white},
mpoint/.style={inner sep = 0.7pt, circle,draw,fill=black},
ypoint/.style={inner sep = 3pt, circle,draw,fill=yellow},
xpoint/.style={inner sep = 3pt, circle,draw,fill=red},
snakeit/.style={decorate, decoration=snake}
]

\begin{scope}[yscale=1, xscale = 1]
	\node[anchor=south](sdinf) at (-3,4) {$\sd_\infty$};
	\node[anchor=south](sd) at (0,2) {$\sd$};
	\node[anchor=south](sdsimp) at (0,0.5) {$\sd_\simp$};
	\node[anchor=south](disamb) at (3,0.5) {$\disamb$};

	\draw(sdinf) to coordinate[below left, pos=0.4](fintext) (sd);
	\draw (sd)--(sdsimp);
	\draw[snakeit, double] (sdsimp)--(disamb);
		
	\node[anchor=south](vc) at (1,-1.5) {$\VC - 1$};
	\node[anchor=south](vc*s) at (-1,-1.5) {$\VC^{*\an} - 2$};
	\node[anchor=south](vc*) at (-1,-2.5) {$\VC^{*} - 2$};
	\node[anchor=south](2vc*) at (-2,0) {$2\VC^{*} - 1$};
		
	\draw(vc)--(sdsimp) (vc*s)--(sdsimp) (vc*)--(vc*s)--(2vc*);

	\node[anchor=south](sco) at (-1, 4) {$\SCO - 1$};
	\node[anchor=south](sr) at (-1, 5.25) {$\sr - 1$};
	\node[anchor=south](2sr) at (1.5, 6.5) {$2~\sr - 1$};
	\node[anchor=south](vcextr) at (1.5, 4) {$2\VC^\extr - 1$};
	\node[anchor=south](lreps) at (4.5, 4) {$2\List_{\varepsilon <1/2} - 3$};
	\node[anchor=south](lr) at (4.5, 5.25) {$2\List - 3$};
		
	\draw (sd)--(vcextr);
	\draw (lr)--(lreps);
	\draw (sd)--(sco)--(sr)--(2sr) (vcextr)--(2sr);
	\draw [blue] (lreps) to coordinate[pos=0.5](x)(sd);
	\node[below, text width=1.3cm] at ($(fintext.south) - (0.1,0.1)$) {\scriptsize{Collapses \\[-0.05cm] for finite\\[-0.2cm] classes}};
	\node[blue, below right] at (x) {\scriptsize{Unless $\sd \leq 0$}};
\end{scope}

\end{tikzpicture}	
	\caption{Hasse diagram for spherical dimensions and related values. For any class $\H$, the respective values are non-decreasing with respect to this partial order. The wavy line indicates rough equivalence. The parameter $\disamb$ is defined in the proof of~\Cref{th-sd-disamb}; roughly, how big of a sphere can a class disambiguate.}
	\label{fig-INEQUALITIES}       
\end{figure}
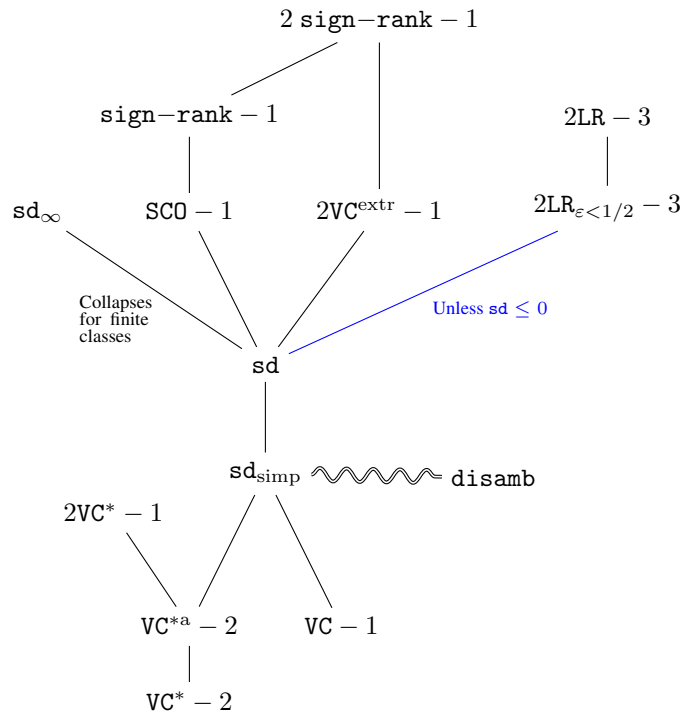

We finish the section by giving in \Cref{fig-INEQUALITIES} below the diagram summarizing the results regarding spherical dimensions and the parameters upper or lower-bounded by it. 

\subsection*{Paper overview}
The paper brings together several not-so-close domains, so it is inevitably definition-heavy. In \Cref{sec-prelim}, we outline the basic definitions from learning theory (realizable distributions, loss function, VC dimension) and topology (simplicial complexes, antipodal spaces). In~\Cref{sec-sd}, we introduce and discuss several versions of the spherical dimension. And in \Cref{sec-applications} we discuss in more detail the four applications of the spherical dimension, outlined above: disambiguations, extremal classes, list replicability, and reductions to SCO.

The rest of the paper is moved to the appendices, although even there we continue to introduce additional terminology whenever it is either more in-depth or domain-specific. This way, in \Cref{sec-background} we elaborate upon barycentric subdivisions and joins. Then, in \Cref{sec-proofs-basic}, we prove some of the technical results, such as Assouad's bounds for antipodal shattering (\Cref{sec-bounds}) and classification of low-VC classes vie spherical dimension (\Cref{sec-low-dim}), and also elaborate on the, more or less straightforward, application of spherical dimension to list replicability (\Cref{sec-lr-proofs}) and reductions to SCO (\Cref{sec-sco-proofs}).
Finally, Sections~\ref{sec-disamb-proofs} and~\ref{sec-extremal-proofs} contain proofs of main theorems regarding disambiguations (\Cref{sec-disamb}) and embeddings into extremal classes (\Cref{sec-extremal}). 
 
 \subsection*{Acknowledgments}
The authors wish to thank Ron Holzman, Noga Alon, and  Roy Meshulam for fruitful discussions of the topics addressed in this paper and of related subjects.

\section{Preliminaries}\label{sec-prelim}
Notationwise, for an integer $n$, we write $[n]$ to denote the set $\{1, \dots, n\}$. We also write $A\sqcup B$ to denote a disjoint union of the sets $A$ and $B$. 

\subsection{Learning theory toolbox}\label{sec-learning-toolbox}
Here we pay the technical dept from \Cref{sec-prelim} and formally define all the notions outlined in it. 

We usually consider a class $\H\subseteq Y^X$, where $X$ is called the \emph{domain} and $Y = \{-, +\}$ the \emph{label space} of $\H$. A function $h\in Y^X$ is called a \emph{hypothesis}, and $h\in \H$ is called a \emph{concept} of $\H$.  Unless stated otherwise, we always consider the class $\H$ to be nonempty, that is, containing at least one hypothesis. For a hypothesis $h$ and a distribution $\mu$ over $X\times Y$, we define the \emph{$0/1$~population loss} $L_\mu(h)$ of $h$ with respect to $\mu$ as
	$$L_\mu(h) = \Pr_{(x,y)\sim\mu}[h(x)\neq y].$$
The distribution $\mu$ is called \emph{realizable} by $\H$ if 
	$$\inf_{h\in \H} L_\mu(h) = 0.$$
In case $X$ is finite, the definition of realizability can be rectified as: $\mu$ is realizable by $\H$ if there is $h\in \H$ such that $h(x) = y$ for all $(x, y)\in\supp(\mu)$. Here $\supp(\mu)$ is the \emph{support} of~$\mu$, $\supp(\mu) = \{(x, y)~|~\Pr_\mu(x,y) > 0\}$.
	
For a subset $E\subseteq X\times Y$, we define $-E = \{(x, -y)~|~(x,y)\in E\}$. For a distribution $\mu$ over $X \times Y$, we define $-\mu$ as a distribution defined by 
$P_{-\mu} \left( E \right) = P_{\mu}(-E)$
for all measurable $E$. In particular, if $X$ is finite, this is equivalent to saying that
$P_{-\mu}(x,y) = P_{\mu}(x,-y)$,
for all $x,y\in X\times Y$. We call $\mu\mapsto -\mu$ an \emph{antipodality map} on the set of distributions over $X\times Y$. 
	
Let $\Delta(\H)$ be the space of all distributions realizable by $\H$, equipped with the \emph{total variation distance}
\begin{align*}
 \TV(\mu,\nu)=\sup_{E} |\mu(E)-\nu(E)|,
\end{align*}
where the supremum is over all measurable events $E$. It is easy to see that $\Delta(\H)$ is closed and the antipodality map $\mu \mapsto -\mu$ is continuous with respect to $\TV$. Let $\Delta^\ant(\H) = \Delta(\H) \cap -\Delta(\H)$. Trivially, $\Delta^\ant$ is a closed subset of $\Delta$, which is closed under the antipodality map, that is, such that $-\mu\in \Delta^\ant$ whenever $\mu\in \Delta^\ant$. The following easy property turns out to be quite handy in this setup: For any distribution $\mu$ and hypothesis $h$, it holds
\begin{align}\label{eq-loss}
L_\mu(h) + L_{-\mu}(h) = 1.
\end{align}
Informally, no hypothesis can be any good for both the distribution and its antipodal. In particular, it easily implies that the antipodality is \emph{fixed-point free} on $\Delta$, that is, for no $\mu\in \Delta$, $\mu = -\mu$.

For a hypothesis $h$ over $X$ and $S\subseteq X$, a \emph{restriction} $h|_S$ of~$h$ to $S$ is a hypothesis $h'$ over $S$ defined as $h'(s) = h(s)$ for all $x\in S$. Similarly, a restriction of $\H$ to $S$ is defined as $\H|_S = \{h|_S~|~h\in \H\}$.
A class $\H$ over~$X$ \emph{shatters} a set $S\subseteq X$ if for every \emph{pattern} $s$ on $S$, that is, for every $s\colon S\rightarrow \{-, +\}$, there is $h\in \H$ such that $h|_S = s$ (by convention, the empty set is always shattered). The VC dimension of $\H$, denoted by $\VC(\H)$, is the largest size of a set shattered by $\H$. 

For a class $\H$ over $X$, the \emph{dual class} $\H^*$ of $\H$ is a class on a domain $\H$ whose functions are parametrized by $X$, and, for $x\in X \sim x\in \H^*$ and $h\in \H$, it holds $x(h) = h(x)$. The dual VC dimension of $\H$, $\VC^*(\H)$, is defined as the VC dimension of $\H^*$, that is, $\VC^*(\H) = \VC(\H^*)$. It also can be educational to define $\VC^*$ explicitly. A subset $H\subseteq \H$ is dually-shattered if for every \emph{dual pattern} $r\colon H\rightarrow \{-, +\}$ there is $x\in X$ such that $r(h) = h(x)$ for all $h\in H$. Just as before, $\VC^*(\H)$ is the maximal size of the set, dually shattered by $\H$. 

Let us now define a relaxation of the VC dimension that works better with respect to our topological approach. We say that a class $\H$ on a domain $X$ \emph{antipodally shatters}\footnote{
Antipodal shattering, under the same name, and antipodal VC, under the name \emph{dual sign-rank}, was introduced in \cite{AMY16}. Their reasons for it, however, seem different from ours: As one can guess from the name, they defined a parameter dual to the sign-rank. However, ``dual'' here has a meaning different from ours.  
} 
 $S\subseteq X$ if for every binary pattern $s$ on $S$ there is $h\in \H$ such that either $h|_S = s$ or $h|_S = -s$. We define \emph{antipodal VC dimension} $\VC^\an(\H)$ as the maximal size of a set, antipodally shattered by $\H$. The \emph{dual antipodal VC dimension} $\VC^{*\an}(\H)$ is defined dually. We note that, as a matter of fact, the VC dimension is good for us as it is, but dual antipodal VC behaves better than dual VC, so we will mostly use only the latter relaxation.
 
 While $\VC^\an$ and $\VC^{*\an}$ clearly are relaxations of $\VC$ and $\VC^*$, in particular, $\VC \leq \VC^\an$ and $\VC^*\leq \VC^{*\an}$, it is noteworthy that Assouad's bounds $\left\lfloor\log \VC \right\rfloor \leq \VC^* \leq 2^{\VC + 1} - 1$ (see~\cite{assouad83}) can be extended to the antipodal VC dimensions:
\begin{lemma}[Assouad's bounds for antipodal shattering]\label{prop-dual-semi-VC}
	For a (partial) class $\H$, it holds
	$$\left\lfloor\log \VC(\H) \right\rfloor \leq \left\lfloor\log \VC^\an(\H) \right\rfloor \leq \VC^*(\H) \leq \VC^{*\an}(\H)\leq 2^{\VC(\H) + 1} - 1,$$
and $$\VC^*(\H) \leq \VC^{*\an}(\H) \leq 2\VC^*(\H) + 1.$$	
Moreover, all inequalities between the neighboring values above are sharp (i.e., for each of these inequalities there exist classes for which they hold with equality).
\end{lemma}

\subsection{Topological toolbox}\label{sec-topology-toolbox}
As it is customary, we consider all maps between topological spaces to be continuous without saying it explicitly. Topological spaces $X$ and $Y$ are called \emph{homeomorphic}, denoted $X\cong Y$, if there is an isomorphism $f\colon X\rightarrow Y$ such that both $f$ and $f^{-1}$ are continuous. We usually think of homeomorphic spaces as of the same space. We will now briefly review some basic definitions from simplicial geometry, following Chapter~1 in~\cite{matousek}. All sets in the definitions of simplicial complexes below are assumed to be finite.

An \emph{abstract simplicial complex} $K$ is a pair $(V, S)$, where $V = V(K)$ is a set of \emph{vertices} of $K$, and $S = S(K) \subseteq 2^V$ is a \emph{downward closed} set of simplices of $K$. Here downward closed means that for any $s\in S(K)$ and any $t\subseteq s$, it holds $t\in S(K)$. An $s\in S(K)$ is called an \emph{abstract simplex}, and its \emph{dimension} is $\dim(s) = |s| - 1$; that is, for example, a $2$-dimensional simplex contains $3$ vertices. A simplex $s\in S(K)$ is called \emph{maximal} if it is not a proper subset of any other simplex in $K$. We also call a $k$-dimensional simplex a \emph{$k$-simplex}. The dimension of $K$ is the maximal dimension of its simplices. 

A \emph{geometric simplex} $\sigma$ in the \emph{ambient space} $\R^d$ is a convex hull of an affinely independent set $A\subseteq \R^d$; $A$ is called the set of \emph{vertices} of $\sigma$, $A = V(\sigma)$. In this case, the \emph{dimension} of $\sigma$ is $k = \dim(\sigma) = |A| - 1$ and $\sigma$ is called a \emph{$k$-simplex}. A convex hull of an arbitrary subset of $V(\sigma)$ is called a \emph{face} of $\sigma$, and is itself a simplex. A geometric simplicial complex $\Omega$ is a pair $(V, S)$, where $V\subseteq \R^d$ is a subset of $\R^d$, called the set of \emph{vertices} of $\Omega$, and $S$ is a set of simplices in $\R^d$, called the set of simplices of $\Omega$, such that:
\begin{itemize}
	\item For every simplex $\sigma \in S(\Omega)$, $V(\sigma) \subseteq V(\Omega)$;
	\item Each face of every simplex $\sigma\in S(\Omega)$ is a simplex of $\Omega$;
	\item The intersection $\sigma_1 \cap \sigma_2$ of any two simplices in $S(\Omega)$ is a face of both $\sigma_1$ and $\sigma_2$.
\end{itemize}
The \emph{dimension} of $\Omega$ is the maximal dimension of its simplices. The union of all simplices of $\Omega$, denoted $\|\Omega\|$, is called the \emph{polyhedron} of $\Omega$. 

A geometric simplicial complex $\Omega$ is called a \emph{geometric realization} of an abstract simplicial complex $K$ if there is an isomorphism $f\colon V(K)\rightarrow V(\Omega)$ such that $\{f(s)~|~s\in S(K)\} = \{V(s)~|~s\in S(\Omega)\}$. With some ambiguity, we then call the topological space $\|\Omega\|$ a \emph{geometric realization} of $K$ and, for $s\in K$, define a \emph{geometric realization} of $s$, denoted $\|s\|$, as a convex hull of the set $\{f(v)~|~v\in s\}$; note that $\|s\| \in S(\Omega)$. It is well-known that for any two geometric realizations $\Omega_1$ and $\Omega_2$ of $K$, $\|\Omega_1\|$ is homeomorphic to $\|\Omega_2\|$, which enables us to talk about \underline{the} geometric realization $\|K\|$ of $K$.

A notion of the geometric simplicial complex is somewhat transitory, enabling us to unify the combinatorial information captured by the abstract simplicial complex $K$ with the geometric information, captured by the space $\|K\|$. Thus, we will usually say simply \emph{simplex} and \emph{simplicial complex} to mean \emph{abstract simplex} and \emph{abstract simplicial complex}. For simplicial complexes $K$ and $L$, a \emph{simplicial map} is a map $f\colon V(K)\rightarrow V(L)$ that maps simplices into simplices. Then $f$ is canonically extended ``by linearity'' to a continuous map $f\colon \|K\|\rightarrow \|L\|$, where, as it is common, we denote the extended map by the same name $f$. It is also well-known and easy to check that the extension $f\colon \|K\|\rightarrow \|L\|$ is an embedding (that is, one-to-one) if and only if $f\colon V(K)\rightarrow V(L)$ is.

For a simplicial complex $K$ and $x\in \|K\|$, let $s\in S(K)$ be a simplex such that $x\in \|s\|$; trivially, for every $x$ there is always at least one such $s$. Then the \emph{barycentric coordinates} of $x$ in $s$ is a unique function $\alpha\colon V(s) \rightarrow [0,1]$ such that $\sum_{v\in V(s)} \alpha(v) = 1$ and $\sum_{v\in V(s)} \alpha(v) \cdot v = x$, where in the second sum $v$'s are assumed to lie in the ambient space $\R^d$. It is easy to see that for $x\in \|K\|$ there is a unique simplex $s\in S(K)$ such that $x\in \|s\|$ and $x\notin \|t\|$ for any proper subsimplex $t$ of~$s$. This $s$ is called the \emph{support} of $x$, denoted $s = \supp(x)$. It is also easy to see that $s$ is the unique simplex containing $x$, such that all barycentric coordinates of $x$ in $s$ are nonzero.

For a topological space $X$, a map $f\colon X\rightarrow X$ is called an \emph{antipodality map} if it is continuous; fixed-point free, that is, $f(x) \neq x$ for all $x\in X$; and involutive, that is, $f(f(x)) = x$ for all $x\in X$. 
An \emph{antipodal space} is a pair $(X, \nu)$ where $X$ is a topological space and $\nu\colon X\rightarrow X$ is an antipodality map. For antipodal spaces $(X, \nu)$ and $(Y, \omega)$, a continuous map $f\colon X\rightarrow Y$ is called \emph{equivariant} if it commutes with antipodalities, that is, if $f\circ \nu(x) = \omega\circ f(x)$ for all $x\in X$. 
We say that $(X, \nu)$ and $(Y, \omega)$ are homeomorphic \emph{as antipodal spaces}, denoted $(X, \nu) \cong (Y, \omega)$, if there is an equivariant homeomorphism between them, that is, if there is an isomorphism $f\colon X\rightarrow Y$ such that both $f$ and $f^{-1}$ are continuous, and $f\circ \nu(x) = \omega \circ f(x)$ for all $x\in X$. We note that this definition is a verbal repetition of Definition~5.2.1 in~\cite{matousek}, where antipodal space is called free $\ZZ_2$-space.

\section{Spherical dimension}\label{sec-sd}
\subsection{Antipodal spheres}\label{sec-spheres}
By an $n$-dimensional sphere $\S^n$ we understand the standard $n$-dimensional unit sphere $\S^n = \{v\in \R^{n+1}~|~\|v\| = 1\}$, which we typically treat as an antipodal space with its standard antipodality map $v\mapsto -v$.

For a simplicial complex $\Q$, we say that a vertex map $\nu\colon V(\Q) \rightarrow V(\Q)$ is an antipodality map if it is i) simplicial, ii) an involution, and iii) there is no simplex containing a vertex and its $\nu$-image. Similarly to antipodal space, we define an \emph{antipodal simplicial complex} as a pair $(\Q, \nu)$.
Finally, we say that an antipodal simplicial complex $(\Q^n, \nu)$ is a \emph{simplicial $n$-sphere} if the antipodal space $(\|\Q^n\|, \nu)$, where $\nu$ is extended to $\|\Q^n\|$ by linearity, is homeomorphic, as an antipodal space, to $(\S^n, v\mapsto -v)$. The fact that we require this to be an isomorphism of antipodal spaces is not negligible, as spheres, in general, admit non-standard antipodalities. That is, $\S^n$ can be equipped with an antipodality $\omega$ such that $(\S^n, \omega)\not\cong (\S^n, v\mapsto -v)$, see~\cite{hirsch64}. In particular, it might be the case that $\|\Q^n\| \cong \S^n$, but $(\|\Q^n\|, \nu) \not\cong (\S^n, v\rightarrow - v)$.

\subsection{Simplicial 
structure of 
\texorpdfstring{$\Delta_\H$}{DH}}
\label{sec-simplicial-D}

We say that the class $\H$ is \emph{finitely supported} if its domain $X$ is finite. In this case, $\Delta_\H$ is naturally a geometric simplicial complex with vertices $X\times Y$ and maximal simplices $\H$, where $h\in \H$ is treated as a subset of $X\times Y$: a distribution $\mu\in \Delta_\H$ corresponds to a point in the geometric realization $\|\supp(\mu)\|$ of the simplex $\supp(\mu)$ with barycentric coordinates $\{\mu(x,y)~|~(x,y)\in \supp(\mu)\}$\footnote{Let us note the potential ambiguity: here $\supp(\mu)$ as defined in \Cref{sec-learning-toolbox}, for $\mu$ treated as a realizable distribution, coincides with $\supp(\mu)$ as defined in \Cref{sec-topology-toolbox}, for $\mu$ treated as a point in the geometric simplicial complex $\Delta_\H$.}. In particular, the dimension of all maximal simplices of $\Delta_\H$, and, consequently, the dimension of $\Delta_\H$ itself, is $|X| - 1$. Moreover, $\Delta_\H$ consists of a finite number of simplices, and hence compact. Instead of doing a somewhat tedious formal proof of this, rather obvious, fact, we take a ``proof by picture'' approach and illustrate the simplicial structure of $\Delta_\H$ in \Cref{fig-DeltaH2} below.

\begin{figure}[!hbt]
	\centering
		\begin{tikzpicture}
	[pt/.style={inner sep = 1.0pt, circle, draw, fill=black},
	bpt/.style={inner sep = 1.5pt, circle, draw=blue, fill=blue},]
		\begin{scope}[xscale=1.2]	
		\coordinate(3p) at (5, 0) {};
		\coordinate(1m) at (3,0) {};	
		\coordinate(2m) at (5,2) {};	
		\coordinate(1p) at (7,0) {};	
		\coordinate(2p) at (5,-2) {};
		
		\draw [draw=black, fill=yellow, fill opacity=0.2] (3p)--(1m)--(2m) -- cycle;
		\draw [draw=black, fill=yellow, fill opacity=0.2] (3p)--(1p)--(2m) -- cycle;
		\draw [draw=black, fill=yellow, fill opacity=0.2] (3p)--(1p)--(2p) -- cycle;
		\draw [draw=black, fill=yellow, fill opacity=0.2] (3p)--(1m)--(2p) -- cycle;
		
		\draw [ultra thick, blue] (1m)--(2m)--(1p)--(2p)--(1m); 
		
		\node[left] at (1m) {$(1, -)$};
		\node[left] at (2m.north) {$(2, -)$};
		\node[right] at (1p) {$(1, +)$};
		\node[right] at (2p.south) {$(2, +)$};
		\node[below right] at (3p) {\small$(3, -)$};
		\node[pt] at (3p){};
		\node[bpt] at (1p){};
		\node[bpt] at (2p){};
		\node[bpt] at (1m){};
		\node[bpt] at (2m){};
		
		\node at (4.4, 0.7) {\small$---$};
		\node at (5.6, 0.7) {\small$+--$};
		\node at (4.4, -0.7) {\small$-+-$};
		\node at (5.6, -0.7) {\small$++-$};
		
		\path (1m) to node(pD)[pt, pos=0.33]{} (2p);	
		\node(D) at (2, -1.5) {$\mu: \substack{(1, -)\text{  w.p. } 2/3 \\(2, +)\text{  w.p. } 1/3}$};
		\draw[gray, -stealth] (D.east)--(pD);
		\end{scope}
	\end{tikzpicture}
	\caption{Simplicial complex $\Delta_\H$ for a class $\H$ on the domain $\{1,2,3\}$ with $4$ concepts: $---$, $-+-$, $+--$, and $++-$. The subcomplex $\Delta^\ant_\H$ is the boundary of $\Delta(\H)$, drawn in thick blue lines.
	}
	\label{fig-DeltaH2}       
\end{figure}
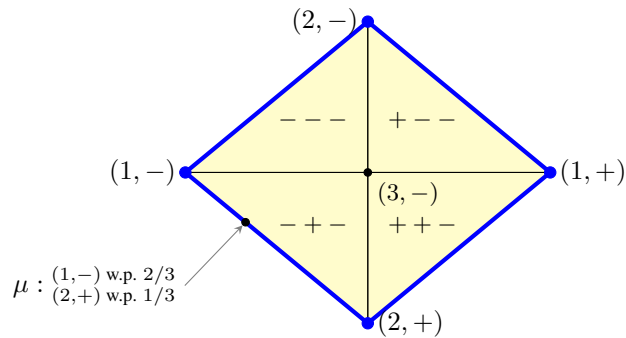 

 It is also easy to see that $\Delta^\ant$ is a (potentially empty) subcomplex of $\Delta_\H$. However, maximal simplices of $\Delta^\ant_\H$ no longer, in general, correspond to concepts of $H$, in particular, the dimension of $\Delta^\ant_\H$ can be much smaller than $|X| - 1$ (as in \Cref{fig-DeltaH2} above). We also note that while typically we denote an abstract simplicial complex by $K$ and its geometric realization by $\|K\|$, for historical consistency we do not use these ``geometrization brackets'' for $\Delta_\H$, and, with some ambiguity, use the same notation for it as for an abstract and geometrical simplicial complex.

\subsection{Spherical dimension}
We are now ready to define all the versions of the spherical dimensions, extending Definition~\ref{def-sd}:
\begin{definition}[Three spherical dimensions]\label{def-sd-zoo} We say that a class $\H$ \emph{admits an infinitely supported $n$-sphere} if there is an equivariant map from $\S^n$ to $\Delta^\ant_\H$ with their standard antipodalities. 

Furthermore, $\H$ admits a (\emph{finitely supported}) $n$-sphere if there is such map to $\Delta^\ant_{\H'}$, where $\H' = \H|_{X'}$ for some finite $X'\subseteq X$. Finally, it admits a \emph{simplicial} $n$-sphere if this equivariant map is simplicial from a simplicial $n$-sphere to $\Delta^\ant_{\H'}$, treated as a simplicial complex. 

The \emph{spherical dimension} $\sd(\H)$ of $\H$ is defined as a maximal integer $n$ such that $\H$ admits an $n$-sphere. Additionally, we say that $\sd(\H)=\infty$ if $\H$ admits $n$-spheres for arbitrary large $n$ and $\sd(\H)=-1$ if $\Delta_\H^\ant = \emptyset$, that is, if $\H$ does not even admit a $0$-sphere. The infinitely supported and simplicial spherical dimensions $\sd_\infty(\H)$ and $\sd_\simp(\H)$ are defined similarly using respective $n$-spheres. 
\end{definition}

The notion of the spherical dimension is rather natural; in particular, in terms of~\cite{matousek}, Section~$5$, $\sd(\H)$ is a \emph{$\ZZ_2$-coindex} of $\Delta^\ant(\H)$.
A typical application of $\sd$ then uses the BU theorem or one of its relatives. Below we list our typical toolset for this approach.

\begin{theorem}[Borsuk-Ulam (BU) theorem, \cite{BU}]\label{t-BU}
	For $0\leq m<n$, any map $f\colon \SS^n \rightarrow \SS^m$ or $f\colon \SS^n \rightarrow \R^n$ collapses a pair of antipodal points. 
\end{theorem}

\begin{theorem}[Lyusternik-Shnirel'man (LS) theorem, \cite{LS30}]\label{t-LS}~\\
Let $\mathcal{F}$ be an 
antipodal-free\footnote{That is, for all $F\in \F$, $x\in F$ implies $-x\notin F$.} cover of the $n$-dimensional sphere $\S^n$, for $n\geq 0$, such that
every set $A \in\mathcal{F}$ is either open or closed. Then, $\lvert \mathcal{F}\rvert \geq n+2$.
This is sharp, there is such a cover of size $n+2$.
\end{theorem}

The above two classical theorems are equivalent to each other, 
and the statements of BU in terms of  $f\colon \SS^n \rightarrow \SS^m$ and $f\colon \SS^n \rightarrow \R^n$ are equivalent as well. Also, the following ``local'' versions are quite useful.

\begin{theorem}[Local BU, main theorem in \cite{jan2000periodic}]\label{thm:Jaw}
Let $\PPP$ be a finite 
simplicial complex of dimension at most $d$.
If, for $n\geq 1$, $f : \SS^n \to \PPP$ is continuous and $2d \leq n$ 
then there is $x \in \SS^n$ so that
$f(x) = f(-x)$. 
\end{theorem}

\begin{theorem}[Local LS, Theorem A in~\cite{chase2024local}]\label{t:localLS}
Let, for $n\geq 1$\footnote{Both \Cref{thm:Jaw} and \Cref{t:localLS} are incorrect for $n=0$: The identity map on $\SS^0$ is non-collapsing, and $\SS^0$ can be covered by a non-overlapping open family, in which case the overlap is $1 < \lceil(0 + 3)/2\rceil = 2$. This detail was overlooked in the corresponding statement in~\cite{chase2024local}, which leads to minor incorrectness in their Theorem~D. We elaborate upon it in~\Cref{sec-lr}.}, $\mathcal{F}$ be a finite antipodal-free cover of the $n$-dimensional sphere $\S^n$ such that either all sets in $\F$ are open, or all are closed. Then there are at least $\lceil (n+3)/2\rceil$ distinct sets in $\F$ with nonempty intersection, and this bound is sharp.
\end{theorem}
An alternative way of stating a conclusion of \Cref{t:localLS} is to say that the \emph{overlap-degree} of $\F$ is at least $\lceil (n+3)/2\rceil$. Also note that, unlike in the original LS, the statement is no longer true if some sets in $\F$ are open and some are closed. In fact, in~\cite{chase2024local} it is shown that for open/closed case the overlap is at least $\lceil t/2 \rceil$ for $t = \lceil (n+3)/2\rceil$, and that this bound is sharp. We also note that \Cref{t:localLS} (Local LS), except for the bounds in open/closed case, follows, in a straightforward way, from \Cref{thm:Jaw} (Local BU). 
We finally note that the list of the four theorems above is not exhaustive, for example, in~\cite{reductions} the authors used a version of BU for closed convex relations (their Theorem~$7$) or, in~\cite{chase2024dual}, a ``local'' version of a topological Radon theorem (their Lemma~$12$).

Recall that out of the three definitions of admitting an infinitely supported / finitely supported / simplicial sphere, we picked the finitely supported as the standard one. In particular, we say that a class admits an $n$-sphere if it admits a finitely supported $n$-sphere. Putting the finitely supported spheres in the center is justified by the fact that most of the important dimensions, like VC and VC$^*$, are finitely supported, that is, witnessed by projections to finite subdomains. Moreover, in most of the cases, for example for list replicability (\Cref{sec-lr}) or extremal classes (\Cref{sec-extremal}), we can apply our tools only to finitely supported classes. 
Note also that the distinction between finitely and infinitely supported spheres vanishes if the class $\H$ itself is finitely supported; in particular, whenever we upper bound spherical dimension for finitely supported classes, we would often state the bound for $\sd$, implying that it trivially extends to a ``formally stronger'' bound on $\sd_\infty$. Finally, we note that in the \Cref{def-sd-zoo} above we defined $\sd$ via finitely supported classes, while in the introduction we defined it for finite classes (which might not be finitely supported). The fact that they are equivalent needs proof, which we give in Corollary~\ref{cor-finite-classes} in \Cref{sec-class-order}.

Further restricting the spheres to be simplicial is important to our results on disambiguations (\Cref{sec-disamb}), and can also be considered natural from the standpoint of simplicial geometry. We also note that the spheres witnessing the 
$\sd\geq \VC - 1$ and $\sd\geq \VC^{*\an} - 2$ lower bounds in Lemma~\ref{lem-vc-lb}, as well as other lower bounds on $\sd$ in Lemma~\ref{lem-vcvc*-lb}, are all simplicial.

Simply based on the fact that admitting a finitely supported sphere is more restrictive than admitting an infinitely supported one, and admitting a simplicial sphere is furthermore restrictive, we have $\sd_\simp \leq \sd\leq \sd_\infty$. 
Although we think that the three spherical dimensions can be separated from each other on the level of individual classes, the difference between them might be insubstantial with respect to Conjectures~\ref{conj-sd-vc-1}--\ref{conj-sd-vc-3}. To this end, we say that parameters $\sd', \sd'' \in \{\sd_\infty, \sd, \sd_\simp\}$ are \emph{roughly equivalent}, if for any integers $a$ and $b$, there is a class $\H'$ with $\VC(\H') \leq a$, $\VC^{*\an}(\H') \leq b$, and $\sd'(\H')\geq n$ if and only if there is a class $\H''$ with $\VC(\H'') \leq a$, $\VC^{*\an}(\H'') \leq b$, and $\sd''(\H'')\geq n$. From this perspective, our main result on disambiguations (\Cref{th-sd-disamb} in \Cref{sec-disamb}) establishes rough equivalence between $\sd_\simp$ and yet another parameter called \emph{disambiguation index} $\disamb$ (although we will not really use this term and will only formally introduce it in the proof of \Cref{th-sd-disamb}).

We finally note two possible modifications to the notion of spherical dimension which we find potentially worth considering, but which we excluded from the main narrative for the sake of brevity. On the one hand, we can restrict these definitions by requiring that the respective equivariant map from $\S^n$ to $\Delta^\ant$ be an embedding (in which case we call the corresponding sphere \emph{embedded}). But, apart from the psychological comfort of thinking about the respective spheres as ``sitting inside'' $\Delta^\ant$, we see no other benefits to this modification. We note, however, that all the spheres in Lemma~\ref{lem-vc-lb} and Lemma~\ref{lem-vcvc*-lb} are embedded.

On the other hand, we can relax the three notions by letting the spheres be homology spheres or have non-standard antipodalities. The benefit of this would be that it corresponds to the natural domain of applicability of the BU-like theorems (Theorems~\ref{t-BU}--\ref{t:localLS}) that are used to infer properties of the respective classes. Most of our results would indeed hold with this relaxation. However, it might be considered rather exotic, dragging in heavier topological machinery, while we do not have examples to substantiate the benefit of this approach.

\section{Applications of spherical dimension}\label{sec-applications}
\subsection{Disambiguations}\label{sec-disamb}
The main question and the background follow~\cite{PCS22}. A \emph{partial class} $\PP$ on $X$ is a subset of $\{-, +, *\}^X$, where $*$ is treated as ``undefined''. The standard involution $y \mapsto -y$ on $\{-, +, *\}$ is defined as $-- = +$, $-+ = -$, and $-* = *$. In this section, we usually refer to usual classes, that is, to subsets of $\{-, +\}^X$, as to \emph{total classes}. The definitions of partial concepts and hypotheses, and distributions realizable by a partial class, are formulated similarly as for total classes. For a partial hypothesis $h\in \{-, +, *\}^X$, the \emph{support} $\supp(h)$ is $\supp(h) = \{x\in X~|~h(x)\neq *\}$. Partial hypotheses are naturally partially ordered by \emph{extension}, where $h_2$ extends $h_1$, denoted $h_1\leq h_2$, if $\supp(h_1)\subseteq \supp(h_2)$ and for all $x\in \supp(h_1)$, $h_1(x) = h_2(x)$. A set $S\subseteq X$ is shattered by $\PP$ if every binary pattern on $S$ is realizable by $\PP$, that is, if for every $s\in \{-, +\}^S$ there is $h\in \PP$ such that $s = h|_S$; as usual, VC dimension of $\PP$ is the maximal size of a set shattered by $\PP$.

A total class $\H$ \emph{disambiguates}\footnote{In~\cite{PCS22} this was called \emph{strongly disambiguates}.} $\PP$ if any partial concept in $\PP$ can be extended to a total concept in $\H$, that is:
$$\text{for any } h\in \PP \text{ there is }\ol{h}\in \H \text{ s.t. }h(x) = \ol{h}(x)\text{ for all } x\in \supp(h).$$

The one of the main results in~\cite{PCS22} is then as follows:
\begin{theorem}[Theorem 1 in~\cite{PCS22}]\label{th-disamb}
	There is a partial class $\PP$ on a countable domain such that $\VC(\PP) = 1$ and for any disambiguation $\H$ of~$\PP$, $\VC(\H) = \infty$.
\end{theorem}
While the above theorem establishes the principal impossibility of disambiguations with finite VC, it is reasonable to ask how big a VC of disambiguation should be for some natural partial classes, such as linear classifiers with margin. For this end, let us define the family of partial classes $\LL_{n, \varepsilon}$, where $\LL$ stands for ``linear classifier'', for integer $n$ and $0< \varepsilon < \pi/2$ as:
\begin{itemize}
	\item The domain of $\LL_{n, \varepsilon}$ is the $n$-dimensional unit sphere $\S^n$;
	\item The partial functions of $\LL_{n, \varepsilon}$ are parametrized by $u\in \S^n$, and we call the respective partial functions $u_\varepsilon$;
	\item For $u,v\in \S^n$, $u_\varepsilon(v) = +$ if $d(u, v) \leq \varepsilon$, $-$ if $d(-u, v) \leq \varepsilon$, and $*$ otherwise.
\end{itemize}
Although the particularities of the distance $d$ on $\S^n$ are not that important, we consider $d$ to be the geodesic metric on $\S^n$. Alternatively, $d(u, v)$ is the angle, in radians, between $u$ and~$v$. Thus, $\varepsilon <\pi/2$ ensures that these classes are well-defined, and it is not hard to verify that for $\varepsilon <\pi/4$, $\VC(\LL_{n, \varepsilon}) = 1$. We additionally note that $\LL_{n, \varepsilon_1}$ disambiguates $\LL_{n, \varepsilon_2}$ whenever $\varepsilon_1 \geq\varepsilon_2$; here disambiguation of a partial class by partial is defined analogously, however normally by disambiguation we mean disambiguation by a total class. Finally, for all $n\leq m$, $\LL_{n, \varepsilon}$ can be obtained from $\LL_{m, \varepsilon}$ by restricting the domain of the latter.


\hyphenation{pa-ra-met-rized}

The family $\LL_{n, \varepsilon}$ is closely related to the family considered in Open Question~$4$ in~\cite{PCS22}, although with some technical differences. In particular, the classes in OQ~$21$ are parametrized with \emph{margin} $\gamma$, which relates to our $\varepsilon$ as $\gamma = \cos(\varepsilon)$. Thus, while OQ~$21$ is concerned with $\gamma \rightarrow 1$, we are interested in $\varepsilon \rightarrow 0$. To better capture the fact that we are interested in the behavior of disambiguations for arbitrary small $\varepsilon$, we say that $\H^n$ disambiguates $\SS^n$ if it disambiguates $\LL_{n, \varepsilon}$ for some $\varepsilon >0$; we note that  $\LL_{n, 0}$ can be trivially disambiguated by two functions, and thus the case $\varepsilon = 0$ is excluded from consideration.

Our main result establishes the rough equivalence between disambiguations of $\SS^n$ and spherical dimension; recall that rough equivalence means that the respective parameters are the same from the perspective of Conjectures~\ref{conj-sd-vc-1}, \ref{conj-sd-vc-2}, and~\ref{conj-sd-vc-3}.
\begin{customthm}{\ref{th-sd-disamb}}[Rough equivalence between disambiguations and spherical dimension]~
For integers $a$, $b$, and~$n$, the following are equivalent:
\begin{itemize}
	\item There is a class with $\VC \leq a$, $\VC^{*\an}\leq b$, and $\sd_\simp \geq n$;
	\item There is a disambiguation of $\SS^n$ with $\VC \leq a$ and $\VC^{*\an}\leq b$.
\end{itemize}
\end{customthm}

\subsection{Extremal classes}\label{sec-extremal}
In this section, we are going to build upon \cite{chase2024dual}. In particular, we refer to Sections~1.1,~1.2, and~2 there for an in-depth discussion of maximum and extremal classes, their relation to sample compression schemes, and the relevant topological notions. Here we repeat, in a minimalistic way, certain definitions and basic properties. 

The key property, enabling the definition of an extremal class, is the Pajor inequality: 
\begin{theorem}[Pajor's Inequality \cite{Pajor:1985}]\label{t:pajor}
    For a class $\H\subseteq\{-,+\}^n$, let 
    \[\str(\H) = \{A\subseteq [n]: A\text{ is shattered by } \H\}.\] 
    Then,
    $\lvert\H\rvert \leq  \lvert \str(\H)\rvert$.
\end{theorem}
Pajor's inequality is perhaps better known in the form of its direct corollary, namely the Sauer-Shelah-Perles inequality, which establishes a polynomial upper bound on the size of classes of bounded VC dimension. Extremal classes are then defined as follows:
\begin{definition}[Extremal Classes]
A finite class $\E\subseteq\{-,+\}^n$ is called \emph{extremal} if it meets Pajor's inequality with equality, that is, if \(\lvert\E\rvert = \lvert\str(\E)\rvert\).
\end{definition}

As long as we are concerned with embedding arbitrary classes into extremal, we define the following natural parameter: for a class $\H$, let $\VC^\extr(\H)$ be the minimal VC dimension of an extremal class, containing $\H$. 
\begin{theorem}\label{t-sd-for-extremal}
	For any finite extremal concept class $\E$, the space $\Delta_\E$ can be mapped into the cubical complex $\CC_\E$, of dimension $\VC(\E)$, without collapsing any antipodal points in $\Delta^\ant_\E$. 
	
    In particular, every extremal concept class $\E$ satisfies $\sd(\E) \leq 2\VC(\E)-1$ and $\VC^*(\E) \leq 2\VC(\E)+1$. Consecutively, for any class $\H$, $\sd(\H) \leq 2\VC^\extr(\H) - 1$, or, equivalently, $\VC^\extr(\H) \geq \lceil(\sd(\H) + 1)/2\rceil$.
\end{theorem}
We refer to \Cref{sec-cubical} for the definition of the cubical complex of the class. The $\VC^*(\E) \leq 2\VC(\E)+1$ was the main result in~\cite{chase2024dual} (Theorem~B). It follows from $\sd(\E) \leq 2\VC(\E)-1$ via $\sd\geq \VC^* - 2$ bound in Lemma~\ref{lem-vc-lb}. However, the bound in terms of $\sd$ has a wider range of applicability because, as noted in Lemma~\ref{lem-vcvc*-lb}, there are classes for which $\sd \gg \VC, \VC^*$. The proof of this will be given in~\Cref{sec-extremal-proofs}. We also note that, structurally, we will prove a much stronger result that the corresponding map is the final map of a \emph{deformation retraction} of $\Delta_\E$ onto~$\CC_\E$, naturally embedded into it.

\subsection{Stability and list replicability}\label{sec-lr}
In this section, we follow the setup from~\cite{chase2024local}. 

A learning rule $\A$ for a class $\H$ is called \emph{$(\varepsilon,L)$-list replicable learner} for $\H$ if for every $\delta>0$ there exists $n=n(\varepsilon,L,\delta)$ such that for every distribution $\mu$ that is realizable by~$\H$, there exist hypotheses $h_1,h_2,\ldots,  h_L$ such that $\Pr_{S\sim \mu^n}[\A(S) \in\{h_1,\ldots, h_L\}] \geq 1-\delta$ and for all $\ell \in [L]$, $L_\mu(h_\ell)< \varepsilon$. 

We then define the \emph{$\varepsilon$-list replicability number} $\List(\H, \varepsilon)$ as minimal $L$ for which $\H$ has a $(\varepsilon,L)$-learner, and the \emph{list replicability number} $\List(\H)$ as $\List(\H) = \lim_{\varepsilon \rightarrow 0} \List(\H, \varepsilon)$; note that, trivially, $\List(\H, \varepsilon)$ is non-increasing in $\varepsilon$. The following theorem refines one of the main results in~\cite{chase2024local}, namely their Theorem D:
\begin{theorem}\label{th-lb-on-lr}
Let $\H$ be a concept class such that $\sd(\H) \geq 1$. Then, for any $0 < \varepsilon < 1/2$, it holds 
$$\List(\H, \varepsilon)\geq \lceil(\sd(\H) + 3)/2\rceil.$$
In particular, $\List(\H, \varepsilon)\geq 1 + \lceil \VC(\H)/2\rceil$ whenever $\VC(\H) \geq 2$, and $\List(\H, \varepsilon)\geq 1 + \lfloor \VC^{*\an}(\H)/2\rfloor$ (with no additional condition). 

The only exception for the bounds in terms of $\VC$ and $\sd$ are the classes of the form $\H = \{h_-, h_+\}$ such that $h_-(x) = -h_+(x)$ for all $x\in X$, for example, $\C_1$, for which $\List(\H, \varepsilon) = 1$,  $1 + \lceil \VC(\H)/2\rceil = 1 + \lceil 1/2\rceil = 2$, and $\lceil(\sd(\H) + 3)/2\rceil = \lceil(0 + 3)/2\rceil = 2$.
\end{theorem}
We finally note that the condition $\VC(\H) \geq 2$ was overlooked in~\cite{chase2024local}.

Curiously enough, Theorem~E in~\cite{chase2024local} proves the upper bound $\List(\U_n)\leq \lceil (n+1)/2 \rceil$ by constructing an explicit list-replicable learner for this class. Together with $\List\geq 1 + \lfloor \VC^\star/2\rfloor$ this implies that $\List(\U_n) = \lceil (n+1)/2 \rceil$ and also gives near-tight bound on $\sd(\U_n)$ from Lemma~\ref{lem-vc-lb}.

\subsection{Reductions to stochastic convex optimization}
In this section, we follow the setup from~\cite{reductions}. 

A $d$-dimensional SCO problem can be defined by an abstract sample space \( Z \), a convex set \( W \subseteq \R^d \), and a collection of convex loss functions \( \{ \ell_z \}_{z \in Z} \) from \( W \) to $\R^+$. The learner receives a sample of \( n \) examples from a distribution $\zeta$ over \( Z \) and its goal is to output \( w \in W \) that minimizes the average loss  
$
L_\zeta(w) = \Ex_{z \sim \zeta} \ell_z(w).
$


\begin{definition}[Reduction to SCO]\label{def:red}
Let $\H$ be a concept class over domain $X$, and let $\alpha > 0$ and $\beta\geq 0$. An $(\alpha,\beta)$-reduction from $\H$ to a $d$-dimensional SCO problem $(Z,W)$
consists of two maps $r_\inp:X\times\{-,+\}\to Z$ and
$r_\outp:W\to \{-,+\}^X$ such that for every realizable distribution $\mu\in \Delta(\H)$ we have 
\[
    L_{r_\inp(\mu)}(w)\leq \inf_{w'\in W}L_{r_\inp(\mu)}(w')+\alpha\implies L_\mu(r_\out(w))\leq \beta,
    \]
    where $r_\inp(\mu)$ is the push-forward measure induced by sampling $z\sim \mu$ and mapping it to $r_\inp(z)$.

	We define the SCO dimension of $\H$, denoted $\SCO(\H)$ to be the smallest number $d$ such that there is an $(\alpha,\beta)$-reduction of $\H$ to a $d$-dimensional SCO problem, for some $\alpha>0$ and $0<\beta<1/2$. 
\end{definition}
The above definition is a compilation of Definition~1 and Example 4 (SCO learning task), Definition~6 (reductions), and Definition 12 (SCO-dimension) in \cite{reductions}. Informally, it aims at capturing the situation when we can design a weak learner for $\H$ by transforming it to an SCO problem, solving the latter (with some accuracy), and then reinterpreting the result of the SCO as a learner's hypothesis. Then their Theorem~1 can be refined as 
\begin{theorem}\label{t:reductions}
    For any class $\H$ we have \[
    \sd(\H) + 1\leq \mathtt{SCO}(\H).
    \]
\end{theorem}

The \emph{sign-rank} is one of the most popular notions in the context of studying the expressivity of kernel methods. In the definition below we follow Definition~11 in~\cite{reductions}, where the sign-rank was studied as a special kind of reduction to a class defined by homogenous hyperplanes in $\R^d$; this, in turn, follows Definition~1 in \cite{KamathMS20}.

\begin{definition}[Sign-rank]\label{def-sign-rank}
For a class $\H\subseteq \{\pm1\}^X$, the \emph{sign-rank} of $\H$ is the smallest $d$ for which there are maps $\varphi\colon X\rightarrow \R^d$ and $w\colon \H\rightarrow \R^d$ such that for all $h\in \H$ and $x\in X$, it holds $\sign \langle w(h), \varphi(x) \rangle = h(x)$.
\end{definition}
We also note that the canonical definition of the sign-rank comes from~\cite{paturi86}, where it is defined as a minimal rank of a real matrix, whose sign pattern coincides with the class. The fact that it coincides with the above definition is well-known. In~\cite{reductions} it was shown that SCO-dimension is upper bounded by the sign-rank, which extends \Cref{t:reductions} to
\[
\sd(\H) + 1\leq \SCO(\H) \leq \sr(\H).
\]

This has some interesting applications for bounding the spherical dimension of products of classes using the following result:
\begin{theorem}\label{prop-sign-rank-prod}
    For any two concept classes $\H_1$, $\H_2$ we have
    \[
    \sr(\H_1\times \H_2)\leq \sr(\H_1)+\sr(\H_2).
    \]
\end{theorem}
We also note that it is well-known that $\sr(\U_n) = n$, which implies $\sd(\U_n)\leq n-1$. While this bound is slightly worse than $\sd(\U_n) \leq n-2 + \delta_n$ in Lemma~\ref{lem-vc-lb} (which follows from list-replicability), it extends, by Proposition~\ref{prop-sign-rank-prod} above, to $\U_n^m$, giving $\sd(\U^m_n) \leq mn$ bound in Lemma~\ref{lem-vcvc*-lb}.

Note also that the embedding witnessing the sign-rank naturally provides an embedding into extremal class (see, for example, Section~3.1 in \cite{chase2024dual}), and so 
	$$\VC^\extr(\H) \leq \sr(\H).$$

\appendix

\section{Further topological background}\label{sec-background}

\subsection{Barycentric subdivisions}\label{sec-barycentric}
We refer to Definition~1.7.2 and the subsequent discussion in~\cite{matousek}.

For an abstract simplicial complex $\Q$ and a simplex $s\in \Q$, the \emph{relative interior} of $s$ is a subset of $\|\Q\|$ defined as $\|s\| - \bigcup_{t\subsetneq s} \|s\|$ (Definition~1.3.4 in~\cite{matousek}). Relative interiors form a partition of $\|\Q\|$ and so for every $x\in \|\Q\|$ there is a unique simplex $s$ such that $x$ is in the relative interior of $s$; it is easy to see that this simplex is exactly the support of $x$, as defined in \Cref{sec-topology-toolbox}.

For an abstract simplicial complex $\Q$, the \emph{first barycentric subdivision} $\Q_1$ of $\Q$ is defined as an abstract simplicial complex whose vertices are non-empty simplices of $\Q$, $V(\Q_1) = S^{>0}(\Q)$, and whose simplices are chains of simplices of $S^{>0}(\Q)$ (where chain is with respect to the order on $S^{>0}(\Q)$ by inclusion).   

For a nonempty simplex $s\in S(\Q)$, let us define $v_s\in \|s\|$ as the (geometric) \emph{barycenter} of $\|s\|$, that is, $s_v = \frac{1}{|s|} \sum_{v\in s} v$. The map $\chi\colon \|\Q_1\| \rightarrow \|\Q\|$ defined on $V(\Q_1) = S^{>0}(\Q)$ as $s\mapsto v_s$ and the extended by linearity to $\|\Q\|$ establishes a homeomorphism between $\|\Q_1\|$ and $\|\Q\|$. In particular, the geometric barycentric subdivision preserves the topology of the space, $\|\Q_1\| \cong \|\Q\|$.

Furthermore, if $(\Q, \nu)$ is an antipodal simplicial complex (as defined in \Cref{sec-sd}), then $\nu$ can be extended from $V(\Q)$ to $\nu_1$ on $V(\Q_1)$ by letting $\nu_1(s) = \nu(s) = \{\nu(v)~|~v\in s\}$, for any $s\in V(\Q_1) = S^{>0}(\Q)$. It is well-known and easy to check that $\nu_1$ is an antipodality map, that is, satisfies i), ii), and iii), and that $\nu_1$, restricted to $V(\Q)$ coincides with $\nu$ (where we assume that $V(\Q)$ is embedded into $V(\Q_1)$ by $v\mapsto \{v\}$). Moreover, $\nu_1$, extended by linearity to $\|\Q_1\|$ coincides, via the homeomorphism $\chi$, with $\nu$ extended by linearity to $\|\Q\|$, that is, $\chi\circ\nu_1(x) = \nu\circ\chi(x)$ for all $x\in \|\Q_1\|$. In particular, $(\|\Q_1\|, \nu_1) \cong (\|\Q\|, \nu)$. Because of this, we will call $\nu_1$ simply by $\nu$. This is illustrated in \Cref{fig-BARYCENTRIC2} below.

\begin{figure}[!hbt]
	\centering
	\begin{tikzpicture} 
[
pt/.style={inner sep = 0.0pt, circle, draw, fill=black},
point/.style={inner sep = 1.2pt, circle,draw,fill=black},
spoint/.style={inner sep = 1.2pt, circle,draw,fill=white},
mpoint/.style={inner sep = 0.7pt, circle,draw,fill=black},
ypoint/.style={inner sep = 3pt, circle,draw,fill=yellow},
xpoint/.style={inner sep = 3pt, circle,draw,fill=red},
FIT/.style args = {#1}{rounded rectangle, draw,  fit=#1, rotate fit=45, yscale=0.5},
FITR/.style args = {#1}{rounded rectangle, draw,  fit=#1, rotate fit=-45, yscale=0.5},
FIT1/.style args = {#1}{rounded rectangle, draw,  fit=#1, rotate fit=45, scale=2},
vecArrow/.style={
		thick, decoration={markings,mark=at position
		   1 with {\arrow[thick]{open triangle 60}}},
		   double distance=1.4pt, shorten >= 5.5pt,
		   preaction = {decorate},
		   postaction = {draw,line width=0.4pt, white,shorten >= 4.5pt}
	}
]

\begin{scope}[yscale=1.2, xscale = 0.6]
	\begin{scope}[yscale = 0.7]
		\node(a) at (-4,0) {$1$};
		\node(b) at (0,4) {$2$};
		\node(c) at (4,0) {$3$};
		\node[mpoint](ac) at (0,0) {};
		\node[mpoint](ab) at (-2,2) {};
		\node[mpoint](bc) at (2,2) {};
		\node[mpoint](abc) at (0,1.5) {};
	
		\draw [dotted] (a)--(ab)--(b)--(bc)--(c)--(ac)--(a);
		\draw [dotted](ab)--(abc) (bc)--(abc);
		\draw [dotted] (b)--(abc)  (c)--(abc);
		
		\draw [dotted] (a)--(abc);
		\draw [dotted] (ac)--(abc);
		
		\node [below] at (ac) {$13$};
		\node [left] at (ab) {$12$};
		\node [right] at (bc) {$23$};
		\node [above] at (abc) {$123$};

	\end{scope}

	\begin{scope}[xshift = 10cm, xscale = 1.3, yscale = 0.7]
		\node[mpoint](1) at (0,0) {};
		\node[mpoint](2) at (0,4) {};
		\node[mpoint](-1) at (4,4) {};
		\node[mpoint](-2) at (4,0) {};
		\draw(1)--(2)--(-1)--(-2)--(1);
		\node[left] at (1) {$1$};
		\node[left] at (2) {$2$};
		\node[right] at (-1) {$1^*$};
		\node[right] at (-2) {$2^*$};

		\node[mpoint](12) at (0,2) {};
		\node[mpoint](1s2) at (2,4) {};
		\node[mpoint](1s2s) at (4,2) {};
		\node[mpoint](12s) at (2,0) {};

		\node[left] at (12) {$12$};
		\node[above] at (1s2) {$1^*2$};
		\node[right] at (1s2s) {$1^*2^*$};
		\node[below] at (12s) {$12^*$};
	\end{scope}
\end{scope}

\end{tikzpicture}	
	\caption{Barycentric subdivision of a $2$-dimensional simplex $\{1,2,3\}$ and a barycentric subdivision of a $1$-dimensional simplicial complex 
	$\{12, 1^*2, 1^*2^*,12^*\}$ that extends the antipodality map $v \mapsto v^*$.}
	\label{fig-BARYCENTRIC2}       
\end{figure}
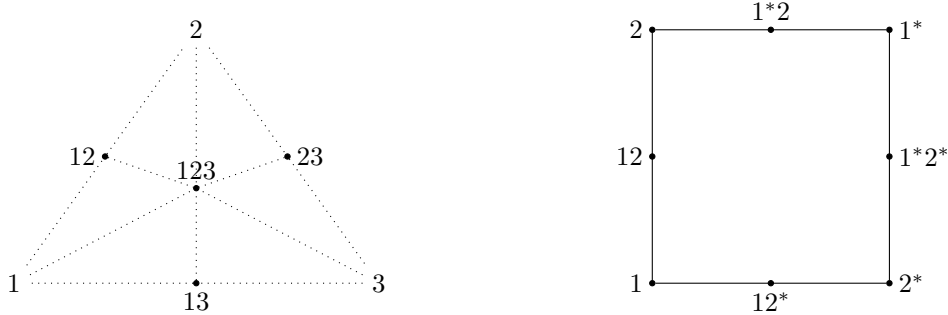 

One important, in general, as well as for us, example of a barycentric subdivision is the barycentric subdivision of the boundary of a single simplex. 

\hyphenation{sub-di-vi-sion}

\begin{example}[Barycentric subdivision of the boundary of a simplex]\label{ex-barycentric}~\\
The barycentric subdivision $\B^n$ of the boundary of an $(n+1)$-dimensional simplex $\Lambda_{n+1}$ is a simplicial complex whose vertices are all nontrivial subsets of $[n+2]$, that is all except for $\emptyset$ and $[n+2]$, and whose simplices are chains of nontrivial subsets of $[n+2]$. Thus defined, $\|\B^n\|$ is homeomorphic to $\S^n$.

Moreover, the complementation map $S \mapsto [n+2] - S$ on $V(\B^n)$ is an antipodality map for $\B^n$, and $(\|\B^n\|, S \mapsto [n+2] - S) \cong (\S^n, x\mapsto -x)$.
\end{example}	

Finally, let us note that $\|(\Q, \nu)\| \cong \|(\Q_1, \nu)\|$ enables us to iterate barycentric subdivisions without changing the topology of the space. This is convenient in the following sense: Assuming some metric on $\|\Q\|$, let us define the \emph{(simplicial) diameter} of $\Q$ as $\max_{s\in S(\Q)} \max_{x,y\in \|s\|} d(x, y)$, denoted $\diam(\Q)$.

\begin{proposition}\label{prop-barycentric-embedding}
	Let $\Q_k$, for $k\geq 0$, be $k$'th barycentric subdivision of a finite simplicial complex $\Q$, where $\Q = \Q_0$. Then the sequence $\diam(\Q_k)$, for $k\geq 0$, is non-increasing and goes to $0$.
\end{proposition}
This property is Exercise~3 in Section~1 in \cite{matousek}. Proof of Proposition~2.21 in~\cite{hatcher} gives an explicit rate of convergence. 

\subsection{Joins}\label{sec-joins}
We follow Section~4.2 in~\cite{matousek} in the development of the respective terminology. In particular, the definitions of joins of simplicial complexes and topological spaces are Definitions~4.2.1 and~4.2.3 there, and Example~5.2.3 defines the join of antipodal spaces.

For simplicial complexes $K$ and $L$, the \emph{join} of $K$ and $L$, denoted $K * L$, is the simplicial complex with vertex set $V(K)\sqcup V(L)$ and with the set of simplices $\{s \sqcup k~|~s\in S, k\in K\}$. For topological spaces $X$ and $Y$, their \emph{join}, denoted $X*Y$ is the topological space $X\times Y \times [0,1]/\approx$, where $\approx$ is an equivalence relation given by $(x,y,0)\approx(x', y, 0)$ and $(x,y,1)\approx(x, y', 1)$, for all $x, x'\in X$ and $y, y'\in Y$. The point $v = (x, y, \alpha)\in X*Y$ is often written as a formal linear combination $v=\alpha x + \beta y$, for $\beta = 1-\alpha$. This notation is consistent with the fact that $0 x + 1 y$ and $0 x + 1 y'$ define the same point in the join, and we will call $\alpha, x$ and $\beta, y$ the \emph{barycentric coordinates} of $v$ in $X*Y$. It is known and easy to check that the definitions of joins for simplicial complexes and for topological spaces are consistent with each other, namely, $\|K\| * \|L\| \cong \|K * L\|$. 

Joins can also be defined for maps: for $f_1\colon X_1 \rightarrow Y_1$ and $f_2\colon X_2 \rightarrow Y_2$, the map $f_1 * f_2 \colon X_1 * X_2 \rightarrow Y_1 * Y_2$ is defined as $\alpha x_1 + \beta x_2 \mapsto \alpha f_1(x_1) + \beta f_2(x_2)$. In a similar way, joins can be defined for simplicial maps and this definition is consistent with the definition for the continuous maps above. This enables us to define the join of antipodal spaces $(X_1, \nu_1)$ and $(X_2, \nu_2)$ as $(X_1 * X_2, \nu_1 * \nu_2)$. It is easy to check that $\nu_1 * \nu_2$ will indeed be a fixed-point free involution whenever $\nu_1$ and $\nu_2$ are. In the same way, this can also be done for joins of antipodal simplicial complexes. We also note that the join operation is trivially associative, and can thus be easily generalized to joins of any finite number of spaces or simplicial complexes.

The $n$-fold joins, also called \emph{join powers}, give us another important example of simplicial spheres. The following example is based on Definition~1.4.1 and Examples~4.2.2 and~5.2.4 from~\cite{matousek}.
\begin{example}[Boundary of a crosspolytope]\label{ex-crosspolytope}
Let $\DD_0$ be the simplicial complex of a $2$-point discrete space, that is, $V(\DD_0) = \{-, +\}$ and $S(\DD_0) = \{\emptyset, -, +\}$. We also assume $\DD_0$ to be antipodal with the vertex antipodality map $\nu_0$ exchanging $-$ and $+$. 

Let $\DD_n$ be the $(n+1)$-fold join of $\DD_0$, denoted $\DD_n = \DD_0^{*(n+1)}$. Then $\DD_n$ can be explicitly described as a simplicial complex with vertices $V(\DD_n) = \{(i, -), (i, +)~|~i\in [n+1]\}$, whose simplices are precisely the subsets $S\subseteq [n+1]\times \{-, +\}$ such that for no $i\in [n+1]$ both $(i, -)$ and $(i, +)$ are in~$S$. The vertex antipodality map $\nu_n$ is defined by sign swap, that is, $\nu_n(i, -) = (i, +)$ and $\nu_n(i, +) = (i, -)$.

This construction is known as the boundary of an $(n+1)$-dimensional \emph{crosspolytope}. In particular, $\|\DD_n\| \cong (\S^{n}, x\mapsto -x)$.

\begin{figure}[!hbt]
	\centering
	\begin{tikzpicture} 
[
pt/.style={inner sep = 0.0pt, circle, draw, fill=black},
point/.style={inner sep = 1.2pt, circle,draw,fill=black},
spoint/.style={inner sep = 1.2pt, circle,draw,fill=white},
mpoint/.style={inner sep = 0.7pt, circle,draw,fill=black},
ypoint/.style={inner sep = 3pt, circle,draw,fill=yellow},
xpoint/.style={inner sep = 3pt, circle,draw,fill=red},
FIT/.style args = {#1}{rounded rectangle, draw,  fit=#1, rotate fit=45, yscale=0.5},
FITR/.style args = {#1}{rounded rectangle, draw,  fit=#1, rotate fit=-45, yscale=0.5},
FIT1/.style args = {#1}{rounded rectangle, draw,  fit=#1, rotate fit=45, scale=2},
vecArrow/.style={
		thick, decoration={markings,mark=at position
		   1 with {\arrow[thick]{open triangle 60}}},
		   double distance=1.4pt, shorten >= 5.5pt,
		   preaction = {decorate},
		   postaction = {draw,line width=0.4pt, white,shorten >= 4.5pt}
	}
]

\begin{scope}[yscale=1, xscale = 0.7]
	\begin{scope}[yscale = 0.7]
		\node[point](-1) at (-1, 0) {};
		\node[point](+1) at (1, 0) {};

		\node[below] at (-1){$-1$};
		\node[below] at (+1){$+1$};
		\node at (0, 2.25) {$\DD_0$};
		
		\node at (3, 0){\Large$*$};

		\node[point](-2) at (6, -1) {};
		\node[point](+2) at (6, 1) {};

		\node[left] at (-2){$-2$};
		\node[left] at (+2){$+2$};
		\node at (5.5, 2.25) {$\DD_0$};

		\node at (8, 0){\Large$=$};
		\node[point](-1) at (10.5, 0) {};
		\node[point](+1) at (13.5, 0) {};
		\node[point](-2) at (12, -1.5) {};
		\node[point](+2) at (12, 1.5) {};

		\node[left] at (-1){$-1$};
		\node[right] at (+1){$+1$};
		\node[below] at (-2){$-2$};
		\node[above] at (+2){$+2$};
		\draw (-1)--(+2)--(+1)--(-2)--(-1);
		\node at (12, 0) {$\DD_1$};
	\end{scope}
\end{scope}

\end{tikzpicture}	
	\caption{Construction of $\DD_1$ as a join of $\DD_0$ with $\DD_0$. Note that for the join, we need to make the vertex sets of the two copies of $\DD_0$ disjoint, so we rename $-$ and $+$ to $-1$, $+1$ for the first copy and $-2$, $+2$ for the second one.}
	\label{fig-CROSSPOLYTOPE}       
\end{figure}
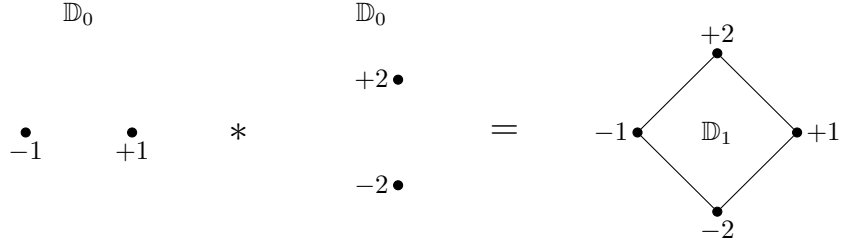 
\end{example}	

A particularly useful property, which can also be considered a corollary of the above~\Cref{ex-crosspolytope}, is the following:
\begin{proposition}\label{prop-join-spheres}
	For antipodal spheres $\SS^n$ and $\SS^m$, $\SS^n * \SS^m \cong \SS^{n+m+1}$. 
\end{proposition}

\section{Proofs for basic results}\label{sec-proofs-basic}

\subsection{Order and equivalence of classes}\label{sec-class-order}
If a class $\H_1$ is obtained from $\H_2$ by restricting to a subdomain or by removing some hypotheses, it is natural to think of $\H_2$ as being, in a sense, ``simpler''. Let us briefly formalize this intuition as follows: For classes $\H_1$ over $X_1$ and $\H_2$ over $X_2$ we write $\H_1 \leq \H_2$ if there are functions $\varphi \colon X_1 \rightarrow X_2$ and $\sigma\colon \H_1 \rightarrow \H_2$ such that $\sigma(h)(\varphi(x)) = h(x)$ for all $h\in \H_1$ and $x\in X_1$. We write $\H_1 \equiv \H_2$ if $\H_1 \leq \H_2$ and $\H_2 \leq \H_1$. It is easy to see that $\leq$ is a preorder and $\equiv$ is the respective equivalence relation. 

It is easy to see that VC, VC$^*$, and VC$^{*\an}$ dimensions are all monotone with respect to this order. While it might be clear on the spot that the same is true for the spherical dimensions, the proof of it is also easy:
\begin{proposition}\label{prop-sd-is-monotone}
	If $\H_1 \leq H_2$, then $\sd_\infty(\H_1) \leq \sd_\infty(H_2)$, $\sd(\H_1) \leq \sd(H_2)$, and $\sd_\simp(\H_1) \leq \sd_\simp(H_2)$.
\end{proposition}
\begin{proof}
	Let $\varphi \colon X_1 \rightarrow X_2$ and $\sigma\colon \H_1 \rightarrow \H_2$ be the functions witnessing $\H_1 \leq \H_2$.
	And let $f\colon \SS^n \rightarrow \Delta_{\H_1}$ be the equivariant map witnessing $\sd_\infty(\H_1) \geq n$. Let $\ol{\varphi}$ be the lifting of $\varphi$ to the set of (not necessarily realizable) distributions over $X_1\times Y$. That is, for a distribution $\varphi$ over $X_1\times Y$ and an event $E\subseteq X_2\times Y$, we define $\Pr_{\ol{\varphi}(\mu)} (E) = \Pr_\mu(\varphi^{-1}(E))$. It is easy to see that thus defined, $\ol{\varphi}(\mu)$ is indeed a probability distribution over $X_2\times Y$. 
	Moreover, we claim that $\ol{\varphi}(\mu)$ is realizable by $\H_2$ whenever $\mu$ is realizable by $\H_1$, and so $\varphi$ maps $\Delta_{\H_1}$ into $\Delta_{\H_2}$.
	
	Indeed, let, for arbitrary $\varepsilon > 0$, $h_1 \in \H_1$ be a concept with $L_\mu(h_1) = \Pr_{(x,y)\sim \mu}\left[h_1(x)\neq y\right]\leq \varepsilon$. Then
	\begin{align*}
		L_{\ol{\varphi}(\mu)}(\sigma(h_1))
			&= \Pr_{(x,y)\sim \ol{\varphi}(\mu)}\left[\sigma(h_1)(x)\neq y\right]
			\\&= \Pr_{(x,y)\sim \mu}\left[\sigma(h_1)(\varphi(x))\neq y\right]
			\\&= \Pr_{(x,y)\sim \mu}\left[h_1(x)\neq y\right] = L_\mu(h_1)  \leq \varepsilon,
	\end{align*}
	as needed. Finally, we are going to show that the map  $\ol{\varphi}\circ \Delta_{\H_1}\rightarrow \Delta_{\H_2}$ is equivariant. It then easily follows that $\ol{\varphi} \circ f$ witnesses $\sd_\infty(\H_2)\geq n$. Indeed, for every $E\subseteq X_2\times Y$ we have:
	\begin{align*}
		\Pr_{\ol{\varphi}(-\mu)} (E) &= \Pr_{-\mu}(\varphi^{-1}(E))
			\\&=\Pr_{(x, y) \sim -\mu}\left[(\varphi(x), y) \in E\right]
			=\Pr_{(x, y) \sim \mu}\left[(\varphi(x), -y) \in E\right]
			\\&=\Pr_{(x, y) \sim \mu}\left[(\varphi(x), y) \in -E\right]
			=\Pr_{-\ol{\varphi}(\mu)} (E),
	\end{align*} 
	and so $\ol{\varphi}(-\mu) = -\ol{\varphi}(\mu)$, ax needed. This finishes the proof that $\sd_\infty(\H_1) \leq \sd_\infty(\H_2)$.
	
	The regular $\sd$, that is, the finitely supported case, follows by observing that for any finitely supported restriction $\H_1'$ of $\H_1$, there is a finitely supported restriction $\H_2'$ of $\H_2$; if $\H_1'$ is restricted to $S$, then it is enough to take a restriction of $\H_2$ to $\varphi(s)$.
	
	Finally, the simplicial case $\sd_\simp$ follows by observing that the equivariant map $\ol{\varphi} \colon \Delta_{\H_1}\rightarrow \Delta_{\H_2}$ is trivially simplicial.
\end{proof}

\begin{corollary}\label{cor-finite-classes}
	For a finite class $\H$, $\sd_\infty(\H) = \sd(\H)$.
\end{corollary}
Note that it resolves some tensions in the definition of $\sd$: In Definition~\ref{def-sd}, we effectively defined it as $\sd_\infty$, but only for finite classes. While for finitely supported $\H$, $\sd_\infty(\H) = \sd(\H)$, this definition aligns with how $\sd$ was later defined in Definition~\ref{def-sd-zoo}, but only for finitely supported $\H$. For finite classes, this equality follows from the present corollary.
\begin{proof}
	Let $\H'$ be a class on the domain of equivalence classes over $X$ for the equivalence relation $x\sim y$ if $x(h) = y(h)$. It is easy to see that $\H'\equiv \H$ and that $\H'$ is finitely supported. But then:
	\begin{align*}
		\sd(\H) = \sd(\H') = \sd_\infty(\H') = \sd_\infty(\H).
	\end{align*}
\end{proof}.

We further note that such ordering on classes will come in handy in \Cref{sec-disamb-proofs}, where we will often see geometrically constructed classes that are naturally defined with a great deal of redundancy.

We find it convenient to explicate some parameters of the minimal classes witnessing $\VC\geq n$ and $\VC^*\geq n$, that is for the classes $\C_n$ of \emph{binary hypercubes} and \emph{universal classes} $\U_n = \C^*_n$. Formally, $\C_n$  is a class of all $2^n$ concepts on $[n]$, and $\U_n$ is its dual, but can also be described explicitly as a class of $n$ indicator functions on the domain of all $2^n$ subsets of $[n]$. 

\Cref{fig-TABLE} provides some exact values for the families $\C_n$ and $\U_n$, which are sometimes convenient for sanity checks. Most will be substantiated in the full statement of Lemma~\ref{lem-vc-lb} in \Cref{sec-bounds}.
\begin{figure}[!hbt]
\begin{center}
\begin{tabular}{c|cccc|ccccc}
$n$ & $|\C^n|$ & $\VC(\C^n)$ & $\VC^{*}(\C^n)$ & $\sd(\C_n)$
	& $|\U^n|$ & $\VC(\U^n)$ & $\VC^{*}(\U^n)$ & $\sd(\U_n)$ & $\sd(\U_n)$\\
  &  $2^n$ & $n$ & $\lfloor \log n\rfloor$ & $n-1$
  & $n$ & $\lfloor \log n\rfloor$ & $n$ & $\geq n-2$ & $\leq$\\
 \hline
1  & 2 & 1 & 0 & 0 & 1 & 0 & 1 & -1 & -1\\ 
2  & 4 & 2 & 1 & 1 & 2 & 1 & 2 & 0& 1\\ 
3  & 8 & 3 & 1 & 2 & 3 & 1 & 3 & 1& 1\\ 
\hline
4  & 16 & 4 & 2 & 3 & 4 & 2 & 4 & 2& 3\\ 
5  & 32 & 5 & 2 & 4 & 5 & 2 & 5 & 3& 3\\ 
6  & 64 & 6 & 2 & 5 & 6 & 2 & 6 & 4& 5
\end{tabular}
\end{center}
	\caption{Certain parameters of classes $\C_n$ and $\U_n$.}
	\label{fig-TABLE}       
\end{figure}

\subsection{Certain bounds on spherical dimension and related parameters}\label{sec-bounds}
\begin{customlem}{\ref{prop-dual-semi-VC}}[Assouad's bounds for antipodal shattering]
	For a (partial) class $\H$, it holds
	$$\left\lfloor\log \VC(\H) \right\rfloor \leq \left\lfloor\log \VC^\an(\H) \right\rfloor \leq \VC^*(\H) \leq \VC^{*\an}(\H)\leq 2^{\VC(\H) + 1} - 1,$$
and $$\VC^*(\H) \leq \VC^{*\an}(\H) \leq 2\VC^*(\H) + 1.$$	
Moreover, all inequalities between the neighboring values above are sharp.
\end{customlem}
We note that the bound $\VC \leq \VC^{*\an} \leq 2\VC^* + 1$ was already proven in \cite{AMY16} (Appendix~E). The refinement of the Assouad's bounds in terms of antipodal shattering, to our knowledge, is new.
\begin{proof} 
	In the first chain of inequalities, the first and the third are trivial, and the second and the fourth are equivalent ``by duality'' by noting that for integers $a$ and $b$, $\lfloor \log a \rfloor \leq b$ is equivalent to $a\leq 2^{b+1} - 1$. We will thus prove the second inequality.
	
	So, assuming $\H$ antipodally shatters a set $S$ of size $2^d$, we need to prove that it dually shatters a $d$-set of concepts $H\subseteq \H$. 
	Let us index the elements of $S$ by the $2^d$ patterns on $[d]$, $S = \{s_t~|~t\in [d]\rightarrow \{-, +\}\}$. Let $\tau_i\colon S\rightarrow \{-, +\}$, for $i\in [d]$, be a pattern on $S$ defined as $\tau_i(s_t) = t(i)$, as illustrated in the table below. Let $h_i\in \H$ be the concept witnessing either $\tau_i$ or $-\tau_i$, and let $q\colon [d]\rightarrow \{-, +\}$ on $i$ be $+$ if it is the former, and $-$ if it is the latter. That is, $h_i(s_t) = q(i)\cdot \tau_i(s_t)$, for $s_t\in S$.  

\begin{center}
\begin{tabular}{c|cccccccc}
$S$ & {\tiny$---$} & {\tiny$--+$} & {\tiny$-+-$} & {\tiny$-++$}
	& {\tiny$+--$} & {\tiny$+-+$} & {\tiny$++-$} & {\tiny$+++$}\\
\hline
$\tau_1$  & $-$ & $-$ & $-$ & $-$
	& $+$ & $+$ & $+$ & $+$\\ 
$\tau_2$  & $-$ & $-$ & $+$ & $+$
	& $-$ & $-$ & $+$ & $+$\\ 
$\tau_3$  & $-$ & $+$ & $-$ & $+$
	& $-$ & $+$ & $-$ & $+$
\end{tabular}
\end{center}	
	
We claim that $S$ dually shatters $H=\{h_i~|~i\in [d]\}$. Let $r\colon [d]\rightarrow \{-, +\}$ be a dual pattern on $H$. Then $s_{q\cdot r}$, where $q\cdot r\colon [d]\rightarrow \{-, +\}$ is a pointwise product, witnesses $r$. Indeed, $h_i(s_{q\cdot r}) = q(i)\cdot \tau_i(s_{q\cdot r}) = q(i)\cdot q(i) \cdot r(i) = r(i)$, for all $i\in [d]$, as needed. The sharpness of the bounds trivially follows from the sharpness of the original bounds by Assouad.
	
In the second set of inequalities, we only need to prove $\VC^{*\an}(\H) \leq 2\VC^*(\H) + 1$. We prove it in the form $\VC^\an(\H) \leq 2\VC(\H) + 1$: these statements are clearly equivalent, just for applications we care about $\VC^{*\an}$ and not $\VC^\an$. So let $\VC(\H)=d$ and let $S$ be a set of size $n$, antipodally shattered by $\H$. It is clear that then $S$ is shattered by $\H'=\H\cup-\H$, where $-\H=\{-h\;:\;h\in H\}$. Denote $n=\VC^\an(\H)$, $d=\VC(\H)$. Then, using the Sauer–Shelah bound and some straightforward manipulation with binomial coefficients, we get 
    \[
   \sum_{k=0}^n\binom{n}{k} = 2^n=|\H'|_S|\leq 2|\H|_S|
   	\leq 2\sum_{k=0}^d \binom{n}{k} = \sum_{k=0}^d\binom{n}{k} + \sum_{k=n-d}^n \binom{n}{k}.
    \]
    But this implies $n-d\leq d+1 \sim n\leq 2d + 1$, implying $\VC^\an(\H) \leq 2\VC(\H) + 1$, as needed.
    
The sharpness of the bound is witnessed by a class of all $(\leq d)$-subsets f a $(d+1)$-set $S$. This class antipodally shatters $S$, but shatters only $(\leq d)$-sets.
\end{proof}

\begin{customlem}{\ref{lem-vc-lb}}
	For a class $\H$, $\sd_\simp(\H)\geq \VC(\H) - 1$ and $\sd_\simp(\H)\geq \VC^{*\an}(\H) - 2$. Moreover:
	\begin{itemize}
		\item $\VC(\C_n) = n$, $\VC^*(\C_n) = \lfloor \log n \rfloor$, and $\sd(\C_n) = n-1$;
		\item $\VC(\U_n) = \lfloor \log n \rfloor$, $\VC^*(\U_n) = n$, and $n-2 \leq \sd(\U_n) \leq n-2 + \delta_n$, where $\delta_n = 1$ for $n$ even and $0$ for $n$ odd. In particular, $\sd(\U_n) = n-2$ for $n$ odd.
	\end{itemize}
	The statements about $\C_n$ and $\U_n$ are true for all three spherical dimensions $\sd_\infty$, $\sd$, or $\sd_\simp$.
\end{customlem}
\begin{proof}
	For the class $\C_n$ the simplices of $\Delta_{\C_n}$ are precisely the subsets $F\subseteq X\times Y$ such that for no $x\in X$ both $(x, -)$ and $(x, +)$ are in $F$. Note that this is precisely the boundary of $n$-dimensional crosspolytope from \Cref{ex-crosspolytope}. In particular, $\Delta_{\C_n} = \Delta^\ant_{\C_n} \cong (\S^{n-1}, x\mapsto -x)$, hence $\sd_\simp(\C_n)\geq n -1$. This also witnesses that $\sd_\simp(\H)\geq \VC(\H) - 1$ for an arbitrary class~$\H$. The fact that $\sd(\C^n)\leq n-1$ follows by the Borsuk-Ulam theorem (\Cref{t-BU}) as, up to homeomorphism, $\Delta^\ant_{\C_n}$ is exactly $\SS^{n-1}$.
	
	To prove that $\sd_\simp(\H)\geq \VC^{*\an}(\H) - 2$, we can assume, without losing generality, that $\H$ is a minimal class witnessing $\VC^{*\an}(\H)\geq n$, for $n\geq 2$. Then the domain $X$ can be associated with the subset of $2^{[n]}$ such that for every $S \subseteq [n]$ exactly one of $S$ and $-S = [n] - S$ is in $X$, and~$\H$ is a set of $n$ concepts $h_i$ for $i=1, \dots, n$, where $h_i = [i\in S]$ is the indicator function of $i$ in $S$, that is, $h_i(S) = +$ if $i\in S$ and $-$ otherwise.
	
	Let $\B^{n-2}$ be the barycentric subdivision of the boundary of $(n-1)$-dimensional simplex from \Cref{ex-barycentric}. Recall that the vertices of $\B^{n-2}$ are all subsets of $[n]$, except for $\emptyset$ and $[n]$, simplices are chains of such subsets, and the antipodality map is the complementation $S\mapsto [n]-S$. Let us define the map $f\colon V(\B^{n-2})\rightarrow V(\Delta_\H) \subseteq X\times Y$ as $f(S) = ([S\in X]S, [S\in X])$. For example, for $n=3$ and the set $12 = \{1,2\}$, $f(12) = (12, +)$ if $12\in X$ and $(-12, -) = (3, -)$ otherwise. Note that $f$ is trivially one-to-one, moreover, it commutes with the antipodality: $f(-S) = ([-S\in X](-S), [-S\in X]) = ([S\in X]S, -[S\in X]) = -f(S)$. Thus, for $f$ to witness an embedded simplicial $(n-2)$-sphere in $\Delta^\ant_\H$ it is enough to show that it is simplicial. For that, let $\sigma = S_1 \subsetneq, \dots, \subsetneq S_k$ be a chain of subsets of $[n]$ such that $S_1\neq \emptyset$ and $S_k\neq [n]$; in other words, $\sigma$ is a simplex of $\B^{n-2}$. Let $i\in S_1$. We claim that $h_i$ witnesses $f(\sigma) \in S(\Delta_\H)$, that is, that $h_i(x) = y$ for all $(x, y) = f(S_j)$, $j=1, \dots, k$. But indeed, $f(S_j) = ([S_j\in X]S_j, [S_j\in X])$, and $h_i([S_j\in X]S_j) = \bigl[i\in[S_j\in X]S_j\bigr] = [S_j\in X]$, as needed, where we have used the fact that $i\in S_j$.
	
	The $\sd_\simp(\H)\geq \VC^{*\an}(\H) - 2$ bound implies $\sd_\simp(\U_n)\geq n - 2$. The upper-bound $\sd(\U_n) \leq n - 2 + \delta_n$ follows from	
	$$\lceil(\sd(\U_n) + 3)/2\rceil \leq \List(\U_n, \varepsilon) \leq  \List(\U_n) \leq \lceil (n+1)/2 \rceil,$$
	where the first inequality is by~\Cref{th-lb-on-lr} and the last one is by Theorem~E in~\cite{chase2024local}, where $\delta_n$ appears from removing the rounding. Note also that \Cref{th-lb-on-lr} does not apply whenever $\sd(\U_n) \leq 0$, which can only happen for $n=1$ or $2$. 
	For $n=1$, $\U_1$ has just one concept and hence, by Lemma~\ref{lem-trivial}, $\sd(\U_n) = -1 = 1-2 + \delta_1$, as needed. And, for $n=2$, the bound is $\sd(\U_2) \leq 1$ and it is true by the above argument whenever $\sd(\U_2)\geq 1$, which trivially implies that it is true in general.
\end{proof}

Let us also slightly refine Conjecture~\ref{conj-sd-vc-2}, by explicating the intuition that we expect that the smallest VC dimension for a class with $\sd \geq n$ to be achieved by the family $\U_n$, and similarly by $\C_n$ for VC$^*$.
\begin{customconjecture}{\ref{conj-sd-vc-2}}
	Spherical dimension is at most exponential in both $\VC$ and $\VC^*$.
	Moreover
	\begin{itemize}
		\item The optimal upper-bound on $\sd$ in terms of $\VC$ is witnessed by the family of universal classes $\U_n$. That is, $\sd(\H) \leq \max \{\sd(\U_n)~|~\VC(\U_n)\leq \VC(\H)\} = 2^{\VC(\H) + 1} - 3$.
		\item Similarly, the optimal upper-bound on $\sd$ in terms of $\VC^*$ is witnessed by the family of cubes $\C_n$. That is, $\sd(\H) \leq \max \{\sd(\C_n)~|~\VC^*(\C_n)\leq \VC^*(\H)\} = 2^{\VC^*(\H) + 1} - 2$.
		 	\end{itemize}
\end{customconjecture}
\noindent Here we also implicitly conjecture that the lower bound on $\sd(\U_n) \geq n-2$ in Lemma~\ref{lem-vc-lb} is sharp.

\subsection{Classification of low-VC classes via spherical dimension}\label{sec-low-dim}

\begin{proposition}\label{prop-thresholds}
 Let $\T_d=\{h_i ~|~ i\in 0, \dots, d\}$, for $d\geq 1$, be the class of thresholds over $X=[d]$, that is, $h_i(j) = +$ if $j \leq i$ and $-$ otherwise; in particular, $h_0$ and $h_d$ are all $-$ and all $+$ concepts respectively. Then $\sd_\simp(\T_d) = \sd(\T_d)=0$.
\end{proposition}
Look also at \Cref{fig-EMBEDDING} in \Cref{sec-cubical-bary} for an illustration for the $\Delta(\T_3)$ and $\Delta^\ant(\T_3)$.
\begin{proof}
    Note that for all $x\neq y\in [d]$ either $\{(x,+),(y,-)\}$ or $\{(x,-),(y,+)\}$ is not realizable by $\T_d$. Hence the only elements in $\Delta_{\T_d}^\ant$ are those distributions that are supported by the all $+$ or the all $-$ function. Let
    \begin{align*}
        A&=\{\mu\in \Delta_{\T_d} \;:\; L_\mu(h_0)=0\},
        \\-A&=\{\mu\in \Delta_{\T_d} \;:\; L_\mu(h_d)=0\}.
    \end{align*}
    Then both $A$ and $-A$ are nonempty, closed, and disjoint, and $\Delta_{\T_d}^\ant=A\sqcup (-A)$; in particular, $A$ and $-A$ are the two connected components of $\Delta_{\T_d}^\ant$. As $\Delta_{\T_d}$ is nonempty, $\sd_\simp(\T_d) \geq 0$. At the same time, as for every $x\in \Delta_{\T_d}^\ant$, $-x$ lives in the other connected component of $\Delta_{\T_d}^\ant$ (if $x\in A$ then $-x \in -A$ and vice versa), $\Delta_{\T_d}^\ant$ cannot contain an equivariant image of a connected antipodal space, from which $\sd(\T_d)\leq 0$, as needed. 
\end{proof}

\begin{proposition}\label{prop-embed-threshold}
    Let $\H$ be a concept class over a finite domain $X$ such that $\sd(\H)\leq 0$ and  $\VC(\H)\leq 1$. Then $\H$ can be embedded into thresholds up to a bit-flip. That is, there is a bit-flip function $p:X\to\{-,+\}$ such that the class $p\cdot \H = \{p\cdot h~|~h\in \H\}$ can be embedded into a class of thresholds. 
\end{proposition}
We note that the bit-flip changes neither $\VC$, nor $\VC^{*\an}$, nor any of the $\sd$'s.
\begin{proof}
	Assume $\H$ is such as in the statement. By bit-flipping it, we can assume that the all $+$ concept is in $\H$. We further assume that for every $x\in X$ there is a concept $h\in \H$ such that $h(x) = -$; otherwise, we can restrict $\H$ to a subdomain for which the above is satisfied, embed it into a threshold there, and then it can be easily seen that it can be easily extended to a threshold on a full domain $X$. We can also safely assume that for each $x_1,x_2\in X$ there is some $h\in \H$ such that $h(x_1)\neq h(x_2)$. Otherwise, $x_1$ is equivalent to $x_2$ and we can think of them as of the same point.
     
    Define a partial order on $X$ by $x\leq z$ if for all $h\in \H$ we have that $h(x)=-$ implies $h(z)=-$; in other words, that the sequence $\{(x, -), (z, +)\}$ is not realizable by $\H$. It is easy to see that this relation is indeed reflexive and transitive, and the assumption that there is always some $h\in \H$ with $h(x)\neq h(z)$ guarantees that it will also be antisymmetric. Hence, it is indeed a partial order. Moreover, it is easy to see that $\H$ can be embedded in a threshold whenever this order is linear, that is, if any two elements are comparable.

    Observe that if two elements $x$ and $z$ are incomparable with respect to this order, then the sequence $\{(x, -), (z, -)\}$ is not realizable by $\H$. Indeed, as $\VC(\H)\leq 1$, one of the four sequences $\{(x, +), (z, +)\}$, $\{(x, -), (z, +)\}$, $\{(x, +), (z, -)\}$, or $\{(x, -), (z, -)\}$ is not realizable. However, the first is realizable by the existence of all $+$ concept, and the second and the third imply $x\leq z$ and $z\leq x$ respectively. So, if $x$ and $z$ are incomparable, this leaves out only the last option, as needed. Let us now prove several other properties of this order. 

    First: There are no two incomparable elements with a common lower bound. Alternatively, if for $z_1, z_2\in X$ there is $x\in X$ such that $x\leq z_1, z_2$, then $z_1$ and $z_2$ are comparable. Indeed, let us take $h\in \H$ such that $h(x)=-$. Then, by the definition of the order, $h(z_1) = h(z_2) = -$, so the sequence $\{(z_1, -), (z_2, -)\}$ is realizable and hence, by the above, $z_1$ and $z_2$ are comparable.

    Second: There are no two incomparable elements with a common upper bound. Note that despite the similarity, the proof of this statement relies on the $\sd(\H)\leq 0$ assumption, not used in the previous statement. So suppose there are $z_1, z_2$, and $x$ such that $z_1$ and $z_2$ are incomparable and $z_1, z_2\leq x$. We claim that then the simplicial $1$-dimensional sphere in~\Cref{fig-SINGLETONS} (left) is realizable by $\H$ and thus witnesses $\sd_\simp(\H)\geq 1$. Indeed, the edges $\{(x, +), (z_1, +)\}$ and $\{(x, +), (z_2, +)\}$ are realized by all $+$ concept. Edges $\{(z_1, +), (z_2, -)\}$ and $\{(z_1, -), (z_2, +)\}$ are realized by the assumption that $z_1$ and $z_2$ are incomparable. And the edge $\{(x, -), (z_1, -)\}$ is realized by a function $h$ such that $h(z_1)=-$: as $z_1\leq x$, this implies that $h(x)=-$; the edge $\{(x, -), (z_2, -)\}$ is handled similarly.

    Third: there are no three incomparable elements. Indeed, if $x$, $z$, and $w$ are such incomparable elements then, similarly to the argument above, the simplicial $1$-dimensional sphere in~\Cref{fig-SINGLETONS} (right) is realizable by $\H$ and thus witnesses $\sd_\simp(\H)\geq 1$.
    
    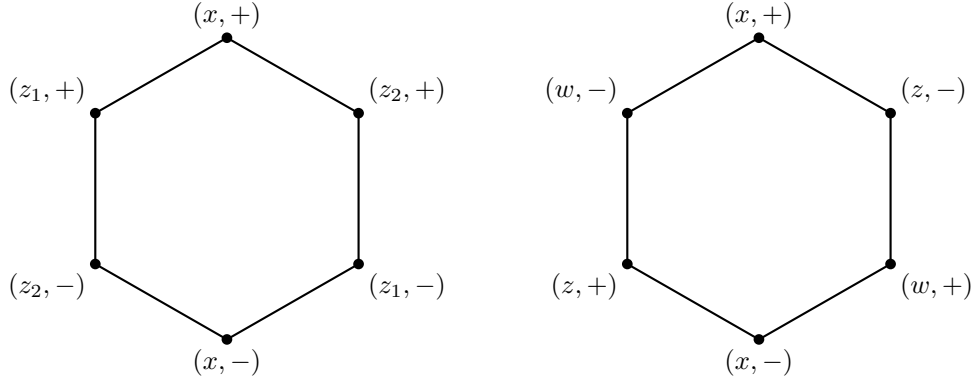
\begin{figure}[!hbt]
	\centering
	 \usetikzlibrary{calc}
\begin{tikzpicture}
    \begin{scope}[xshift = -7cm]        
    \coordinate (A) at (90:2cm);   
    \coordinate (B) at (30:2cm);  
    \coordinate (C) at (-30:2cm); 
    \coordinate (D) at (-90:2cm); 
    \coordinate (E) at (-150:2cm);
    \coordinate (F) at (150:2cm); 

    \draw[thick] (A) -- (B) -- (C) -- (D) -- (E) -- (F) -- cycle;

    \foreach \point in {A, B, C, D, E, F} {
        \fill[black] (\point) circle (2pt);
    }

    \node[above] at (A) {$(x,+)$};
    \node[above right] at (B) {$(z_2,+)$};
    \node[below right] at (C) {$(z_1,-)$};
    \node[below] at (D) {$(x,-)$};
    \node[below left] at (E) {$(z_2,-)$};
    \node[above left] at (F) {$(z_1,+)$};
    \end{scope}

    \begin{scope}        
    \coordinate (A) at (90:2cm);   
    \coordinate (B) at (30:2cm);  
    \coordinate (C) at (-30:2cm); 
    \coordinate (D) at (-90:2cm); 
    \coordinate (E) at (-150:2cm);
    \coordinate (F) at (150:2cm); 

    \draw[thick] (A) -- (B) -- (C) -- (D) -- (E) -- (F) -- cycle;

    \foreach \point in {A, B, C, D, E, F} {
        \fill[black] (\point) circle (2pt);
    }

    \node[above] at (A) {$(x,+)$};
    \node[above right] at (B) {$(z,-)$};
    \node[below right] at (C) {$(w,+)$};
    \node[below] at (D) {$(x,-)$};
    \node[below left] at (E) {$(z,+)$};
    \node[above left] at (F) {$(w,-)$};
    \end{scope}

\end{tikzpicture}
	\caption{One-dimensional simplicial spheres in $\Delta_{\H}^\ant$, witnessed by either two incomparable elements $z_1$ and $z_2$ with a common upper bound $x$ (left), or three incomparable elements $x$, $z$, and $w$ (right).}
	\label{fig-SINGLETONS}       
	\end{figure}  

    Now note that the third property implies that there are at most two minimal elements in $X$. If there is just one such minimal element, then the first property implies that all elements are comparable, in which case the order is linear, and hence $\H$ can be embedded into a threshold, as needed. Then the second property implies that each $x\in X$ is comparable with exactly one of $x_1$ and $z_1$. This implies that our domain consists of two disjoint chains $X=\{x_i\}_{i=1}^n\sqcup \{z_j\}_{j=1}^m$, where $x_i\leq x_{i+1}$, $z_{j}\leq z_{j+1}$, and $x_i$ and $z_j$ are incomparable for all $i,j$. In particular, $x_n$ and $z_m$ are incomparable, which implies that the sequence $\{(x_1,-),(z_m,-)\}$ is not realizable; alternatively, $h(x_n)=-$ implies $h(z_m)=+$ for all $h\in \H$. 

    Now, using the contrapositive of the definition of our partial order, it is clear that $h(z_m)=+$ implies $h(z_{j})=+$ for all $j$, hence we can define a bit-flip function $p: X\to \{-,+\}$ by $p(x_i)=+$ and $p(z_j)=-$ for all $i$ and $j$. Then it is easy to see that after applying this bit-flip, the elements of $p\cdot\H$ will be linearly ordered as
        \[
        x_1\leq x_2 \dots \leq x_n\leq z_m\leq z_{m-1}\dots z_1.
        \]
    In particular, $p\cdot\H$ can be embedded into a threshold, as needed.
\end{proof}

\begin{customlem}{\ref{lem-trivial}}
	Let $\H$ be a nonempty class over a finite domain. Then one of the following mutually exclusive cases holds:
	\begin{itemize}
		\item $\H$ is a class with one hypothesis ($\sd = -1$, $\VC=0$);
		\item $\H$  has at least two hypotheses and, up to a bit-flip, is a subclass of thresholds ($\sd = 0$, $\VC=1$);
		\item $\H$ is a $\VC=1$ class that is not, up to a bit-flip, a subclass of thresholds ($\sd = 1$, $\VC=1$);
		\item $\VC(\H)\geq 2$ and $\sd(\H)\geq 1$.
	\end{itemize}
	Moreover, in the above classification, $\sd$ is any from $\sd$ or $\sd_\simp$.
	
	Moreover, the first bullet holds even without the finite domain assumption, that is, $\sd_\infty(\H) = -1$ if and only if $\H$ has at most one hypothesis.	
\end{customlem}
\begin{proof}
We start by noting that $\sd=-1$, for any $\sd$, is equivalent to $\Delta_\H^\ant$ being empty, which, in turn, is equivalent to $\H$ having at most one hypothesis: indeed, if $\H$ contains $h_1\neq h_2$, then there is $x\in X$ such that $h_1(x)\neq h_2(x)$, in which case $(x, -), (x, +) \in \Delta_\H^\ant$ and witness $\sd_\simp \geq 0$. 

So assume from now on that $\H$ is finite and has at least two hypotheses. Note that if $\VC(\H) \geq 2$ then, by Lemma~\ref{lem-vc-lb}, $\sd_\simp \geq 1$, and so this situation falls under the last bullet in the classification. So let us now assume that $\VC(\H) = 1$. We now need to check that the second and the third bullets exhaust the possible situations. That is, that $\sd(\H)$ is either $0$ or $1$, and that $\sd(\H)=0$ if and only if $\H$ can be embedded into thresholds.

Note that $\VC(\H) = 1$, again by Lemma~\ref{lem-vc-lb}, implies that $\sd_\simp\geq 0$. At the same time, it is well-known and easy to prove that $\VC(\H) = 1$ implies that $\H$ can be embedded into an extremal class $\E$ with $\VC(\E) = 1$. Then, by (yet to be proven) \Cref{t-sd-for-extremal}, $\sd(\H) \leq 2\VC(\E) - 1 = 1$. Thus, indeed, $\sd(\H)$ is either $0$ or $1$. 

Finally, Propositions~\ref{prop-thresholds} and~\ref{prop-embed-threshold} above establish that, in one direction, $\sd(\H)\leq 0$ whenever $\H$ is embeddable into a threshold; here we also use monotonicity of $\sd$ and the fact that bit-flip does not change $\sd$ or $\sd_\simp$. And, in the other direction, $\H$ can be embedded, up to a bit-flip, into a threshold whenever $\sd(\H)\leq 0$. 
\end{proof}

\subsection{Stability and list-replicability}\label{sec-lr-proofs}

The proof of~\Cref{th-lb-on-lr} relies on the following statement:
\begin{proposition}[Corollary~3 in~\cite{chase2024local}]\label{prop-closed-cover}
	For a finite class $\H$ over $X$, $\List(\H, \varepsilon)$ is the minimal integer $L$ such that there exist a closed cover $\F = \left\{A_h~|~h\in \{-, +\}^X\right\}$ of $\Delta_\H$ of overlap-degree at most $L$, such that $L_\mu(h) \leq \varepsilon$ for all $h\in \{-, +\}^X$ and $\mu\in A_h$.
\end{proposition}

\begin{customthm}{\ref{th-lb-on-lr}}
Let $\H$ be a concept class such that $\sd(\H) \geq 1$. Then, for any $0 < \varepsilon < 1/2$, it holds 
$$\List(\H, \varepsilon)\geq \lceil(\sd(\H) + 3)/2\rceil.$$
In particular, $\List(\H, \varepsilon)\geq 1 + \lceil \VC(\H)/2\rceil$ whenever $\VC(\H) \geq 2$, and $\List(\H, \varepsilon)\geq 1 + \lfloor \VC^{*\an}(\H)/2\rfloor$ (with no additional condition). 

The only exception for the bounds in terms of $\VC$ and $\sd$ are the classes of the form $\H = {h_-, h_+}$ such that $h_-(x) = -h_+(x)$ for all $x\in X$, for example, $\C_1$, for which $\List(\H, \varepsilon) = 1$,  $1 + \lceil \VC(\H)/2\rceil = 1 + \lceil 1/2\rceil = 2$, and $\lceil(\sd(\H) + 3)/2\rceil = \lceil(0 + 3)/2\rceil = 2$.
\end{customthm}
\begin{proof}
	The main statement of the theorem is a straightforward application of Local LS (\Cref{t:localLS}) to Proposition~\ref{prop-closed-cover}, which goes as follows. 
	Let $\beta \colon \S^d \rightarrow \Delta^\ant_{\H'}$, where $d = \sd(\H)\geq 1$, $\H'$ is the restriction of $\H$ to a finite domain, and $\beta$ an equivariant map, such that $\H'$ and $\beta$ witness $\sd(\H) = d$. 
	Let $\F$ be the closed cover from Proposition~\ref{prop-closed-cover} such that $\List(\H', \varepsilon)$ is the overlap degree of $\F$. Note that the property $L_\mu(h) + L_{-\mu}(h) = 1$ (\Cref{eq-loss}) implies that all sets in $\F$ are antipodal-free. Let $\G = \beta^{-1}[\F]$. Then $\G$ is a closed antipodal cover of $\S^n$, whose overlap equals to the one of $\F$. Then, by \Cref{t:localLS}, $\List(\H, \varepsilon) \geq \List(\H', \varepsilon) \geq \lceil(d + 3)/2\rceil$, as needed.
	
	The bounds in terms of $\VC$ and $\VC^{*\an}$ follow from Lemma~\ref{lem-vc-lb}, where the $\VC^{*\an}$ bound is true with an additional condition $\VC^{*\an}(\H) \geq 3$, which we will lift in a short while. The remaining claims are easy case studies. Note that, by definition $\List(\H, \varepsilon)\geq 1$. Let us prove the following easy statement: \emph{For a class $\H$, $\List(\H, \varepsilon)=1$ if and only if $\H$ has just one concept or $\H = \{h_-, h_+\}$ from the statement of the theorem.} The fact that $\List(\H, \varepsilon)=1$ for the two classes described above is trivial. So assume $\H$ does not fall under these two cases. In all other situations, there are concepts $h_1, h_2 \in \H$ and $x_1, x_2\in X$ such that $h_1(x_1)\neq h_2(x_1)$ and $h_1(x_1)= h_2(x_1)$.  Assuming, for simplicity, that $h_1(x_1) = h_1(x_2) = h_2(x_2) = +$ and $h_2(x_1) = -$, we observe that for $\H'$ = $\H|_{\{x_1, x_2\}}$, $\Delta_{\H'}$ contains simplices $\{(x_1, -), (x_2, +)\}$ and $\{(x_1, +), (x_2, +)\}$, which means that $\Delta_{\H'}$ contains a path between antipodal vertices $(x_1, -)$ and $(x_1, +)$. If $\F$ is the covering of $\Delta_{\H'}$ from Proposition~\ref{prop-closed-cover}, then no $F$ in $\F$ can cover both $(x_1, -)$ and $(x_1, +)$, and so, essentially by the definition of connectedness, the overlap of $\F$ is at least $2$. Hence, $\List(\H, \varepsilon)\geq \List(\H', \varepsilon)\geq 2$, as needed.
	
	For the VC bound $\List(\H, \varepsilon)\geq 1 + \lceil \VC(\H)/2\rceil$, it is true for $\VC \geq 2$. It is also easy to see that, in the remaining cases, it can only be invalidated if $\VC(\H) = 1$ and $\List(\H, \varepsilon) = 1$. By the above statement, it can only be the case when $\H = \{h_-, h_+\}$, and indeed, this case is an exception.
	
	The $\VC^{*\an}$ bound $\List(\H, \varepsilon)\geq 1 + \lfloor \VC(\H)/2\rfloor$ is true for $\VC^{*\an} \geq 3$, and again, can only be invalidated if $\VC^{*\an}(\H) = 2$ and $\List(\H, \varepsilon) = 1$. However, this is impossible, as in both two cases $\H=\{h\}$ or $\H = \{h_-, h_+\}$ for which $\List(\H, \varepsilon) = 1$, the corresponding class does not dually antipodally shatter two concepts.
	
	Finally, the main bound $\List(\H, \varepsilon)\geq \lceil(\sd(\H) + 3)/2\rceil$ is true for $\sd >0$ and can only be invalidated if $\sd(\H) = 0$ and $\List(\H, \varepsilon) = 1$. Again, just as in the VC bound, it can only happen when $\H = \{h_-, h_+\}$, and this case is an exception.
\end{proof}

\subsection{SCO reductions}\label{sec-sco-proofs}

The proof of \Cref{t:reductions} closely follows the proof of Theorem~$1$ in~\cite{reductions}, so we will only do a brief walkthrough. It relies on the following version of the Borsuk-Ulam theorem:

\begin{theorem}[Theorem~$9$ in \cite{reductions}]\label{t:BU-open}
Let $W$ be a convex set in $\mathbb{R}^d$ and let $G\subseteq \mathbb{S}^n \times W$. Additionally, suppose that:
\begin{itemize}
    \item For any $u\in \mathbb{S}^n$, the set $G_u = \{w\in W~|~(u,w)\in G\}$ is nonempty;
    \item For any $u\in \mathbb{S}^n$, the convex hulls of $G_u$ and $G_{-u}$ are disjoint;
    \item For any $w\in W$, the set $G^w =\{u\in \mathbb{S}^n ~|~ (u,w)\in G\}$ is open.
\end{itemize}	
Then $d\geq n+1$.
\end{theorem}     

Also, as a technical statement, we use the following: 
\begin{lemma}[Lemma~$21$ in \cite{reductions}]\label{l:Inf-Lin-Contin}
 Let $I\subseteq \R^d$ be a nonempty collection of pointwise positive vectors in $\mathbb{R}^d$. Let $X=\{x\in [0,1]^d\;:\; \sum_{i=1}^d x_i=1\}$ and define $F:X\to \mathbb{R}$ by
 \[
 F(x)=\inf_{\alpha\in I}\langle x, \alpha\rangle.
 \]
 Then $F$ is continuous.
\end{lemma}

\begin{customthm}{\ref{t:reductions}}
    For any class $\H$ we have \[
    \sd(\H) + 1\leq \mathtt{SCO}(\H).
    \]
\end{customthm}
\begin{proof}
    Let $\mathtt{SCO}(\H)=d$. 
    Recall that it means that there are abstract set $Z$, convex set $W\subseteq \R^d$, and a collection of convex loss functions \( \{ \ell_z \}_{z \in Z} \) from \( W \) to $\R^+$, and a pair of functions $r_\inp: X\times\{-,+\}\to Z$ and $r_\outp: W\to \{-,+\}^X$, defining an $(\alpha, \beta)$ reduction, for $\alpha > 0$ and $0< \beta < 1/2$. That is, $r_\inp$ and $r_\out$ are such that 
    and for every realizable distribution $\mu\in \Delta_\H$ we have 
    \[
        L_{r_\inp(\mu)}(w)\leq \inf_{w'\in W}L_{r_\inp(\mu)}(w')+\alpha\implies L_\mu(r_\out(w))\leq \beta.
        \]
    
    Let also $\sd(\H)=n$ and $f\colon \S^n \rightarrow \Delta_\H^\ant$ be a finitely supported $n$-sphere witnessing it. Note that the fact that the sphere is finitely supported enables us to assume, without losing generality, that the sets $X$, and, consequtively, $Z$, are finite.
    Let us now define $G\subseteq \S^n \times W$ as:
    \begin{align*}
    	G &= \left\{ (u, w)~|~ \text{$w$ is $\alpha/2$-optimal for $r_\inp(f(u))$}\right\}
    	\\&=\left\{(u, w)~|~ L_{r_\inp(f(u))}(w)< \inf_{w'\in W}L_{r_\inp(f(u))}(w')+\alpha/2\right\}.
    \end{align*}
    
    Note that if we prove that $G$ satisfies the conditions in \Cref{t:BU-open}, the application of this theorem yields the desired bound $n+1\leq d$. Also, out of the three conditions on $G$, the first one is trivially satisfied by construction, and the second one follows by observing that each $G_u$ is convex and if $w\in G_u \cap G_{-u}$ then the hypothesis $r_\out(w)$ is $\beta$-optimal for both distributions $f(v)$ and $-f(v)$, contradicting \Cref{eq-loss}.
    
    Finally, for $w\in W$, $G^w$ is defined as $G^w = \left\{u\in \S^n~|~F_w(f(u)) < \alpha/2 \right\}$, where
    $$F_w(\mu) = L_{r_\inp(\mu)}(w) - \inf_{w'\in W}L_{r_\inp(\mu)}(w').$$
    Thus, $G^w$ is open whenever $F_w$ is continuous, for which, in turn, it is enough to check that the map $\OPT(\eta) = \inf_{w'\in W}L_{\eta}(w')$ is continuous; here $\eta = r_\inp(\mu)$ is a probability distribution over $Z$. The latter follows by applying Lemma~\ref{l:Inf-Lin-Contin} to $\OPT$ rewritten as 
    \begin{align*}
    	\OPT(\eta) &= \inf_{w'\in W} \Ex_{z\sim \eta} l_z(w')
	    	= \inf_{w'\in W} \sum_{z\in Z} \Pr_\eta(z) \cdot l_z(w').
    \end{align*}

\end{proof}

\begin{customthm}{\ref{prop-sign-rank-prod}}
    For any two concept classes $\H_1$, $\H_2$ we have
    \[
    \sr(\H_1\times \H_2)\leq \sr(\H_1)+\sr(\H_2).
    \]
\end{customthm}
\begin{proof}
    For $i\in \{1,2\}$, let $n_i$ be the sign-ranks of $\H_i$, and let
    $\varphi_i: X_i\to  \mathbb{R}^{n_i} $ and $w_i:\H_i\to \R^{n_i}$ be the respective representations maps, witnessing it (as per Definition~\ref{def-sign-rank}). Thus, for all $x_i\in X_i$ and $h_i\in \H_i$ we have \[
     h_i(x_i)=\sign\langle w_i(h_i), \varphi_i(x_i)\rangle.
    \]
    Now define the representation maps $\varphi: X_1\sqcup X_2\to \R^n\times \R^m=\R^{n+m}$, $w:\H_1\times \H_2\to \R^n\times\R^m=\R^{n+m}$  by 
    \[
\varphi(x)=\begin{cases}
	\big(\varphi_1(x),0\big) &x\in X_1, \\
	\big(0,\varphi_2(x)\big) &x\in X_2,
	\end{cases}	
    \]
    and 
    \[
    w(h_1\times h_2)=(w(h_1),w(h_2)).
    \]
    A simple calculation shows that we indeed have 
    \[
    h(x)=\mathtt{sign} \langle w(h), \varphi(x)\rangle.
    \]
    for all $h\in \H_1\times \H_2$, $x\in X_1\sqcup X_2$, which implies the desired result.
\end{proof}

\subsection{Products of classes}\label{sec-products}
For classes $\H_1$ and $\H_2$ on disjoint domains $X_1$ and $X_2$, the \emph{product} $\H_1 \times \H_2$ is defined as:
$$\H_1 \times \H_2 = \{h_1\times h_2~|~h_1\in \H_1, h_2\in \H_2\},$$
where $h = h_1\times h_2$ is a function on $X_1\sqcup X_2$ such that $h(x) = h_1(x)$ for $x\in X_1$ and $h_2(x)$ for $x\in X_2$. In simplicial geometry, joins naturally correspond to direct products, see~\Cref{sec-joins}. In particular, this extends to our setup:
\begin{lemma}\label{lem-products-classes}
	For classes $\H_1$ and $\H_2$, $\Delta_{\H_1 \times \H_2} = \Delta_{\H_1} * \Delta_{\H_2}$, where the join is a geometric join of antipodal spaces. Moreover, if the domains of $\H_1$ and $\H_2$ are finite, in which case $\Delta_{\H_1}$ and $\Delta_{\H_2}$ have an additional structure of simplicial complexes, then the above equality is also in the sense of the join of antipodal simplicial complexes.
\end{lemma}
\begin{proof}
	The correspondence is established by representing $\mu$ on $X_1\sqcup X_2$, realizable by $\H_1 \times \H_2$, in the barycentric coordinates of the join as
	$$\mu = \mu(X_1) \cdot \frac{\mu|_{X_1}}{\mu(X_1)} + \mu(X_2) \cdot \frac{\mu|_{X_2}}{\mu(X_2)}.$$
	
	Note that, unless $\mu(X_1) = 0$, the measure $\mu|_{X_1}/\mu(X_1)$ is a distribution, realizable by $X_1$. And the case $\mu(X_1) = 0$, and, similarly, $\mu(X_2) = 0$, aligns with the definition of the join, as in this case the corresponding coefficient $\mu(X_1)$ is $0$, and hence the term  $\mu|_{X_1}/\mu(X_1)$ does not matter. The check for the simplicial case is trivial.
\end{proof}

\begin{customlem}{\ref{lem-vcvc*-lb}}
	For classes $\H_1$ and $\H_2$, $\VC(\H_1 \times \H_2) = \VC(\H_1) + \VC(\H_2)$ and $\sd(\H_1 \times \H_2) \geq \sd(\H_1) + \sd(\H_2) + 1$, where $\sd$ can be any from $\sd_\infty$, $\sd$, or $\sd_\simp$. 
	
	In particular, for an $m$'th direct power $\U^m_n$ of the universal class $\U_n$, we have:
	\begin{itemize}
		\item $\VC(\U^m_n) = m\lfloor\log n\rfloor$;
		\item $n\leq \VC^*(\U^m_n) \leq n + \lfloor\log m\rfloor$. Moreover, for $n\geq 3$, the lower bound can be improved to $n - 2 + \lfloor\log m\rfloor\leq \VC^*(\U^m_n)$;
		\item $m(n-1) - 1 \leq \sd_\simp(\U^m_n) \leq \sd(\U^m_n) \leq mn$.
	\end{itemize}
	In particular, for the class $\U_n^n$, $\VC(\U^n_n) \sim \VC^*(\U^n_n) \sim n$ (up to a log factor) and $\sd_\simp(\U^n_n) \gtrsim n^2$. More generally, for $\alpha, \beta \in \R^+$, we can define a family of classes $\U_{\beta n}^{\alpha n}$ (with some rounding), that would witness the asymptotical behavior $\VC \sim n^\alpha$, $\VC^* \sim n^\beta$, and $\sd_\simp \sim n^{\alpha + \beta}$.
\end{customlem}
\begin{proof}
The equality $\VC(\H_1 \times \H_2) = \VC(\H_1) + \VC(\H_2)$ is trivial, and the inequality $\sd_\infty(\H_1 \times \H_2) \geq \sd_\infty(\H_1) + \sd_\infty(\H_2) + 1$ follows from the statement $\Delta_{\H_1 \times \H_2} = \Delta_{\H_1} * \Delta_{\H_2}$ in Lemma~\ref{lem-products-classes}, together with the discussion in~\Cref{sec-joins}, in particular, from $\SS^n * \SS^m \cong \SS^{n+m+1}$ in Proposition~\ref{prop-join-spheres}. Indeed, if $\beta_1\colon \S^{d_1} \rightarrow \Delta_{\H_1}$ and $\beta_2\colon \S^{d_2} \rightarrow \Delta_{\H_2}$ witness $\sd_\infty(\H_1)\geq d_1$ and $\sd_\infty(\H_2)\geq d_2$ respectively, then $\beta_1 * \beta_1 \colon \S^{d_1 + d_2 +1} \rightarrow \Delta_{\H_1\times \H_2}$ witness $\sd_\infty(\H_1 \times \H_2)\geq d_1 + d_2 + 1$. 

The fact that it applies to all three $\sd$'s also requires only some straightforward checks. The $\sd$ case is by taking some finitely supported witnesses $\H_1'$ and $\H_2'$, and the $\sd_\simp$ is by the discussion in~\Cref{sec-joins} that the geometric join is consistent with the simplicial one. 

The equality $\VC(\U^m_n) = m\lfloor\log n\rfloor$ and the inequality $\sd_\simp(\U^m_n)\geq m(n-2) + (m-1) = m(n-1) - 1$ follow from the above and from the bounds $\VC(\U_n) = \lfloor \log n \rfloor$ and $\sd_\simp(\U_n) \geq n-2$ in Lemma~\ref{lem-vc-lb}. The upper bound $\sd(\U^m_n) \leq mn$ follows from Proposition~\ref{prop-sign-rank-prod} as 
\begin{align*}
	\sd(\U^m_n) \leq \sr(\U^m_n) \leq m\cdot \sr(\U_n) = mn.
\end{align*}

The upper bound $\VC^*(\U^m_n) \leq n + \lfloor\log m\rfloor$ follows by observing that in order to dually shatter~$d$ functions, the class should have at least $2^d$ elements in the domain, so $2^{\VC^*(\U^m_n)} \leq m2^n$. The $n \leq \VC^*(\U^m_n)$ lower bound holds because $\U_n$ is a restriction of $\U^m_n$. 

The $n - 2 + \lfloor\log m\rfloor\leq \VC^*(\U^m_n)$ lower bound follows from the following construction. Recall that the natural parametrization of $\U_n$ is to consider its domain to be all subsets of $[n]$, and its functions $h_i$ to be characteristic functions of an element, $h_i(S) = [i\in S]$, for $i\in [n]$. Let $\U_n'$ be the same class, but additionally containing all $-$ and all $+$ concepts. Trivially, $\U'_{n-2}$ is a subclass of $\U_n$, so assume that we are dealing with $\U'_{n-2}$ instead of $\U_n$. Let $d$ be maximal such that $2^d \leq m$; assume, without losing generality, that $m = 2^d$, and associate $m$ with the set of all subsets of $[d]$, making the domain of $(\U'_{n-2})^m$ to be $2^{[n-2]}\times 2^{[d]}$. Now, for $i \in [n-2 + d]$, let us define the function $g_i$ as:
\begin{itemize}
	\item For $i \in [n-2]$, $g_i = h_i^m$, that is, $g_i(S, P) = [i\in S]$, for all $S\subseteq[n-2]$ and $P\subseteq [d]$;
	\item For $j \in [d]$, $g_{n-2 + j}(S, P) = [j\in P]$. That is, $g_{n-2 + j}(\cdot, P) \equiv -$ whenever $j\notin P$ and $\equiv +$ whenever $j\in P$. As both $\equiv -$ and $\equiv +$ are in $\U'_n$, $g_{n-2 + j}\in (\U'_n)^m$.
\end{itemize}
It is then easy to see that $(\U'_n)^m$ dually shatter $g_i$, for $i\in [n-2 + d]$, as needed.
\end{proof}

\section{Proofs for disambiguations}\label{sec-disamb-proofs}
Although we typically denote a generic class by $\H$, in the context of disambiguations our usual notation for it would be $\D$. In particular, we use $\H$ in  \Cref{sec-disamb-antipodal}, which does not deal specifically with disambiguation and mostly uses $\D$ through the rest of the section.

\subsection{Antipodal domains}\label{sec-disamb-antipodal}
We say that a set $X$ is an \emph{antipodal domain} if it is equipped with a fixed-point free involution $\nu$, which we call \emph{antipodal map}; if the $\nu$ is clear from the context, we will denote $\nu(x)$ by $-x$. This is an obvious relaxation of antipodal space introduced in~\Cref{sec-sd}, in particular, we consider any antipodal space as an antipodal domain with the same antipodality. However, in this definition, we do not assume any topology on $X$ and, naturally, do not require $\nu$ to be continuous. 

For an antipodal domain $X$, we say that a total hypothesis $h$ on $X$ is \emph{antipodal} if $h(-x) = -h(x)$, for all $x\in X$; a class is antipodal if all hypotheses in it are antipodal. We say that $r\colon X\rightarrow X$ is a \emph{representation map} if for every $x\in X$, either $r(x) = r(-x) = x$, or $r(x) = r(-x) = -x$. For a total hypothesis $h$ on $X$ and a representation map $r$ we define the \emph{symmetrization} of $h$ by $r$, denoted $h^r\colon X\rightarrow \{-, +\}$, as 
	$$h^r(x) = \begin{cases}
		h(x) &\text{if } h(x) = -h(-x);\\
		h(x) &\text{if } h(x) \neq -h(-x) \text{ and } x=r(x);\\
		-h(x) &\text{if } h(x) \neq -h(-x) \text{ and } x=-r(x).
	\end{cases}$$
The point of symmetrization is to make a hypothesis antipodal, so it would make sense to call it antipodalization; however, this would sound ugly. Furthermore, for a total class $\H$ and a representation map $r$ over $X$, we define the \emph{symmetrization} of $\H$ by $r$, denoted $\H^r$, as $\H^r = \{h^r~|~h\in \H\}$; crucially, all concepts of $\H$ are symmetrized via the same representation map. Finally, we say that $\H'$ is a symmetrization of $\H$ if $\H' = \H^r$ for some symmetrization map $r$.

It is easy to see that a symmetrization of $\H$ is antipodal and that antipodal classes are invariant under symmetrization. What is also important for us is that $\VC$ and $\VC^{*\an}$ behave well under symmetrization.
\begin{proposition}[Symmetrization does not increase $\VC$ and $\VC^{*\an}$]\label{lem-antipodal-vc}~\\ \vspace{-0.5cm}
	\begin{enumerate}
		\item An antipodal class $\H$ dually shatters $H\subseteq \H$ if and only if it dually antipodally shatters $H$. In particular, $\VC^*(\H)= \VC^{*\an}(\H)$;
		\item Let $\H'$ be a symmetrization of $\H$. Then $\VC(\H')\leq \VC(\H)$ and $\VC^{*\an}(\H') \leq \VC^{*\an}(\H)$.
	\end{enumerate}
\end{proposition}
\emph{Note.} It is not, in general, true that $\VC^*(\H')\leq \VC^{*}(\H)$. To an extent, we introduced dual antipodal shattering as a way of circumventing this problem.
\begin{proof} 
The first statement is trivial: Indeed, for any $x\in X$ that witnesses pattern $t$ on $H\subseteq \H$, by the definition of antipodal class, $-x$  witnesses pattern $-t$. 

The rest of the proof is for the second statement. So let $\H'$ be a symmetrization of $\H$ with a representation map $r$. We will use the following statement, which easily follows from the definition of symmetrization: \emph{If $r(x) = x$, then $h^r(x) = h(x)$, for any $h\in\{-, +\}^X$ and $x\in X$}.

Let $S\subseteq X$ be a set witnessing $\VC(\H')\geq d$, that is, $S = \{s_1, \dots, s_d\}$ is a $d$-set shattered by $\H'$. Let $S' = \{s'_1, \dots, s'_d\}$ be such that $s'_i = s_i$ or $-s_i$, for all $i\in [d]$. We claim that $\H'$ shatters any such $S'$. Indeed, let $t\in \{-, +\}^d$ be an arbitrary pattern on $S$, and let $t'\in \{-, +\}^d$ be defined as $t'(i) = t(i)$ if $s'_i = s_i$ and $-t(i)$ otherwise. Now, if $h'_t\in \H'$ is a concept witnessing $t$ on $S$, that is, such that $h'_t(s_i) = t(i)$, for all $i\in [d]$, then, by construction, $h'_{t'}(s_i') = [s_i=s_i'] \cdot h'_{t'}(s_i) = [s_i=s_i'] t'(i) = [s_i=s_i']^2 t(i) = t(i)$, where $[s_i=s_i']$ is $+$ if $s_i=s_i'$ and $-$ otherwise. Thus, $h'_{t'}$ witnesses the pattern $t$ on $X'$. Trivially, the map $t\mapsto t'$ is a bijection on $\{-, +\}^d$, and so $\H'$ witnesses all the patterns on $S'$, that is, shatters $S'$. 

By the above, by taking an appropriate $S'$ instead of $S$, we can, without losing generality, assume that for every $i\in [d]$, $r(s_i) = s_i$. We claim that then $\H$ shatters $S$. Indeed, if $h'_t\in \H'$ is the concept witnessing $t$ on $S$, and let $h_t\in \H$ be such that $h'_t = h^r_t$. As $r(s_i) = s_i$, by the outlined statement above, it follows that $h'_t(s_i) = h_t(s_i) = t(i)$, for all $i\in [d]$. Thus, $h_t\in \H$ witnesses $t$ on $S$, and so $\H$ shatters $S$, implying $\VC(\H) \geq d$. As this is true for any $d\leq \VC(\H')$, we have $\VC(\H)\geq \VC(\H')$, as needed. 

Regarding the dual dimensions, note first that $\VC^{*\an}(\H') = \VC^*(\H')$ by Proposition~\ref{prop-dual-semi-VC}. Now, suppose $\H'$ dually shatters some $H'\in \H'$. Note that $\H$ as a set can be identified with $\H$, and we will denote by $H$ the set $H'$, treated as a subset of $\H$; in particular, $H' = \{h^r~|~h\in H\}$. 
Let $S\subseteq X$ be the set witnessing all dual patterns for $\H'$ on $H'$ 
and let $S' = \{r(s)~|~s\in S\}$. Then for every dual pattern $t$ on $H'$, there is $s'\in S'$ witnessing for $\H'$ either $t$ or $-t$ on it. 
However, every $s\in S'$ satisfies $s=r(s)$, and so, by the outlined statement, $h\equiv h' = h^r$ on $S'$, for every $h\in H$. Thus, $S'$ witnesses the dual antipodal shattering of $H$ by $\H_n$. This immediately entails $\VC^{*\an}(\H)\geq \VC^{*\an}(\H')$, as needed. 
\end{proof}

For a set $X$, we define the \emph{antipodal extension} $X^a$ of $X$ as a set $X^a = X \times \{-, +\}$ together with a fixed-point free involution $(x, y)\mapsto (x, -y)$. We also often consider $X^a = X^- \sqcup X^+$, where $X^y = X\times \{y\}$, and in this construction we typically identify $X^+$ with $X$. We also usually assume that $X^a$ is equipped with a standard representation map $r(x, y) = (x, +)$. For a hypothesis $h$ on $X$, the \emph{antipodal extension} $h^a$ of $h$ is a function $h^a\colon X^a \rightarrow Y$ defined as $h^a(x, +) = h(x)$ and $h^a(x, -) = -h(x)$; in other words, $h^a(x, y) = y\cdot h(x)$. Finally, an antipodal extension of a class $\H$ over $X$ is an antipodal class $\H^a = \{h^a~|~h\in \H\}$.

In the other direction, for an antipodal class $\H$ over antipodal domain $X$ with a representation map $r$, we define an \emph{restriction to representatives} $\H^r$ of $\H$ as a restriction (in a usual sense) of $\H$ to $\Img(r)$.

While symmetrization enforces antipodality on a class on an already antipodal domain, extensions are used to introduce antipodality to both the domain and the class. Unlike symmetrizations, they do not lose information, and pairing them with restrictions enables us to consider usual classes as antipodal ones and vice versa. All in all, these two operations are rather straightforward and used for technical convenience.
\begin{proposition}[Antipodal extension preserves $\VC$ and $\VC^{*\an}$]\label{prop-regular-to-antipodal}~
	\begin{enumerate}
		\item For a class $\H$, $(\H^a)^r = \H$; 
		\item For an antipodal class $\H$, $(\H^r)^a = \H$;
		\item For a class $\H$, $\VC(\H) = \VC(\H^a)$ and $\VC^{*\an}(\H) = \VC^{*\an}(\H^a)$.	
	\end{enumerate}
\end{proposition}
Note that restrictions with different representation maps can produce different classes. However, the equivalence of classes that arise this way can be easily characterized and, all in all does not produce anything deep.
\begin{proof}
(1) and (2) are trivial. In (3) $\VC(\H)\leq \VC(\H^a)$ and $\VC^{*\an}(\H)\leq \VC^{*\an}(\H^a)$ are also trivial, and we only need to prove the other direction. Let $S^a\subseteq X^a$ be shattered by $\H^a$. Then, obviously, $S^a$ does not contain a pair of antipodal points. Let $S^+ = S^a \cap X^+$ and $S^- = S^a\cap X^-$. It is now easy to see that $S^a$ is shattered by $\H^a$ if and only if $S^+ \sqcup -S^- \subseteq X$ is shattered by $\H$. That proves that $\VC(\H)= \VC(\H^a)$.

Similarly, if $H\subseteq \H^a$ is dually antipodally shattered and it is witnessed by $S^a\subseteq X^a$, then $S = S^a \cup -S^a$ witnesses dual shattering of $H$ by $\H^a$ and $S\cap X$ witnesses dual antipodal shattering of $H$ by $\H$. Hence, $\VC^{*\an}(\H) = \VC^{*\an}(\H^a)$.
\end{proof}

\subsection{Disambiguations and covers}\label{sec-disamb-covers}
For a partial function $h$ on $\SS^n$, let $h^+ = h^{-1}(+) \subseteq \SS^n$, and similarly for $h^-$ and $h^*$. We define the \emph{interior} $U_h$ of $h$ as the (topological) interior of $h^+ \cap (-h^-)$. It is easy to see that, by construction, $U_h$ is open and antipodal-free. Note that $U_h$ can also be explicitly defined as
\begin{align*}
	U_h = \{x\in \SS^n~|~&\text{ there is $\varepsilon>0$ s.t. $h\equiv +$ on $U_\varepsilon(x)$} \\
									&\text{ and $h\equiv -$ on $U_\varepsilon(-x)$}\},
\end{align*}
where, as usual, $U_\varepsilon(x)$ stands for an open $\varepsilon$-ball around $x$. Finally, for a class $\D$ over $\SS^n$, we define $\U(\D)$ as $\U(\D) = \{U_h~|~h\in \D\}$. Although formally this definition makes sense for partial classes, we will only use it for total.

\begin{proposition}\label{xprop-covers}
	For a total class $\D$ on $\SS^n$
	\begin{enumerate}
		\item $\D$ disambiguates $\SS^n$ if and only if $\U(\D)$ is an (open, antipodal-free) cover;
		\item If $\D$ disambiguates $\SS^n$ then there is a finite subclass $\D'$ of $\D$ that disambiguates $\SS^n$;
		\item If $\D$ disambiguates $\SS^n$ then $|\D| \geq n+2$.
		\item If $\D$ disambiguates $\SS^n$ then any symmetrization of $\D$ disambiguates $\SS^n$.
	\end{enumerate}
\end{proposition}
In view of (4) and also because, by Lemma~\ref{lem-antipodal-vc}, symmetrization does not increase $\VC$ and $\VC^{*\an}$, we can concentrate on antipodal classes as long as we are interested in those two parameters.
\begin{proof} Note that for any total class $\D$ on $\SS^n$, by construction, $\U(\D)$ is a family of open antipodal-free sets. If, additionally, $\D$ disambiguates $\SS^n$, then it disambiguates $\LL_{n, \varepsilon}$ for some $\varepsilon >0$. For $u\in \SS^n$, let $\ol{u}$ be the concept of $\D$ extending $u_\varepsilon$. Then, trivially, $u\in U_\varepsilon(u)\subseteq U(\ol{u})$, and hence $\U(\D)$ is a cover.

In the other direction, if $\U(\D)$ is a cover, let us, by compactness of $\SS^n$, pick a finite subclass $\D'$ of $\D$ such that $\U(\D')$ is a cover. Again by compactness, there is $\varepsilon >0$ such that for every $u\in \S^n$ there is $U\in \U(\D')$, $U = U_h$ for $h\in \D'$, such that $U_\varepsilon(u) \subseteq U_h$. Thus, $\D'$ disambiguates $\LL_{n, \varepsilon'}$ for any $0<\varepsilon'<\varepsilon$. 

The second statement directly follows from the first one, and the third is by a direct application of the Lusternik–Schnirelmann theorem (\Cref{t-LS}). Finally, the fourth statement follows from an easy observation that for any symmetrization $h'$ of a concept $h$, $U(h')\supseteq U(h)$.
\end{proof}

\subsection{Disambiguations of simplicial spheres}\label{sec-disamb-simplicial}
Let $(\Q^n, v\!\mapsto\!-v)$ be a simplicial $n$-sphere (introduced in \Cref{sec-sd}).
We say that a class $\D$ on the antipodal domain $V(\Q^n)$ \emph{disambiguates} $\Q^n$ if for every $s\in S(\Q^n)$ there is $h\in \D$ such that $h(v) = +$ for all $v\in s$ and $h(v)=-$ for all $v\in -s$. Note that if $\D$ is antipodal, then it suffices to require that the respective $h\equiv +$ only on $s$.

Similarly to how we defined simplicial diameter of $\Q$ in~\Cref{sec-background}, for a simplicial $n$-sphere $\Q^n$ and an equivariant homeomorphism $\iota \colon \|\Q^n\| \rightarrow \SS^n$ we define the \emph{(simplicial) diameter of $\iota$}, denoted $\diam_\iota(\Q^n)$, as 
	$$\diam_\iota(\Q^n) = \max_{s\in S(\Q^n)} \max_{x,y\in \|s\|} d(\iota(x), \iota(y)).$$

\begin{lemma}[Equivalence of simplicial and regular disambiguations]\label{th-vc-to-covers-refined} ~
	\begin{enumerate}
		\item For every disambiguation $\D$ of $\SS^n$ there is $\varepsilon >0$ such that for every simplicial $n$-sphere $\Q^n$ together with an equivariant homeomorphism $\iota\colon \|\Q^n\|\rightarrow \SS^n$ such that $\diam_\iota(\Q^n) < \varepsilon$ there is a disambiguation $\D'$ of $\Q^n$ such that $\D' \leq \D$ and $\D'$ is antipodal whenever $\D$ is;
		\item Let $\D$ be a disambiguation of a simplicial $n$-sphere  $\Q^n$. Then for the barycentric subdivision $\Q^n_1$ of $\Q^n$, there is a disambiguation $\D'$ of $\|\Q^n_1\|$ such that $\D' \equiv \D$ and $\D'$ is antipodal whenever $\D$ is.
	\end{enumerate}
\end{lemma}
Here the relations $\leq$ and $\equiv$ between classes are as defined at the beginning of the section. Note that Proposition~\ref{prop-barycentric-embedding} in \Cref{sec-background} plus compactness imply that $\diam_\iota(\Q^n_k) \xrightarrow{k} 0$, for any simplicial $n$-sphere $\Q^n$.  This implies that for any simplicial $n$-sphere $\Q^n$, for example for the barycentric subdivision of the boundary of $(n+1)$-simplex $\B^n$ (\Cref{ex-barycentric}), a sufficiently high barycentric subdivision of $\Q^n$ would satisfy the condition from (1). We also note that several other equivalences can be formulated along the lines of \Cref{th-vc-to-covers-refined}. In particular, the disambiguation $\D'$ in (2) turns out to be simplicial, that is, constant on the relative interiors of simplices of $\Q^n_1$. Here we stick to the bare minimum needed for us to proceed. 
\begin{proof}
  (\emph{1}). Let $\U = \U(\D)$; as $\D$ disambiguates $\SS^n$, $\U$ is a cover of $\SS^n$. By a standard compactness argument (also used in the proof of Proposition~\ref{xprop-covers}), there is $\varepsilon > 0$ such that for every $x\in \SS^n$, $U_\varepsilon(x) \subseteq U$ for some $U\in \U$. For this $\varepsilon$, if $\diam(\iota) < \varepsilon$, let $\D'$ be the restriction of $\D$ to $V(\Q^n)$, that is, $\D' = \{h'~|~h\in \D\}$, where $h'(v) = h(\iota(v))$. By construction, $\D' \leq \D$ and it is antipodal whenever $\F$ is. We only now need to check that $\D'$ disambiguates $\Q^n$.
   
  Let $s$ be an arbitrary simplex of $\Q^n$, and let $v\in s$. By our choice of $\varepsilon$, there is $h\in \H$ such that $U_\varepsilon(\iota(v)) \subseteq U(h)$. As  
  $\diam(\iota) < \varepsilon$, for every $u\in s$, $d(\iota(v), \iota(u)) < \varepsilon$, so $\iota(u) \in U_\varepsilon(\iota(v)) \subseteq U(h)$, for every $u\in s$. By the definition of $U(f)$, this means that $h'(u) = h(\iota(u)) = +$ for every $u\in s$ and $h'(u) = h(\iota(u)) = -$ for every $u\in -s$. Thus, $h'$ witnesses the disambiguation on the simplex $s$. As $s$ is arbitrary, $\D'$ disambiguates $\Q^n$, as needed.
  
	(\emph{2}). Recall that, by the definition of (abstract) barycentric subdivision, vertices of $\Q^n_1$ are nonempty simplices of $\Q^n$, $V(\Q^n_1) = S^{>0}(\Q^n)$, and simplices of $\Q^n_1$ are chains of nonempty simplices of $\Q^n$. As $V(\Q^n_1)$ are treated both like vertices and like simplices, this can be confusing, so we illustrate the construction and the proof in \Cref{fig-BARYCENTRIC} below. Let us define $\rho \colon S^{>0}(\Q^n) \rightarrow V(\Q^n)$ in an arbitrary way, satisfying two properties:
	\begin{itemize}
		\item For a simplex $s$, $\rho(s) \in s$;
		\item For a simplex $s$, $\rho(-s) = -\rho(s)$.
	\end{itemize}
	
	For $h\in \D$, let us define $h'$ on $\|\Q^n_1\|$ as follows. 
	For $x\in \|\Q^n_1\|$, let $s\in S(\Q^n_1)$ be the support of $x$, $s=\supp(x)$, that is, the smallest simplex $s\in S(\Q^n_1)$ such that $x\in \|s\|$. Let $v$ be the minimal, as a simplex in $S^{>0}(\Q^n)$, vertex of~$s$. Finally, for $h\in \D$, we put $h'(x) = h(\rho(v))$. It is easy to see that for $\D' = \{h'~|~h\in \D\}$, $\D'\equiv \D$ and $\D'$ is antipodal whenever $\D$ is. Again, we need to check that $\D'$ disambiguates $\|\Q^n_1\|$.

	Let $x\in \|\Q^n_1\|$, $s = \supp(x)$, $v$ the minimal vertex of $s$, $v\in V(\Q^n_1) = S^{>0}(\Q^n)$, and $h\in \D$ be a hypothesis that is constant $+$ on $v$ and constant $-$ on $-v$. It is enough to check that for every maximal simplex $s'$, extending $s$, and for every $x'$ in the relative interior of $s'$ (that is, for those $x'$ for which $s' = \supp(x')$), it holds $h'(x') = +$ and $h'(-x') = -$. As $s$ is a subchain of~$s'$, we have $v'\subseteq v$, where again we treat $v'$ and $v$ as simplices in $S^{>0}(\Q^n)$. So $\rho(v')\in v'\subseteq v$ and, by the choice of $h$, $h'(x') =  h(\rho(v')) = +$, and similarly, $h'(-x') = h(\rho(-v')) = -$, as needed.  

	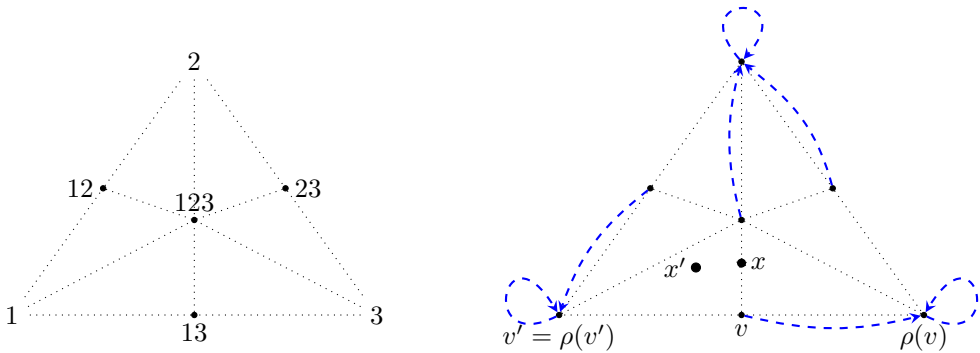
\begin{figure}[!hbt]
		\centering
		\begin{tikzpicture} 
[
pt/.style={inner sep = 0.0pt, circle, draw, fill=black},
point/.style={inner sep = 1.2pt, circle,draw,fill=black},
spoint/.style={inner sep = 1.2pt, circle,draw,fill=white},
mpoint/.style={inner sep = 0.7pt, circle,draw,fill=black},
ypoint/.style={inner sep = 3pt, circle,draw,fill=yellow},
xpoint/.style={inner sep = 3pt, circle,draw,fill=red},
FIT/.style args = {#1}{rounded rectangle, draw,  fit=#1, rotate fit=45, yscale=0.5},
FITR/.style args = {#1}{rounded rectangle, draw,  fit=#1, rotate fit=-45, yscale=0.5},
FIT1/.style args = {#1}{rounded rectangle, draw,  fit=#1, rotate fit=45, scale=2},
vecArrow/.style={
		thick, decoration={markings,mark=at position
		   1 with {\arrow[thick]{open triangle 60}}},
		   double distance=1.4pt, shorten >= 5.5pt,
		   preaction = {decorate},
		   postaction = {draw,line width=0.4pt, white,shorten >= 4.5pt}
	}
]

\begin{scope}[yscale=1.2, xscale = 0.6]
	\begin{scope}[yscale = 0.7]
		\node(a) at (-4,0) {$1$};
		\node(b) at (0,4) {$2$};
		\node(c) at (4,0) {$3$};
		\node[mpoint](ac) at (0,0) {};
		\node[mpoint](ab) at (-2,2) {};
		\node[mpoint](bc) at (2,2) {};
		\node[mpoint](abc) at (0,1.5) {};
	
		\draw [dotted] (a)--(ab)--(b)--(bc)--(c)--(ac)--(a);
		\draw [dotted](ab)--(abc) (bc)--(abc);
		\draw [dotted] (b)--(abc)  (c)--(abc);
		
		\draw [dotted] (a)--(abc);
		\draw [dotted] (ac)--(abc);
		
		\node [below] at (ac) {$13$};
		\node [left] at (ab) {$12$};
		\node [right] at (bc) {$23$};
		\node [above] at (abc) {$123$};

	\end{scope}

	\begin{scope}[yscale = 0.7, xshift = 12cm]
		\node[mpoint](a) at (-4,0) {};
		\node[mpoint](b) at (0,4) {};
		\node[mpoint](c) at (4,0) {};
		\node[mpoint](ac) at (0,0) {};
		\node[mpoint](ab) at (-2,2) {};
		\node[mpoint](bc) at (2,2) {};
		\node[mpoint](abc) at (0,1.5) {};
	
		\draw [dotted] (a)--(ab)--(b)--(bc)--(c)--(ac)--(a);
		\draw [dotted](ab)--(abc) (bc)--(abc);
		\draw [dotted] (b)--(abc)  (c)--(abc);
		
		\draw [dotted] (a)--(abc);
		\draw [dotted] (ac)--(abc) node[pos=0.55, point](x){};
		\node[right] at (x) {$x$};
		\node [point](x') at (-1, 0.75) {};
		\node[left] at (x') {$x'$};
		
		\node [below] at (ac) {$v$};
		\node [below] at (c) {$\rho(v)$};
		\node [below] at (a) {$v' = \rho(v')$};

		\draw[thick, dashed, blue, -stealth] (abc) to [out = 110, in = -110] (b);
		\draw[thick, dashed, blue, -stealth] (ab) to [out = -150, in = 65] (a);
		\draw[thick, dashed, blue, -stealth] (bc) to [out = 110, in = -30] (b);
		\draw[thick, dashed, blue, -stealth] (ac) to [out = -10, in = -170] (c);
		\draw[thick, dashed, blue, -stealth] (b) to [out = 140, in = 40, looseness = 40] (b);
		\draw[thick, dashed, blue, -stealth] (c) to [out = -20, in = 50, looseness = 80] (c);
		\draw[thick, dashed, blue, -stealth] (a) to [out = -160, in = 130, looseness = 80] (a);
	\end{scope}
\end{scope}

\end{tikzpicture}	
		\caption{A barycentric subdivision of a simplex $\{1,2,3\}$ and an illustration for the proof. In the second picture, the blue lines indicate $\rho$. For a point $x$, $s = \{13,  123\}$, so $v=13$; for $x'$, $s' = \{1, 13, 123\}$ and $v'=1 \subseteq 13 = v$.}
		\label{fig-BARYCENTRIC}       
	\end{figure} 
\end{proof}  
  
\subsection{Equivalence with spherical dimension}\label{sec-disamb-sd}
Let us now draw the main connection between the spherical dimension and disambiguations of spheres.

\begin{customthm}{\ref{th-sd-disamb}}[Rough equivalence of disambiguations and spherical dimension]  ~
\begin{enumerate}
\item Let $\D$ be an antipodal disambiguation of a simplicial $n$-sphere $\Q^n$. Then both $\D$ and an arbitrary restriction to representatives $\D'$ of $\D$ admit an embedded simplicial  $n$-sphere. 
\item Let $\H$ be a class over $X$ admitting a simplicial $n$-sphere $\beta\colon V(\Q^n) \rightarrow V(\Delta^\ant_\H) \subseteq X\times Y$, and let $\H^a$ be its antipodal extension to $X^a = X\times Y$. Then $\beta$-pullback $\H_\beta$ of $\H^a$ is an antipodal disambiguation of $\Q^n$.

Here $\H_\beta$ is a class on $V(\Q^n)$ defined as $\H_\beta = \{h_\beta~|~h\in \H^a\}$, where $h_\beta(v) = h(\beta(v))$, for $v\in V(\Q^n)$. In particular, $\H_\beta \leq \H^a$.
\end{enumerate}
In particular, for integers $a$, $b$, and $n$, the following are equivalent. Below, $\Q^n$ denotes a simplicial $n$-sphere:
\begin{itemize}
	\item There is a class with $\VC \leq a$, $\VC^{*\an}\leq b$, and $\sd_\simp \geq n$;
	\item There is a class with $\VC \leq a$, $\VC^{*\an}\leq b$, and $\sd_\simp \geq n$, witnessed by embedded spheres;
	\item There is an antipodal disambiguation of $\Q^n$ with $\VC \leq a$, $\VC^{*\an}\leq b$;
	\item There is a disambiguation of $\Q^n$ with $\VC \leq a$, $\VC^{*\an}\leq b$;
	\item There is an antipodal disambiguation of $\SS^n$ with $\VC \leq a$, $\VC^{*\an}\leq b$;
	\item There is a disambiguation of $\SS^n$ with $\VC \leq a$, $\VC^{*\an}\leq b$.
\end{itemize}

In particular, if for a class $\H$ we define the disambiguation index $\disamb(\H)$ as the maximal $n$ for which there is a disambiguation $\D$ of $\SS^n$ such that $\D \leq \H$, then the parameters $\sd_\simp$ and $\disamb$ are roughly equivalent (and also roughly equivalent to $\sd_\simp$, witnessed by embedded spheres).
\end{customthm}
We also note that in the ``in particular'' case, the respective statements are applicable not only to $\VC$ and $\VC^{*\an}$, but to all class characteristics that are non-increasing under $\leq$, symmetrization, and antipodal extension.
\begin{proof}
Let us start by recalling that a finite class $\H$ admits an embedded simplicial $n$-sphere if there is a simplicial $n$-sphere $\Q^n$ and a one-to-one equivariant map $\beta\colon V(\Q^n)\rightarrow V(\Delta^\ant_\H) \subseteq X \times Y$ such that $\beta$ maps simplices of $\Q^n$ to simplices of $\Delta^\ant_\H$. 

\emph{(1.)} For $\D$ and $\D'$ as in the statement, let $X = r\left[V(\Q^n)\right] \subseteq V(\Q^n)$, where $r$ is the respective representation map for which $\D' = \D^r$. Note that $\D'$ is defined on $X$ and is not antipodal.
 Let us now define $\beta\colon V(\Q^n) \rightarrow X\times Y$ as follows: For $v\in V(\Q^n)$, $\beta(v) = (v, +)$ if $v = r(v)$ and $(-v, -)$ if $v=-r(v)$. Alternatively, $\beta(v) = \left(r(v), [r(v) = v]\right)$, where, similarly to Lemma~\ref{lem-antipodal-vc}, $[r(v) = v] = +$ if $r(v) = v$ and $-$ if $r(v) = -v$. Trivially, $\beta$ is one-to-one. Moreover, $\beta(-v) = \left(r(-v), [r(-v) = -v]\right) = \left(r(v), [r(v) = -v]\right) = \left(r(v), -[r(v) = v]\right) = -\beta(v)$, and so $\beta$ is also equivariant.

We also claim that $\beta$ is simplicial to $\Delta_{\D'}$. Indeed, let us take an arbitrary $s\in S(\Q^n)$ and let $h_s\in \D$ be such that $h_s \equiv +$ on $s$; note that this automatically implies that $h(-s)\equiv -$, as $\D$ is antipodal. But then for any $v\in s$, $h_s(r(v)) = [r(v) = v]$, and so $\beta(v) = (r(v), h_s(r(v)))$. Thus, $\beta(s) = \{(r(v), h_s(r(v)))~|~v\in s\}$ is realizable by $h_s|_X \in \D'$, and hence belongs to $\Delta_{\D'}$. The fact that $\beta(s)$ is actually in $\Delta^\ant_{\D'}$ follows by noting that $-\beta(s) = \beta(-s) \in \Delta_{\D'}$, where the first equality is by equivariance of $\beta$, and the second inclusion is witnessed by some other function $h_{-s}$. This concludes the proof for $\D'$; the statement for $\D$ follows from that because $\Delta_{\D'}$ is (simplicially) embedded into $\Delta_\D$ with the same antipodality.

This proof bears mixed features of both triviality and evil magic. To rectify this, we illustrate how it works (and fails) in~\Cref{fig-EQUIVALENCE1} below.
\begin{figure}[!hbt]
	\centering
	\begin{tikzpicture} 
[
pt/.style={inner sep = 0.0pt, circle, draw, fill=black},
point/.style={inner sep = 1.2pt, circle,draw,fill=black},
spoint/.style={inner sep = 1.2pt, circle,draw,fill=white},
mpoint/.style={inner sep = 0.7pt, circle,draw,fill=black},
ypoint/.style={inner sep = 3pt, circle,draw,fill=yellow},
xpoint/.style={inner sep = 3pt, circle,draw,fill=red},
FIT/.style args = {#1}{rounded rectangle, draw,  fit=#1, rotate fit=45, yscale=0.5},
FITR/.style args = {#1}{rounded rectangle, draw,  fit=#1, rotate fit=-45, yscale=0.5},
FIT1/.style args = {#1}{rounded rectangle, draw,  fit=#1, rotate fit=45, scale=2},
vecArrow/.style={
		thick, decoration={markings,mark=at position
		   1 with {\arrow[thick]{open triangle 60}}},
		   double distance=1.4pt, shorten >= 5.5pt,
		   preaction = {decorate},
		   postaction = {draw,line width=0.4pt, white,shorten >= 4.5pt}
	}
]

\begin{scope}[yscale=0.5, xscale = 0.5]
	\begin{scope}
		\node[mpoint](1) at (0,0) {};
		\node[mpoint](2) at (0,4) {};
		\node[mpoint](-1) at (4,4) {};
		\node[mpoint](-2) at (4,0) {};
		\draw(1)--(2)--(-1)--(-2)--(1);
		\node[left] at (1) {$1$};
		\node[left] at (2) {$2$};
		\node[right] at (-1) {$-1$};
		\node[right] at (-2) {$-2$};
		\node at (2,2) {$\Q^1$};
	\end{scope}

	\node at (6.5, 2) {\Large$\xrightarrow{\beta}$};

	\begin{scope}[xshift = 10cm]
		\node[mpoint](1) at (0,0) {};
		\node[mpoint](2) at (0,4) {};
		\node[mpoint](-1) at (4,4) {};
		\node[mpoint](-2) at (4,0) {};
		\draw(1)--(2)--(-1)--(-2)--(1);
		\node[left] at (1) {$(1, +)$};
		\node[left] at (2) {$(2, +)$};
		\node[right] at (-1) {$(1, -)$};
		\node[right] at (-2) {$(2, -)$};
		\node at (2,2) {$\Delta_{\D'}$};
	\end{scope}

\end{scope}

\end{tikzpicture}	
	\caption{The map $\beta$ from the above proof. Note that for the function, say, $h_{2, -1}\in \D$, positive on the vertices of the top simplex in $\Q^1$, $h_{2, -1}(1) = -$ and $h_{2, -1}(2) = +$, and so it witnesses the existence of the same top simplex in $\Delta_{\D'}$. And, in the absence of $h_{2, -1}$ in $\D$, the respective simplex would be absent in $\Delta_{\D'}$, breaking the simpliciality of the map.}
	\label{fig-EQUIVALENCE1}       
\end{figure}
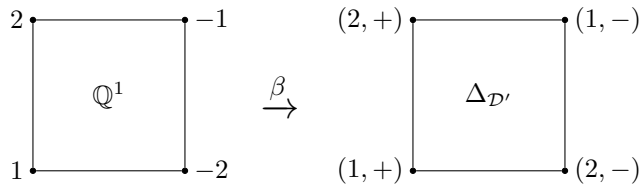 

\emph{(2.)} We only need to check that $\H_\beta$ is antipodal and that it is a disambiguation. For antipodality, quite trivially, $h_\beta(-v) = h(\beta(-v)) = h(-\beta(v)) = -h(\beta(v)) = -h_\beta(v)$, for all $h_\beta \in \H$ and $v\in V(\Q^n)$; here we use antipodality of $\beta$ and antipodality of $h\in \H^a$.

For disambiguation, let $s\in S(\Q^n)$, and let $h_s\in \H$ be the concept witnessing $s\in \Delta_\H$, that is, $h_s(x) = y$ for all $(x, y) \in \beta(s)$. But then $h_s^a(x, y) = +$ for all $(x, y) \in \beta(s)$, which is equivalent to $\left(h^a_s\right)_\beta(v) = +$ for all $v\in s$, and hence $\left(h^a_s\right)_\beta(s) \equiv +$, as needed; note that $\left(h^a_s\right)_\beta(-s) \equiv -$ follows by antipodality.

The final list of equivalences is just tallying up the results from this section, shown in \Cref{fig-SD-AND-DISAMB} below. 
\end{proof}

\begin{figure}[!hbt]
	\centering
	\begin{tikzpicture} 
[
pt/.style={inner sep = 0.0pt, circle, draw, fill=black},
point/.style={inner sep = 1.2pt, circle,draw,fill=black},
spoint/.style={inner sep = 1.2pt, circle,draw,fill=white},
mpoint/.style={inner sep = 0.7pt, circle,draw,fill=black},
ypoint/.style={inner sep = 3pt, circle,draw,fill=yellow},
xpoint/.style={inner sep = 3pt, circle,draw,fill=red},
FIT/.style args = {#1}{rounded rectangle, draw,  fit=#1, rotate fit=45, yscale=0.5},
FITR/.style args = {#1}{rounded rectangle, draw,  fit=#1, rotate fit=-45, yscale=0.5},
FIT1/.style args = {#1}{rounded rectangle, draw,  fit=#1, rotate fit=45, scale=2},
vecArrow/.style={
		thick, decoration={markings,mark=at position
		   1 with {\arrow[thick]{open triangle 60}}},
		   double distance=1.4pt, shorten >= 5.5pt,
		   preaction = {decorate},
		   postaction = {draw,line width=0.4pt, white,shorten >= 4.5pt}
	}
]

\begin{scope}[yscale=0.5, xscale = 0.5]
	\node(QnA) at (0,0) {\parbox{3.1cm}{Antipodal disambiguation of~$\Q^n$}};
	\node(Qn) at (0,6) {\parbox{3.1cm}{Disambiguation of~$\Q^n$}};

	\node(SnA) at (12,0) {\parbox{3.1cm}{Antipodal disambiguation of~$\SS^n$}};
	\node(Sn) at (12,6) {\parbox{3.1cm}{Disambiguation of~$\SS^n$}};


	\node[text width=2.8cm](SdEx) at (0,-6) {\centering $\sd_\simp \geq n$ \\ \scriptsize{(embedded spheres)}};
	\node(Sd) at (12,-6) {$\sd_\simp \geq n$};

	\draw[->](Qn.5) to node[above, pos=0.5] 
		{\scriptsize L~\ref*{th-vc-to-covers-refined}(1)} (Sn.175);
	\draw[<-](Qn.-5) to node[below, pos=0.5] 
		{\scriptsize L~\ref*{th-vc-to-covers-refined}(2)} (Sn.-175);
	\draw[->](QnA.5) to node[above, pos=0.5] 
		{\scriptsize L~\ref*{th-vc-to-covers-refined}(1)} (SnA.175);
	\draw[<-](QnA.-5) to node[below, pos=0.5] 
		{\scriptsize L~\ref*{th-vc-to-covers-refined}(2)} (SnA.-175);

	\draw[->](Sn.-80) to node[right, pos=0.5] 
		{\scriptsize P~\ref*{lem-antipodal-vc} and P~\ref*{xprop-covers}(1)} (SnA.80);
	\draw[<-](Sn.-100) to node[left, pos=0.5] 
		{\scriptsize trivial} (SnA.100);
	\draw[<-](Qn.-90) to node[left, pos=0.5] 
		{\scriptsize trivial} (QnA.90);

	\draw[->](QnA.-90) to node[left, pos=0.5] 
		{\scriptsize T \ref*{th-sd-disamb}(1)} (SdEx.90);
	\draw[->](SdEx.0) to node[above, pos=0.5] 
		{\scriptsize trivial} (Sd.-180);
	\draw[->](Sd.155) to node[above right, pos=0.4] 
		{\scriptsize T \ref*{th-sd-disamb}(2) and P \ref*{prop-regular-to-antipodal}} (QnA.-25);

\end{scope}

\end{tikzpicture}	
	\caption{Implications between statements about classes with $\VC\leq a$ and $\VC^{*\an}\leq b$ from the conclusion of \Cref{th-sd-disamb}.}
	\label{fig-SD-AND-DISAMB}       
\end{figure}
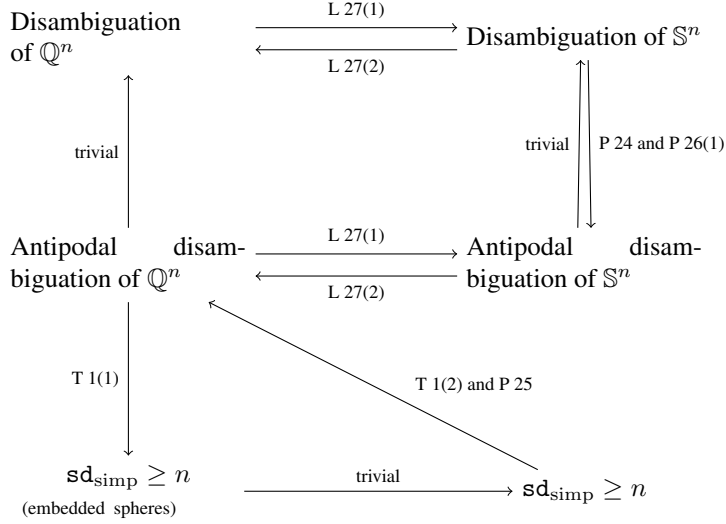 

\section{Proofs for extremal classes}\label{sec-extremal-proofs}

As some proofs in this section are rather involved, here we will drop our convention for not using geometrization brackets $\|\cdot\|$ for the spaces $\Delta$ and $\Delta^\ant$. That is, we will write $\Delta$ and $\Delta^\ant$ for simplicial complexes and $\|\Delta\|$ and $\|\Delta^\ant\|$ for their geometric realizations.

\subsection{Cubical complex of a class}\label{sec-cubical}
Our proof will crucially rely on the notion of a \emph{cubical complex} of a class. Similarly to simplicial complexes, cubical complexes can be defined abstractly; See, for example, Definition~7.32 in~\cite{bridson13}, or~\cite{sageev14} for a more informal walkthrough. Such formal approach, however, is unnesessary meticulous and, following Section~2 in~\cite{chase2024dual}, we will make a shortcut and will construct our cubical complex as a subspace of the usual Euclidean cube (composed out of faces of this cube). 

Our first technical step is to introduce \emph{strong shattering}. It will be convenient for us to use partial hypotheses, so we will use some basic notation and terminology from the beginning of \Cref{sec-disamb}. For a partial hypothesis $h$ over a finite domain $X$, we define the \emph{dimension} of $h$, denoted $d(h)$, as $d(h) = |\{x\in X~|~h(x) = *\}| = |X| - |\supp(h)|$. We define the \emph{discrete cube} of~$h$, denoted $C(h)$, as a set of all $2^{d(h)}$ total hypotheses, extending $h$.
For a class $\H$ over $X$ and $S\subseteq X$, we say that $S$ is \emph{strongly shattered} by $\H$ if there is a partial hypotheses $h$ such that $S = X - \supp(h)$ and $C(h)\subseteq \H$. Trivially, every set strongly shattered by a class is shattered by it.  It is well-known that for extremal classes strong shattering is equivalent to shattering, see, for example, \cite{Bollobas:95}; moreover, this correspondence goes both ways, that is, a class is extremal if and only if it strongly shatters all its shattered sets. This will be crucial for us in the construction of the cubical complex, which revolves around strong shattering. So, although the same definition can be considered for an arbitrary class $\H$, we will now restrict our attention to a finite extremal class $\E$. 

Recall that in~\Cref{sec-disamb} we defined the order on partial hypotheses by extension by saying that $h_1\leq h_2$ if $\supp(h_1)\subseteq \supp(h_2)$ and $h_2|_{\supp(h_1)} = h_1|_{\supp(h_1)}$. It is easy to see that this can be equivalently formulated as $h_1\leq h_2$ if $C(h_1) \supseteq C(h_2)$. From this perspective, it would be natural to flip the order on partial hypotheses to make it coincide with the order on cubes by inclusion; however, we stick with the present one for consistency with~\Cref{sec-disamb}. It is also easy to see that discrete cubes and partial hypotheses over $X$ are in bijective correspondence to each other, so with some ambiguity, we would at times refer to one as to another. In particular, we define the dimension of a cube as the dimension of the corresponding partial hypothesis and refer to cubes of dimension $d$ as to $d$-cubes.

For an extremal class $\E$, we say that a discrete cube $C$ is a \emph{cube of $\E$} if $C\subseteq E$, and we define an (abstract) cubical complex of $\E$, denoted $\CC_\E$, as a collection of all discrete cubes of~$\E$. Note that here we do not allow our cubes to be empty. Let us note several basic facts about the cubical complex $\CC_\E$ that we are going to utilize:
\begin{itemize}
	\item The dimension of $\CC_\E$, denoted $\dim(\CC_\E)$ is defined as a maximal dimension of a cube in $\CC_\E$. By the abovementioned correspondence between shattering and strong shattering for extremal classes, $\dim(\CC_\E) = \VC(\E)$;
	\item $0$-cubes of $\CC_\E$, also called \emph{vertices} of $\CC_\E$, are precisely the concepts of $\E$. We also at times call $1$-cubes by \emph{edges} and $2$-cubes by \emph{squares};
	\item We say that $C_1$ is a \emph{subcube} of $C_2$ if $C_1\subseteq C_2$. As argued, for the respective partial hypotheses $h_1$ and $h_2$, this is equivalent to $h_2$ extending $h_1$, that is, $C_1\subseteq C_2 \sim h_1 \geq h_2$;
	\item Any subcube of a cube of $\CC_\E$ is a cube of $\CC_\E$. Moreover, any two cubes of $\CC_\E$ are either disjoint or intersect by a cube of $\CC_\E$. This property enables us to indeed call $\CC_\E$ a cube \emph{complex}, as opposed to it being just a collection of cubes. 
\end{itemize}

Let $I=[-1, 1]$ and, for a finite domain $X$, let $I^X$ be a $|X|$-dimensional Euclidean cube with coordinates parametrized by $X$. For a discrete cube $C = C(h)$ over $X$ we define the \emph{geometric realization} of $C$, denoted $\|C\|\subseteq I^X$, as 
$$\|C\| = \left\{v\in I^X~|~ v(x) = h(x)\text{ for }x\in \supp(h)\right\}.$$
Finally, we define the geometric cubical complex of $\E$, denoted $\|\CC_\E\|$, or sometimes simply $\|\E\|$, as $\|\E\| = \bigcup_{C\in \CC_\E} \|C\|$. This construction is illustrated in \Cref{fig-CUBE-COMPLEX} below.

\begin{figure}[hbt]
\centering
\begin{tikzpicture}
\begin{scope}[scale=2, xshift=1.5cm]  
    \coordinate (O) at (0,0,0);
    \coordinate (A) at (0,1,0);
    \coordinate (B) at (1,0,0);
    \coordinate (C) at (1,1,0);
    \coordinate (D) at (0,0,1);

    \draw[fill=black] (O) circle (1pt) node[anchor=east] {\scriptsize$---$};
    \draw[fill=black] (A) circle (1pt) node[anchor=east] {\scriptsize$-+-$};
    \draw[fill=black] (B) circle (1pt) node[anchor=west] {\scriptsize$+--$};
    \draw[fill=black] (C) circle (1pt) node[anchor=west] {\scriptsize$++-$};
    \draw[fill=black] (D) circle (1pt) node[anchor=east] {\scriptsize$--+$};

    \draw[fill=gray, fill opacity=0.3] (O) -- (A) -- (C) -- (B) -- cycle;
    \draw (O) -- (D);
\end{scope}

\end{tikzpicture}
\caption{A $2$-dimensional illustration of the cubical complex for the extremal class $\E = \{---, -+-, ++-, +--, --+\}$. It has $5$ vertices, $5$ edges, and $1$ square. The square corresponds to a cube $**-$, and the unique maximal edge, connecting $---$ and $--+$, corresponds to a cube $--*$.}
\label{fig-CUBE-COMPLEX}
\end{figure}
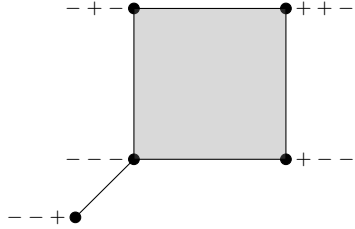

The geometric realization of cubes and complexes enjoy nice properties, similar to those listed for abstract cubical complexes:
\begin{itemize}
	\item $\dim(\|\E\|) = \dim(\CC_\E) = \VC(\E)$;
	\item $C_1 \subseteq C_2$ if and only if $\|C_1\| \subseteq \|C_2\|$; $C_1$ and $C_2$ are disjoint if and only if $\|C_1\|$ and $\|C_2\|$ are disjoint; and, if $C_1$ and $C_2$ are not disjoint, then $\|C_1\| \cap \|C_2\| = \|C_1 \cap C_2\|$.
\end{itemize}

\subsection{Barycentric subdivisions of the cubical complex  and the space of realizable distributions of a class}\label{sec-cubical-bary}

For a finite extremal class $\E$, we define the \emph{barycentric subdivision} of the cubical complex $\CC_\E$, denoted $\CC_{\E, 1}$, as a \underline{simplicial complex} whose vertices are cubes of $\CC_\E$ and whose simplices are chains of cubes of~$\CC_\E$.

The respective geometrization of $\CC_{\E, 1}$ is then defined as follows. Recall that the cube complex $\|\E\|$ is naturally embedded into $I^X$, where $X$ is the domain of $\E$. Let $\iota\colon \{-, +, *\} \rightarrow I$ be defined as $\iota(-) = -1$, $\iota(+) = 1$, and $\iota(*) = 0$. We then extend $\iota$ to the map $\iota\colon \{-, +, *\}^X \rightarrow I^X$, whose domain are all partial hypotheses over $X$, by letting $\iota(h) = \bigl(\iota(h(x))\bigr)_{x\in X}$. 

Note that $\iota$ maps the discrete cubes (interpreted as partial hypotheses), in a way, consistent with the definition of the geometric realization of these cubes. In particular, for any discrete cube $C$, $\|C\|$ is a convex hull of $\iota(C) = \{\iota(h)~|~h\in C\}$; and for any partial hypothesis $h$, $\|h\| = \iota(h)$. Thus, $\iota$ provides a natural embedding of the vertices of $\CC_{\E, 1}$, which are cubes of $\E$, into $\|\E\|$.
It can now be easily checked that, extended by linearity, $\iota$ establishes an isomorphism between the geometric realization $\|\CC_{\E, 1}\|$ (or simply $\|\E_1\|$) of the barycentric subdivision of $\CC_\E$, and the geometric realization $\|\E\|$ of $\CC_\E$. As before, we usually will interpret this as if $\|\E_1\|$ is itself embedded into $I^X$. This is illustrated in \Cref{fig-CUBE-COMPLEX-BARY} below. Once more: $\|\E\|$ and $\|\E_1\|$ are the same topological space and we only use different names for it when we want to emphasize the underlying simplicial complex from which it arises.


\begin{figure}[hbt]
\centering
\begin{tikzpicture}
[
pt/.style={inner sep = 1.3pt, circle, draw, fill=black},
point/.style={inner sep = 1.2pt, circle,draw,fill=black},
spoint/.style={inner sep = 1.2pt, circle,draw,fill=white},
mpoint/.style={inner sep = 0.7pt, circle,draw,fill=black},
ypoint/.style={inner sep = 3pt, circle,draw,fill=yellow},
xpoint/.style={inner sep = 3pt, circle,draw,fill=red},
FIT/.style args = {#1}{rounded rectangle, draw,  fit=#1, rotate fit=45, yscale=0.5},
FITR/.style args = {#1}{rounded rectangle, draw,  fit=#1, rotate fit=-45, yscale=0.5},
FIT1/.style args = {#1}{rounded rectangle, draw,  fit=#1, rotate fit=45, scale=2},
vecArrow/.style={
		thick, decoration={markings,mark=at position
		   1 with {\arrow[thick]{open triangle 60}}},
		   double distance=1.4pt, shorten >= 5.5pt,
		   preaction = {decorate},
		   postaction = {draw,line width=0.4pt, white,shorten >= 4.5pt}
	}
]

\begin{scope}[scale=3]  
    \coordinate (---) at (0,0,0);
    \coordinate (-+-) at (0,1,0);
    \coordinate (+--) at (1,0,0);
    \coordinate (++-) at (1,1,0);
    \coordinate (--+) at (0,0,1);

    \draw[fill=black] (---) circle (0.5pt) node[anchor=east] {\scriptsize$---$};
    \draw[fill=black] (-+-) circle (0.5pt) node[anchor=east] {\scriptsize$-+-$};
    \draw[fill=black] (+--) circle (0.5pt) node[anchor=west] {\scriptsize$+--$};
    \draw[fill=black] (++-) circle (0.5pt) node[anchor=west] {\scriptsize$++-$};
    \draw[fill=black] (--+) circle (0.5pt) node[anchor=east] {\scriptsize$--+$};

    \draw[fill=gray, fill opacity=0.1] (---) -- (-+-) -- (++-) -- (+--) -- cycle;
    \draw (---) -- (--+);
    
    \node[pt] (-0-) at (0, 0.5) {};
    \node[pt] (0+-) at (0.5, 1) {};
    \node[pt] (+0-) at (1, 0.5) {};
    \node[pt] (0--) at (0.5, 0) {};

    \node[pt] (00-) at (0.5, 0.5) {};
    \node[pt] (--0) at (0, 0, 0.5) {};
    
    \node [left] at (-0-) {\scriptsize$-*-$};
    \node [right] at (+0-) {\scriptsize$+*-$};
    \node [above] at (0+-) {\scriptsize$*+-$};
    \node [below] at (0--) {\scriptsize$*--$};
    \node [left] at (--0) {\scriptsize$--*$};
    \node [below] at (00-) {\scriptsize$**-$};
    
    \draw[dotted] (00-)--(---)
    	(00-)--(-0-) (00-)--(-+-) (00-)--(0+-) (00-)--(++-)
    	(00-)--(+0-) (00-)--(+--) (00-)--(0--);
\end{scope}

\end{tikzpicture}
\caption{A barycentric subdivision of the cubical complex in \Cref{fig-BARYCENTRIC}. The corresponding simplicial complex has $11$ vertices, $18$ edges, and $6$ $2$-dimensional simplices.}
\label{fig-CUBE-COMPLEX-BARY}
\end{figure}
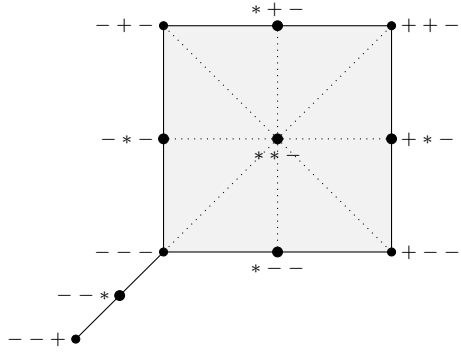

To define the barycentric subdivision of $\Delta_\E$, denoted $\Delta_{\E, 1}$, all we need is to apply the standard definition of the barycentric subdivision from~\Cref{sec-background} to the definition of $\Delta_\E$ in \Cref{sec-topology-toolbox}. Thus, the vertices of $\Delta_{\E, 1}$ are partial hypotheses $h$ with nonempty support that can be extended to a total hypothesis in $\E$ (we say that $h$ is realizable by $\E$), and the simplices are chains of such partial hypotheses. Geometrically, a partial hypothesis $h\in V(\Delta_\E)$ is interpreted as a uniform distribution over its (nonempty) domain, and points in simplices are convex combinations of uniform distributions over nested domains. We also note that because $\Delta_\E^\ant$ is simplicially embedded into $\Delta_\E$, it follows that $\Delta_{\E, 1}^\ant$ is an antipodal simplicial complex, simplicially embedded into $\Delta_{\E, 1}$, with antipodality extended in a standard way to the barycentric subdivision.

With this elaboration, the parallel between the barycentric subdivisions of $\C_\E$ and of $\Delta_\E$ becomes quite pronounced:
\begin{itemize}
	\item The vertex sets in both cases are subsets of the set of all partial hypotheses over $X$. The simplices are also formed in a similar way, namely, as chains of partial hypotheses;
	\item For $\Delta_{\E, 1}$, we pick partial hypotheses that are \emph{nonempty} and realizable by $\E$, and for $\CC_\E$ those that correspond to cubes of $\E$. Trivially, any $h$ corresponding to a cube is also realizable, and so the only partial hypotheses that can be in $V(\CC_{\E, 1}) - V(\Delta_{\E, 1})$ is the completely undefined hypothesis $h_*$, which only happens when $\E$ is the binary hypercube $\C_n$;
	\item If $\E = \C_n$, then $\|\E\| = I^n$ and $\|\Delta_\E\| = \|\Delta^\ant_\E\| \cong \SS^{n-1}$ is its boundary. Moreover, $\Delta_{\E, 1}$ is simplicially embedded into  $\CC_{\E, 1}$;
	\item In all other cases, that is, when $\E \neq \C_n$, $\CC_{\E, 1}$ is simplicially embedded into $\Delta_{\E, 1}$. Moreover, $\CC_{\E, 1}$ is a \emph{full} subcomplex of $\Delta_{\E, 1}$, that is, for any simplex $s \in S(\Delta_{\E, 1})$, if $v\in V(\CC_{\E, 1})$  for all $v\in s$, then $s\in S(\CC_{\E, 1})$.
\end{itemize}
\Cref{fig-EMBEDDING} below illustrates the embedding of $\CC_{\E, 1}$ into $\Delta_{\E, 1}$ for a particular extremal class. 

\begin{figure}[hbt]
\centering
\begin{tikzpicture}
[
pt/.style={inner sep = 0.9pt, circle, draw, fill=black},
ptb/.style={inner sep = 0.9pt, circle, draw=blue, fill=blue},
ptl/.style={inner sep = 1.5pt, circle, draw, fill=black},
ptbl/.style={inner sep = 1.5pt, circle, draw=blue, fill=blue}
]

\begin{scope}[yscale=0.7, xscale = 1]  
    \coordinate (+**) at (0,0);
    \coordinate (*+*) at (4,0);
    \coordinate (**+) at (2,3);
    \coordinate (-**) at (6,3);
    \coordinate (*-*) at (4,6);
    \coordinate (**-) at (8,6);
        
    \draw[fill=blue, fill opacity=0.1] (+**)--(**+)--(*+*)--cycle;
    \draw[fill=blue, fill opacity=0.1] (*-*)--(**-)--(-**)--cycle;
    \draw[fill=gray, fill opacity=0.1] (**+)--(*-*)--(-**)--(*+*)--cycle;

    \draw[thick, blue] (+**) to node[ptb, midway] (+*+){} (**+)
    	to node[ptbl, midway] (*++){} (*+*)
    	to node[ptb, midway] (++*){} (+**);
	\node[ptbl](+++) at (2, 1.1) {};
    \draw[thick, blue] (-**) to node[ptb, midway] (-*-){} (**-)
    	to node[ptb, midway] (*--){} (*-*)
    	to node[ptbl, midway] (--*){} (-**);
	\node[ptbl](---) at (6, 4.9) {};

	\draw(**+) to node[pt, midway](*-+){} (*-*);
	\draw(**+) to node[ptl, midway](-*+){} (-**);
	\draw(*+*) to node[pt, midway](-+*){} (-**);

	\node[ptl](--+) at (4, 4.1) {};
	\node[ptl](-++) at (4, 1.9) {};

	\node[ptb] at (+**) {};
	\node[ptb] at (**+) {};
	\node[ptb] at (*-*) {};
	\node[ptb] at (**-) {};
	\node[ptb] at (-**) {};
	\node[ptb] at (*+*) {};

    \node [left] at (+**) {\scriptsize$+**$};
    \node [left] at (**+) {\scriptsize$**+$};
    \node [left] at (*-*) {\scriptsize$*-*$};
    \node [right] at (**-) {\scriptsize$**-$};
    \node [right] at (-**) {\scriptsize$-**$};
    \node [right] at (*+*) {\scriptsize$*+*$};
    
    \node [below] at (++*) {\scriptsize$++*$};
    \node [below] at (+++) {\scriptsize$+++$};
    \node [above] at (*--) {\scriptsize$*--$};
    \node [above] at (---) {\scriptsize$---$};
    \node [left] at (+*+) {\scriptsize$+*+$};
    \node [left] at (*-+) {\scriptsize$*-+$};
    \node [right] at (-*-) {\scriptsize$-*-$};
    \node [right] at (-+*) {\scriptsize$-+*$};

    \node [right] at (-++) {\scriptsize$-++$};
    \node [left] at (--+) {\scriptsize$--+$};
    \node [above right] at (-*+) {\scriptsize$-*+$};
    \node [below right] at (*++) {\scriptsize$*++$};
    \node [above left] at (--*) {\scriptsize$--*$};
    
    \draw[ultra thick, blue, double] (+++)--(*++) (--*)--(---);
    \draw[ultra thick, double] (*++)--(-++)--(-*+)--(--+)--(--*);
    
    \draw[dotted] (+++)--(*+*) (+++)--(++*) (+++)--(+**) (+++)--(+*+) (+++)--(**+)
    	(---)--(*-*) (---)--(*--) (---)--(**-) (---)--(-*-) (---)--(-**)
    	(-++)--(**+) (-++)--(-**) (-++)--(-+*) (-++)--(*+*)
    	(--+)--(**+) (--+)--(*-+) (--+)--(*-*) (--+)--(-**);
\end{scope}

\end{tikzpicture}
\caption{The barycentric subdivision of the cubical complex $\CC_{\E, 1}$ simplicially embedded into the barycentric subdivision of the space of realizable distributions $\Delta_{\E, 1}$ for the extremal class $\E = \{---, --+, -++, +++\}$. Antipodal regions are indicated in blue. Note that the point, say, $(1, -)\in X \times Y$ from the original definition of~$\Delta_\E$ (see \Cref{sec-topology-toolbox} and \Cref{fig-DeltaH}), when restated in terms of partial hypotheses, becomes $-**$.} 
\label{fig-EMBEDDING}
\end{figure}
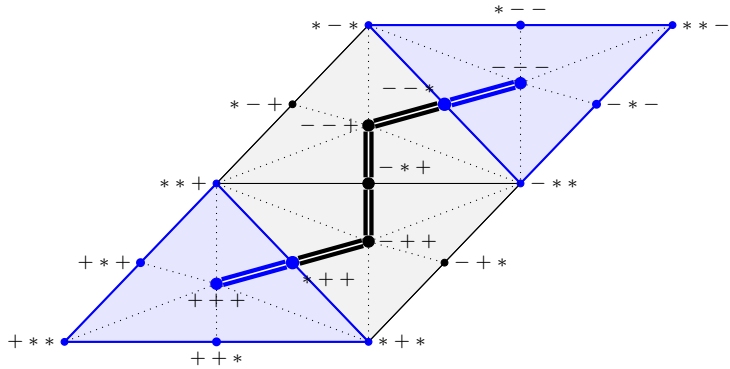

\subsection{Deformation retraction}\label{sec-retraction}
Having defined a simplicial embedding of $\CC_{\E, 1}$ into $\Delta_{\E, 1}$, we are now going to prove that the former is a deformation retract of the latter. So let us briefly recall the respective topological definitions.

For a topological space $B$ and a subspace $A \subseteq B$, a map $f\colon B\rightarrow A$ is called a \emph{retraction} if $f$, restricted to $A$, is an identity, that is, for all $a\in A$, $f(a) = a$. A \emph{deformation retraction} in this setup is a map $F\colon [0, 1] \times B \rightarrow B$ such that $F(0, \cdot)$ is an identity on $B$, $F(1, B) = A$, and $F(t, \cdot)$ is an identity on $A$ for all $t\in [0, 1]$, see Section~1.2 in~\cite{matousek}. The subspace $A$ is called a \emph{(deformation) retract} of $B$ if there is a (deformation) retraction from $B$ to $A$. Note that for a deformation retraction $F$ from $B$ to $A$, the map $F(1, \cdot)$ is a retraction from $B$ to $A$, and so every deformation retract is a retract (the opposite is not true in general). Being a deformation retract is a rather strong property, in particular, this is a special case of homotopy equivalence.

It is quite clear how to define a \emph{simplicial retraction}: For a subcomplex $K$ of $L$ this is simply a simplicial map $f\colon V(L)\rightarrow V(K)$ which is an identity on $V(K)$. At the same time, we do not know of a standard definition of a simplicial deformation retraction. A good reason for that is that $L\times [0, 1]$, which would be a domain of the respective $F$, does not naturally inherit the simplicial structure from $L$. This, however, is a typical obstacle in defining simplicial homotopy, which is usually resolved via prism operators (see proof of Theorem 2.10 in~\cite{hatcher}), so probably a reasonable definition can be given along these lines. In any case, this goes far beyond the aim of the present paper, so we will not try to enforce simpliciality on the deformation retraction that we are going to construct.

Finally, a topological space $B$ is called \emph{contractible} if there exists a \emph{contraction map} $F\colon [0,1]\times B \rightarrow B$ to a point $x\in B$, that is, $F(0, \cdot)$ is an identity on $B$ and $F(1, \cdot) \equiv x$.

\subsection{Restrictions}\label{sec-restrictions}
For a finite extremal class $\E$ on $X$ and a partial hypothesis $h$ on $X$, we define an \emph{$h$-restriction} of $\E$, denoted $\E_h$, as a class on $X$ defined as the class of all total hypotheses of $\E$, extending $h$, that is
$$\E_h = \{g\in \E~|~g\geq h\}.$$
We note that this is closely related (but not identical) to how restrictions are defined in~\cite{Moran:12} (Definition~3.5). In particular, it is known and easy to check that the following properties hold for restrictions:
\begin{itemize}
	\item $\E_h \subseteq \E$ and it is nonempty whenever $h$ is realizable by $\E$. Also, for the completely undefined hypothesis $h_*$, $\E_{h^*} = \E$;
	\item For any $h$, realizable by $\E$, $\CC_{\E_{h}, 1}$ is a full subcomplex of $\CC_{\E, 1}$, spanned by the vertices $\{g\in V(\CC_{\E, 1})~|~g\geq h\}$; 
	\item For any $h \neq h_*$, there is no $g\geq h, -h$; In particular, $\|\E_{h(x), 1}\|$ and $\|\E_{-h(x), 1}\|$ are disjoint. As an easy corollary, for $x\in \|\Delta_{\E, 1}\|$, let us define $h(x)\in V(\Delta_{\E, 1})$ as the minimal vertex of the support of $x$. Then $\|\E_{h(x), 1}\|$ and $\|\E_{h(-x), 1}\|$ are disjoint for all $x\in \|\Delta^\ant_{\E, 1}\|$.
\end{itemize}
The usefulness of restrictions comes from the fact that they enable us to associate vertices of $\Delta_{\E, 1}$, parametrized by realizable partial hypotheses of $\E$, with nonempty subcomplexes of $\CC_{\E, 1}$. This is illustrated in~\Cref{fig-EMBEDDING2} below.

\begin{figure}[hbt]
\centering
\begin{tikzpicture}
[
xpt/.style={inner sep = 0.0pt},
pt/.style={inner sep = 0.9pt, circle, draw, fill=black},
ptb/.style={inner sep = 0.9pt, circle, draw=blue, fill=blue},
ptl/.style={inner sep = 1.5pt, circle, draw, fill=black},
ptbl/.style={inner sep = 1.5pt, circle, draw=blue, fill=blue}
]

\begin{scope}[yscale=0.7, xscale = 1]  
    \coordinate (+**) at (0,0);
    \coordinate (*+*) at (4,0);
    \coordinate (**+) at (2,3);
    \coordinate (-**) at (6,3);
    \coordinate (*-*) at (4,6);
    \coordinate (**-) at (8,6);
        
    \draw[dotted] (+**)--(**+)--(*+*)--cycle;
    \draw[dotted] (*-*)--(**-)--(-**)--cycle;
    \draw[dotted] (**+)--(*-*)--(-**)--(*+*)--cycle;

    \draw[dotted] (+**) to node[pt, midway] (+*+){} (**+)
    	to node[ptl, midway] (*++){} (*+*)
    	to node[pt, midway] (++*){} (+**);
	\node[ptl](+++) at (2, 1.1) {};
    \draw[dotted] (-**) to node[pt, midway] (-*-){} (**-)
    	to node[pt, midway] (*--){} (*-*)
    	to node[ptl, midway] (--*){} (-**);
	\node[ptl](---) at (6, 4.9) {};

	\draw[dotted](**+) to node[pt, midway](*-+){} (*-*);
	\draw[dotted](**+) to node[ptl, midway](-*+){} (-**);
	\draw[dotted](*+*) to node[pt, midway](-+*){} (-**);

	\node[ptl](--+) at (4, 4.1) {};
	\node[ptl](-++) at (4, 1.9) {};

	\node[pt] at (+**) {};
	\node[pt] at (**+) {};
	\node[pt] at (*-*) {};
	\node[pt] at (**-) {};
	\node[pt] at (-**) {};
	\node[pt] at (*+*) {};

    \node [left] at (+**) {\scriptsize$+**$};
    \node [left] at (**+) {\scriptsize$**+$};
    \node [left] at (*-*) {\scriptsize$*-*$};
    \node [right] at (**-) {\scriptsize$**-$};
    \node [right] at (-**) {\scriptsize$-**$};
    \node [right] at (*+*) {\scriptsize$*+*$};
    
    \node [below] at (++*) {\scriptsize$++*$};
    \node [below] at (+++) {\scriptsize$+++$};
    \node [above] at (*--) {\scriptsize$*--$};
    \node [above] at (---) {\scriptsize$---$};
    \node [left] at (+*+) {\scriptsize$+*+$};
    \node [left] at (*-+) {\scriptsize$*-+$};
    \node [right] at (-*-) {\scriptsize$-*-$};
    \node [right] at (-+*) {\scriptsize$-+*$};

    \node [right] at (-++) {\scriptsize$-++$};
    \node [left] at (--+) {\scriptsize$--+$};
    \node [above right] at (-*+) {\scriptsize$-*+$};
    \node [below right] at (*++) {\scriptsize$*++$};
    \node [above left] at (--*) {\scriptsize$--*$};
    
    \draw[ultra thick] (+++) to node [pos=0.33, xpt](1){} node [pos=0.66, xpt](2){} 
    (*++) to node [pos=0.33, xpt](3){} node [pos=0.66, xpt](4){} 
    (-++) to node [pos=0.33, xpt](5){} node [pos=0.66, xpt](6){} 
    (-*+) to node [pos=0.33, xpt](7){} node [pos=0.66, xpt](8){} 
    (--+) to node [pos=0.33, xpt](9){} node [pos=0.66, xpt](10){} 
    (--*) to node [pos=0.33, xpt](11){} node [pos=0.66, xpt](12){} 
    (---);
    
    \draw[dotted] (+++)--(*+*) (+++)--(++*) (+++)--(+**) (+++)--(+*+) (+++)--(**+)
    	(---)--(*-*) (---)--(*--) (---)--(**-) (---)--(-*-) (---)--(-**)
    	(-++)--(**+) (-++)--(-**) (-++)--(-+*) (-++)--(*+*)
    	(--+)--(**+) (--+)--(*-+) (--+)--(*-*) (--+)--(-**);
    	
    \draw [dashed, red, -stealth] (*+*)--(+++);
    \draw [dashed, red, -stealth] (*+*)--(1); 
    \draw [dashed, red, -stealth](*+*)--(2); 
    \draw [dashed, red, -stealth](*+*)--(*++);
    \draw [dashed, red, -stealth](*+*)--(3); 
    \draw [dashed, red, -stealth](*+*)--(4); 
    \draw [dashed, red, -stealth](*+*)--(-++);	

    \draw [dashed, orange, -stealth] (**+)--(+++);
    \draw [dashed, orange, -stealth] (**+)--(1); 
    \draw [dashed, orange, -stealth](**+)--(2); 
    \draw [dashed, orange, -stealth](**+)--(*++);
    \draw [dashed, orange, -stealth](**+)--(3); 
    \draw [dashed, orange, -stealth](**+)--(4); 
    \draw [dashed, orange, -stealth](**+)--(-++);	
    \draw [dashed, orange, -stealth](**+)--(5); 
    \draw [dashed, orange, -stealth](**+)--(6); 
    \draw [dashed, orange, -stealth](**+)--(-*+); 
    \draw [dashed, orange, -stealth](**+)--(7); 
    \draw [dashed, orange, -stealth](**+)--(8); 
    \draw [dashed, orange, -stealth](**+)--(--+); 

    \draw [dashed, green, thick, -stealth](**-)--(---); 
    
    \draw [dashed, blue, thick, -stealth] (--*) to[out=-80, in = -70, looseness = 2] (--+);
    \draw [dashed, blue, thick, -stealth] (--*) to[out=-80, in = -70, looseness = 2] (9);
    \draw [dashed, blue, thick, -stealth] (--*) to[out=-80, in = -70, looseness = 2] (10);
    \draw [dashed, blue, thick, -stealth] (--*) to[out=-70, in = -90, looseness = 2] (---);
    \draw [dashed, blue, thick, -stealth] (--*) to[out=-70, in = -90, looseness = 2] (11);
    \draw [dashed, blue, thick, -stealth] (--*) to[out=-70, in = -90, looseness = 2] (12);
    
\end{scope}

\end{tikzpicture}
\caption{Subcomplexes of $\CC_{\E, 1}$ for $\E$ from \Cref{fig-EMBEDDING}, that correspond to restrictions to partial hypotheses $*+*$, $**+$, $--*$, and $**-$. Note that, as argued above, for antipodal vertices $**+$ and $**-$, the geometric cube complexes for the corresponding restrictions are disjoint.} 
\label{fig-EMBEDDING2}
\end{figure}
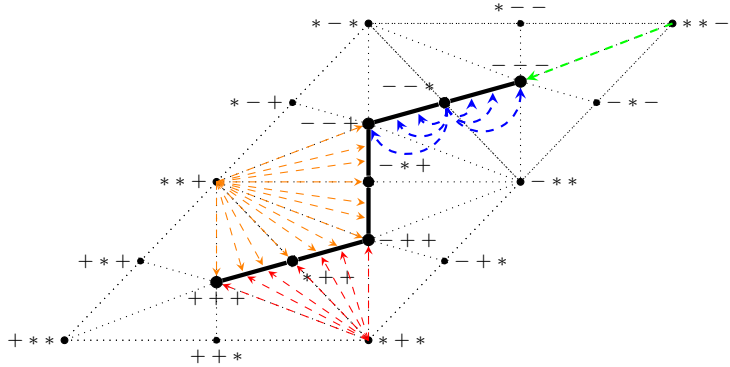

Let us also elaborate on the second bullet. If we define $\CC_{\E, 1}^h$ as a full subcomplex of $\CC_{\E, 1}$, spanned by the vertices $\{g\in V(\CC_{\E, 1})~|~g\geq h\}$, then the claim was that $\CC_{\E, 1}^h = \CC_{\E_h, 1}$; its proof is by a straightforward consideration of what are the discrete cubes of $\E_h$. At the same time, let us similarly define $\Delta^h_{\E, 1}$ as a full subcomplex of $\Delta_{\E, 1}$ spanned by $\{g\in V(\Delta_{\E, 1})~|~g\geq h\}$. One can notice that $\Delta_{\E_h, 1}$ is, in fact, also a full subcomplex of  $\Delta_{\E, 1}$, but spanned by a larger vertex set $\{g\in V(\Delta_{\E, 1})~|~\textrm{there is }w\in V(\Delta_{\E, 1}) \textrm{ s.t. } g, h\leq w\}$. Hence, $\Delta^h_{\E, 1}$ is a full subcomplex of $\Delta_{\E_h, 1}$, but the two \underline{are not}, in general, equivalent. Note also that from the above it is easy to see that $\|\E_{h, 1}\| = \|\Delta^{h}_{\E, 1}\| \cap \|\E_{1}\|$. This is illustrated in \Cref{fig-CUBE-COMPLEX-BARY2} below. 

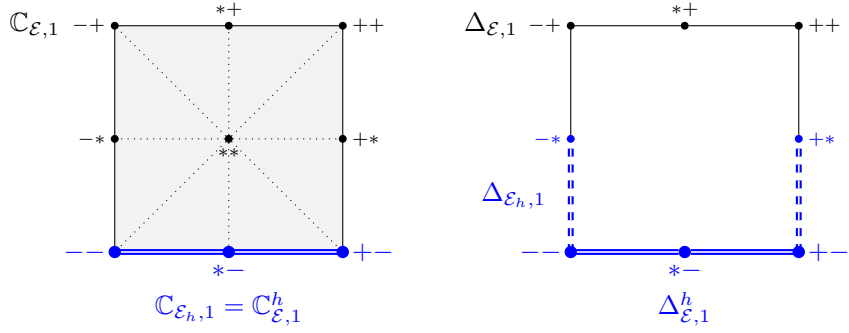
\begin{figure}[hbt]
\centering
\begin{tikzpicture}
[
xpt/.style={inner sep = 0.0pt},
pt/.style={inner sep = 0.9pt, circle, draw, fill=black},
ptb/.style={inner sep = 0.9pt, circle, draw=blue, fill=blue},
ptl/.style={inner sep = 1.5pt, circle, draw, fill=black},
ptbl/.style={inner sep = 1.5pt, circle, draw=blue, fill=blue}
]

\begin{scope}[scale=3]  
\begin{scope}
    \coordinate (--) at (0,0,0);
    \coordinate (-+) at (0,1,0);
    \coordinate (+-) at (1,0,0);
    \coordinate (++) at (1,1,0);
    \coordinate (0-) at (0.5, 0);

    \draw[fill=gray, fill opacity=0.1] (--) -- (-+) -- (++) -- (+-) -- cycle;
	\draw [blue, double, thick] (--)--(0-)--(+-);
    
    \node[pt] (-0) at (0, 0.5) {};
    \node[pt] (0+) at (0.5, 1) {};
    \node[pt] (+0) at (1, 0.5) {};
    \node[ptbl] (0-) at (0.5, 0) {};

    \node[ptbl] at (--) {};
    \node[pt] at (-+) {};
    \node[ptbl] at (+-) {};
    \node[pt] at (++) {};

    \node[pt] (00) at (0.5, 0.5) {};
    
    \node [left] at (-0) {\scriptsize$-*$};
    \node [right] at (+0) {\scriptsize$+*$};
    \node [above] at (0+) {\scriptsize$*+$};
    \node [below] at (0-) {\color{blue}{\bf$*-$}};
    \node [below] at (00) {\scriptsize$**$};
    \node [left] at (--) {\color{blue}{\bf$--$}};
    \node [left] at (-+) {\scriptsize$-+$};
    \node [right] at (+-) {\color{blue}{\bf$+-$}};
    \node [right] at (++) {\scriptsize$++$};

    \draw[dotted] (00)--(--)
    	(00)--(-0) (00)--(-+) (00)--(0+) (00)--(++)
    	(00)--(+0) (00)--(+-) (00)--(0-);
    	
    \node at (-0.35, 1) {$\CC_{\E, 1}$};	
    \node at (0.5, -0.25) {\textcolor{blue}{$\CC_{\E_h, 1} = \CC^h_{\E, 1}$}};	
\end{scope}

\begin{scope}[xshift = 2cm]
    \coordinate (--) at (0,0,0);
    \coordinate (-+) at (0,1,0);
    \coordinate (+-) at (1,0,0);
    \coordinate (++) at (1,1,0);
    \coordinate (0-) at (0.5, 0);

    \node[ptb] (-0) at (0, 0.5) {};
    \node[pt] (0+) at (0.5, 1) {};
    \node[ptb] (+0) at (1, 0.5) {};
    \node[ptbl] (0-) at (0.5, 0) {};
		
    \draw (-0) -- (-+) -- (0+) -- (++) -- (+0);

	\draw [blue, double, thick] (--)--(0-)--(+-);
	\draw [blue, double, dashed, thick] (--)--(-0) (+-)--(+0);

    \node[ptbl] at (--) {};
    \node[pt] at (-+) {};
    \node[ptbl] at (+-) {};
    \node[pt] at (++) {};

    \node [left] at (-0) {\textcolor{blue}{\scriptsize$-*$}};
    \node [right] at (+0) {\textcolor{blue}{\scriptsize$+*$}};
    \node [above] at (0+) {\scriptsize$*+$};
    \node [below] at (0-) {\textcolor{blue}{\bf$*-$}};
    \node [left] at (--) {\textcolor{blue}{\bf$--$}};
    \node [left] at (-+) {\scriptsize$-+$};
    \node [right] at (+-) {\textcolor{blue}{\bf$+-$}};
    \node [right] at (++) {\scriptsize$++$};

    \node at (-0.35, 1) {$\Delta_{\E, 1}$};	
    \node at (0.5, -0.25) {\textcolor{blue}{$\Delta^h_{\E, 1}$}};	
    \node at (-0.25, 0.25) {\textcolor{blue}{$\Delta_{\E_h, 1}$}};	
\end{scope}
\end{scope}

\end{tikzpicture}
\caption{Complexes $\CC_{\E_h, 1} = \CC^h_{\E, 1}$, $\Delta_{\E_h, 1}$, and $\Delta^h_{\E, 1}$ for $\E = \C_2 = \{--, -+, +-, ++\}$, $h=*-$, and $\E_h = \{-+, ++\}$.}
\label{fig-CUBE-COMPLEX-BARY2}
\end{figure}

Although most of the theory developed so far can, in principle, be applicable to arbitrary classes, we now give a crucial fact about cubical complexes of specifically extremal classes.

\begin{proposition}\label{prop-ext-contractible}
	For an extremal class $\E$:
	\begin{itemize}
		\item The cubical complex $\|\E\|$ is contractible, that is there is $x\in \|\E\|$ and a contraction map $F\colon [0, 1] \times \|\E\| \rightarrow \|\E\|$ such that $F(0, \cdot)$ is the identity map and $F(1, \cdot) \equiv x$;
		\item For any $h$, realizable by $\E$, the class $\E_h$ is extremal. In particular, the cubical complex $\|\E_h\|$ is contractible.
	\end{itemize}
\end{proposition}
The above properties are Proposition~4.12 and Theorem~3.1(6) in~\cite{Chalopin:22}. They were also crucial ingredients for the abovementioned $\VC^*\geq 2\VC +1$ bound for extremal classes from~\cite{chase2024dual}, see Proposition~13 there. We also note that in~\cite{Chalopin:22}  they establish \emph{collapsibility}, which is a stronger version of contractibility.

\subsection{Deformation retraction from \texorpdfstring{$\|\Delta_\E\|$}{DE} to \texorpdfstring{$\|\CC_\E\|$}{CE}}\label{sec-retraction-to-cc}
We are now ready to prove the central claim for this section. Note that we prove it in a much stronger way than originally stated in \Cref{sec-extremal}.

\begin{customthm}{\ref{t-sd-for-extremal}} 
	Let $\E$ be a finite extremal class that is \underline{not} a binary hypercube $\C_n$, for any $n$.
	
	Then $\CC_{\E, 1}$ is a full subcomplex of $\Delta_{\E, 1}$ and there is a deformation retraction $F$ from $\|\Delta_{\E, 1}\|$ to $\|\E_1\|$ such that for any $x\in \|\Delta_{\E, 1}\|$ and $t\in [0,1]$, $F(t, x) \in \|\Delta^{h(x)}_{\E, 1}\|$, where
	$h(x)\in V(\Delta_{\E, 1})$ is the minimal vertex of the support of $x$.
	
	
	In particular, for the final map $f_1$ of $F$, $f_1(x) = F(1, x)$, it holds $f_1(x)\in \|\E_{h(x), 1}\|$. In particular, $f_1$ does not collapse antipodal points, that is, $f_1(x) \neq f_1(-x)$ for any $x\in \|\Delta_{\E, 1}^\ant\|$.	
	
	And, if $\E$ is a binary hypercube $\C_n$, then $\|\CC_{\E, 1}\|$ is an $n$-dimensional cube, and $\|\Delta_{\E, 1}\| = \|\Delta^\ant_{\E, 1}\|$ is its boundary, 
	$\|\Delta^\ant_{\E, 1}\| \cong (\S^{n-1}, x\mapsto -x)$. In particular, the natural embedding of $\|\Delta^\ant_{\E, 1}\|$ into $\|\CC_{\E, 1}\|$ does not collapse antipodal points.
	
	In both cases, $\|\Delta_\H\|$ can be mapped into the cubical complex $\|\CC_\E\|$ of dimension $\VC(\E)$ without collapsing any antipodal points in $\|\Delta^\ant_\E\|$.	
    In particular, $\sd(\E) \leq 2\VC(\E)-1$ and $\VC^*(\E) \leq 2\VC(\E)+1$. Consecutively, for any class $\H$, $\sd(\H) \leq 2\VC^\extr(\H) - 1$, or, equivalently, $\VC^\extr(\H) \geq \lceil(\sd(\H) + 1)/2\rceil$.	
\end{customthm}
Before the proof, let us give two cautionary examples. First of all, the situation in \Cref{fig-EMBEDDING}, where $\Delta^\ant \cap \CC$ is nonempty and closed under antipodality, is not, in general, true. \Cref{fig-EMBEDDING3} below shows a simple extremal class for which $\Delta^\ant$ is nonempty, but $\Delta^\ant \cap \CC$ is. In particular, this means that we cannot hope to make the final map $f_1$ above antipodal. We also note that there are easy examples where $\Delta^\ant \cap \CC$ is nonempty, but not closed under antipodality (for example, for $\E = \{+++, -++, --+\}$).

\begin{figure}[hbt]
\centering
\begin{tikzpicture}
[
pt/.style={inner sep = 0.9pt, circle, draw, fill=black},
ptb/.style={inner sep = 0.9pt, circle, draw=blue, fill=blue},
ptl/.style={inner sep = 1.5pt, circle, draw, fill=black},
ptbl/.style={inner sep = 1.5pt, circle, draw=blue, fill=blue}
]

\begin{scope}[yscale=0.7, xscale = 1]  
	\node[ptbl] (*-) at (0,0) {};
	\node[ptl] (+-) at (2,0) {};
	\node[ptl] (+*) at (4,0) {};
	\node[ptl] (++) at (6,0) {};
	\node[ptbl] (*+) at (8,0) {};

	\draw(*-)--(+-) (++)--(*+);
	\draw[ultra thick] (+-)--(+*)--(++);
    
    \node [below] at (*-) {\scriptsize$*-$};
    \node [below] at (+-) {\scriptsize$+-$};
    \node [below] at (+*) {\scriptsize$+*$};
    \node [below] at (++) {\scriptsize$++$};
    \node [below] at (*+) {\scriptsize$*+$};

\end{scope}

\end{tikzpicture}
\caption{The embedding of $\CC_{\E, 1}$ into $\Delta_{\E, 1}$ for the extremal class $\E = \{+-, ++\}$. The antipodal part $\Delta^\ant_{\E, 1}$ consists of two vertices $*-$ and $*+$ and is disjoint from $\CC_{\E, 1}$.} 
\label{fig-EMBEDDING3}
\end{figure}
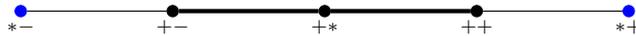

As a second warning, recall that in \Cref{sec-retraction} we argued that although defining simplicial deformation retraction is problematic, the definition of simplicial retraction is straightforward. We thus might ask if the deformation retraction $F$ above can be made such that the final map $f_1$ is simplicial; note that $f_1$ itself is a retraction from $\|\Delta_{\E, 1}\|$ to $\|\E_1\|$. However, the answer to this question is no. This can easily be seen so for the vertex $**+$ for the class in \Cref{fig-EMBEDDING2}. Even without the restriction $f_1(**+) \in V(\Delta_{**+, 1})$ (but assuming that $f_1$ is a retraction, that is, that it keeps $\CC$ fixed), one can can easily see that mapping $**+$ to any vertex of $\CC$ would break simpliciality.

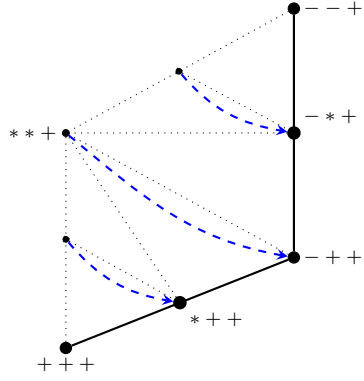
\begin{figure}[hbt]
\centering
\begin{tikzpicture}
[
xpt/.style={inner sep = 0.0pt},
pt/.style={inner sep = 0.9pt, circle, draw, fill=black},
ptb/.style={inner sep = 0.9pt, circle, draw=blue, fill=blue},
ptl/.style={inner sep = 1.5pt, circle, draw, fill=black},
ptbl/.style={inner sep = 1.5pt, circle, draw=blue, fill=blue}
]

\begin{scope}[yscale=1.5, xscale = 1.5]  
	\node[pt] (**+) at (2,3) {};

	\node[ptl](+++) at (2, 1.1) {};
	\node[ptl](-++) at (4, 1.9) {};
	\node[ptl](--+) at (4, 4.1) {};

    \draw[thick] (+++) to node [pos=0.5, ptl](*++){}
    (-++) to node [pos=0.5, ptl](-*+){}
    (--+);

    \node [left] at (**+) {\scriptsize$**+$};
    \node [below] at (+++) {\scriptsize$+++$};
    \node [right] at (-++) {\scriptsize$-++$};
    \node [right] at (--+) {\scriptsize$--+$};
    \node [above right] at (-*+) {\scriptsize$-*+$};
    \node [below right] at (*++) {\scriptsize$*++$};

    \draw[dotted] (+++) to node [pt, midway](1){} (**+) (*++)--(**+) (-++)--(**+) (-*+)--(**+) (--+) to node [pt, midway](2){} (**+);
    \draw[dotted] (1)--(*++) (2)--(-*+);
    
    \draw [dashed, blue, thick, -stealth] (**+) to[out=-40, in = 170, looseness = 1] (-++);
    \draw [dashed, blue, thick, -stealth] (1) to[out=-50, in = 170, looseness = 1] (*++);
    \draw [dashed, blue, thick, -stealth] (2) to[out=-50, in = 170, looseness = 1] (-*+);    
\end{scope}

\end{tikzpicture}
\caption{
Piecewise-linear, but not simplicial, part of the retraction of $\Delta_{\E, 1}$ to $\CC_{\E, 1}$ for $\E$ from \Cref{fig-EMBEDDING}.}
\label{fig-SUBDIVISION}
\end{figure}

However, although for technical reasons we will not prove it, we note that the deformation retraction that we will construct will be \emph{piecewise-linear}, which is the next best thing after simpliciality, see~\cite{bryant02} for some background. Moreover, from the construction, it would be obvious that $f_1$ can be made simplicial from a subdivision of $\Delta_{\E, 1}$. See~\Cref{fig-SUBDIVISION} below for an example of how this will look for the particular abovementioned vertex $**+$.

\begin{proof}
The case $\E = \C_n$ has been discussed in~\Cref{sec-cubical-bary}. In the same section, we also argued that $\CC_{\E, 1}$ is a full subcomplex of $\Delta_{\E, 1}$ whenever $\E \neq \C_n$. Note that $F(t, x) \in \|\Delta^{h(x)}_{\E, 1}\|$ and $F(1, x) \in \|\E_{1}\|$ trivially imply $f_1(x) = F(1, x) \in \|\Delta^{h(x)}_{\E, 1}\| \cap \|\E_{1}\| = \|\E_{h(x), 1}\|$. Also, the statement $f_1(x) \neq f_1(-x)$ for any $x\in \|\Delta_{\E, 1}^\ant\|$ follows from the fact that, as argued in \Cref{sec-restrictions},  $\|\E_{h(x), 1}\|$ and $\|\E_{h(-x), 1}\|$ are disjoint. 

The concluding statement $\sd(\E) \leq 2\VC(\E)-1$ follows by a straightforward application of \Cref{thm:Jaw} (Local BU): The dimension of $\CC_\E$ is $\VC(\E)$ and, as the map does not collapse antipodal points, $2\VC(\E) > \sd(\E) \sim 2\VC(\E) \geq \sd(\E) + 1$. Again, \Cref{thm:Jaw} is not applicable whenever $\sd \leq 0$, but in this case the claimed bound can only be invalidated whenever $\sd = \VC = 0$. But $\VC = 0$ implies the class has only one concept, hence $\sd = -1$, and the bound stands. Finally, $\VC^*(\E) \leq 2\VC(\E)+1$ follows from it as, by Lemma~\ref{lem-vc-lb}, $\sd\geq \VC^* - 2$.

So it only remains to prove the existence of the deformation retraction $F$ with the prescribed properties in case $\E\neq \C_n$.

Let $W$ be a subset of vertices of $V(\Delta_{\E, 1})$ such that $W$ is upwards-closed with respect to the order on partial hypotheses and contains $V(\CC_{\E, 1})$.
And let $\Q_W$ be the full subcomplex of $\Delta_{\E, 1}$, spanned by $W$. We are going to prove, by induction on the poset of such sets $W$, that there is the corresponding deformation retraction $F_W$ from $\|\Q_W\|$ to $\|\E_1\|$. The statement of the lemma then corresponds to $W = V(\Delta_{\E, 1})$; and the base of induction, $W = V(\CC_{\E, 1})$, is given by a constant deformation retraction map $F_{V(\CC_{\E, 1})} \colon \|\Q_{V(\CC_{\E, 1})}\| \times [0, 1] \rightarrow \|\E_1\|$ defined by $F(x, t) = x$, where we use the fact that $\Q_{V(\CC_{\E, 1})} = \CC_{\E, 1}$.

For the induction step, let us take such $W$ and any minimal $w \in W$ such that $w\notin V(\CC_{\E, 1})$. Now assuming, by induction, the existence of $F_{W'}$, for $W' = W - w$, we want to construct the corresponding $F_W$.

For this $w$, let us define the \emph{star} and the \emph{link} of $w$ in $W$ as the subcomplexes $\St_W(w) = \{s \in S(\Q_W)~|~\text{there is } t\in S(\Q_W)\text{ s.t. } s,w\leq t\}$ and $\Lk_W(w) = \{s \in S(\St_W(w))~|~w\notin s\}$. Those notions are quite standard, see~\cite{bryant02}. Recall that joins of simplicial complexes and maps between them were defined in~\Cref{sec-joins}. We utilize the usual fact that $\St_W(w)$ is the join of $w$ and $\Lk_W(w)$, $\St_W(w) = w * \Lk_W(w)$, that is, 
$$S(\St_W(w)) = \left\{s_w \sqcup s_l~|~s_w\in \{\emptyset, w\}, s_l\in S(\Lk_W(w))\right\}.$$
We note that, by construction, $\Lk_W(w)$ is a subcomplex of $\Q_{W'}$, moreover, $\Lk_W(w) = \Q_{W'} \cap \St_W(w)$ and $\Q_W = \Q_{W'} \cup \St_W(w)$. Finally, by the choice of $w$ as minimal in $V(W)$, it easily follows that $\St_W(w) = \Delta^{w}_{\E, 1}$.

We now define an intermediate deformation retraction $G$ from $\|\Q_W\| = \|\Q_{W'}\| \cup w * \|\Lk_W(w)\|$ to $\|\E_1\| \cup w * \|\E_{w,1}\|$ as $G(t, x) = F_{W'}(t, x)$ for $x\in \|\Q_{W'}\|$ and $G(t, x) = (I_w * F_{W'}) (t, x)$ for $x\in w * \|\Lk_W(w)\| = \|\St_W(w)\|$, where $I_w$ is the identity deformation retraction of $w$ to itself. Here $I_w * F_{W'}$ is the join of the respective retractions. The construction of $G$ is illustrated in~\Cref{fig-RETRACTION1} below. 

\begin{figure}[hbt]
\centering
\begin{tikzpicture}
[
xpt/.style={inner sep = 0.0pt},
pt/.style={inner sep = 0.9pt, circle, draw, fill=black},
ptb/.style={inner sep = 0.9pt, circle, draw=blue, fill=blue},
ptl/.style={inner sep = 2pt, circle, draw, fill=black},
ptbl/.style={inner sep = 2pt, circle, draw=blue, fill=blue}
]

\begin{scope}[yscale=1.5, xscale = 1.5]  
	\node[pt] (**+) at (2,3) {};
	\node[ptb] (+*+) at (1,1.5) {};
	\node[ptb] (*-+) at (3,4.5) {};
	\node[pt] (*+*) at (4,0) {};
	\node[pt] (++*) at (2,0) {};
	\node[pt] (-+*) at (5,1.5) {};

	\node[ptbl](+++) at (2, 1.1) {};
	\node[ptbl](-++) at (4, 1.9) {};
	\node[ptbl](--+) at (4, 4.1) {};
	\node[ptl](---) at (6, 4.9) {};
	
	\draw[fill=blue, fill opacity=0.075] (1,1.5)--(2,3)--(3, 4.5)--(4, 4.1)--(4, 1.9)--(2, 1.1)--cycle;

    \draw[ultra thick, double, blue] (+++) to
		node[xpt, pos=0.16](9){} node[xpt, pos=0.33](10) {} 
    	node [pos=0.5, ptbl](*++){}
		node[xpt, pos=0.66](11){} node[xpt, pos=0.83](12) {} 
    (-++) to node [pos=0.5, ptbl](-*+){}
    (--+);
    \draw[ultra thick, double] (--+) to node [pos=0.5, ptl](--*){} (---);

    \node [left] at (**+) {\scriptsize$w = **+$};
    \node [below left ] at (+++) {\scriptsize$+++$};
    \node [above right] at (-++) {\scriptsize$-++$};
    \node [below right] at (--+) {\scriptsize$--+$};
    \node [below right] at (--*) {\scriptsize$--*$};
    \node [below right] at (---) {\scriptsize$---$};
    \node [above right] at (-*+) {\scriptsize$-*+$};
    \node [below right] at (*++) {\scriptsize$*++$};
    \node [left] at (+*+) {\scriptsize$+*+$};
    \node [left] at (*-+) {\scriptsize$*-+$};
    \node [below] at (++*) {\scriptsize$++*$};
    \node [right] at (*+*) {\scriptsize$*+*$};
    \node [right] at (-+*) {\scriptsize$-+*$};
  
    \draw 
    	(+++)
    		to node[xpt, pos=0.33](a5){} node[xpt, pos=0.66](a6){}
    	(**+);
    \draw     	   
    	(*++)--(**+) (-++)--(**+) (-*+)--(**+) 
    	(**+)
    		to node[xpt, pos=0.33](a7){} node[xpt, pos=0.66](a8){}
    	(--+);
    \draw (+*+) 
    		to node[xpt, pos=0.33](a1){} node[xpt, pos=0.66](a2){}
    	(**+)
    		to node[xpt, pos=0.33](a3){} node[xpt, pos=0.66](a4){}
    	(*-+);
    \draw[ultra thick, blue] (+++)--(+*+) (*-+) to node[xpt, pos=0.5](x1){} (--+);
    \draw (+++)--(++*)
    	 to node[xpt, pos=0.33](1){} node[xpt, pos=0.66](2) {} 
    		(*+*)
    	to node[xpt, pos=0.33](3){} node[xpt, pos=0.66](4){} 	
    		(-+*)--(-++)
    	(+++) 
			to node[xpt, pos=0.33](5){} node[xpt, pos=0.66](6) {}     	
    	(*+*) (*+*)--(*++) (*+*)
			to node[xpt, pos=0.33](7){} node[xpt, pos=0.66](8) {}     	
    	(-++);
    
    \draw [dashed, red, thick, -stealth] (++*) to[out=80, in = -80, looseness = 1] (+++);
    \draw [dashed, red, thick, -stealth] (-+*) to[out=175, in = -35, looseness = 1] (-++);
    \draw [dashed, red, thick, -stealth] (*+*) to[out=135, in = -75, looseness = 1] (*++);

	\draw [dashed, red, thick, -stealth] (1) to (5);
	\draw [dashed, red, thick, -stealth] (2) to (6);
	\draw [dashed, red, thick, -stealth] (3) to (7);
	\draw [dashed, red, thick, -stealth] (4) to (8);
	\draw [dashed, red, thick, -stealth] (5) to (9);
	\draw [dashed, red, thick, -stealth] (6) to (10);
	\draw [dashed, red, thick, -stealth] (7) to (11);
	\draw [dashed, red, thick, -stealth] (8) to (12);
	
    \draw [dashed, red, thick, -stealth] (+*+) to[out=0, in = 135, looseness = 1] (+++);
    \draw [dashed, red, thick, -stealth] (*-+) to[out=-45, in = 180, looseness = 1] (--+);
	\draw [dashed, green, thick, -stealth] (a1) to (a5);
	\draw [dashed, green, thick, -stealth] (a2) to (a6);
	\draw [dashed, green, thick, -stealth] (a3) to (a7);
	\draw [dashed, green, thick, -stealth] (a4) to (a8);

	\node[above, anchor=-140] at (x1) {\textcolor{blue}{$\Lk_W(w)$}};
	\node at (3.3,2.7) {\textcolor{blue}{$\St_W(w)$}};
	\node at (4.4,2.7) {\textcolor{blue}{$\E_{w, 1}$}};
\end{scope}

\end{tikzpicture}
\caption{Construction of the deformation retraction $G$ from $\|\Q_W\| = \|\Q_{W'}\| \cup w * \|\Lk_W(w)\|$ to $\|\E_1\| \cup w * \|\E_{w,1}\|$.
Here $\E$ is the extremal class from \Cref{fig-EMBEDDING2}. We assume that, apart from vertices of $\CC_\E$, $W'$ also contains vertices $++*$, $-+*$, $*+*$, $+*+$, and $*-+$, and $w=**+$. Filled blue region represents $\|\St_W(w)\|$, the thick blue lines are $\|\Lk_W(w)\| = \|\St_W(w)\| \cap \|\Q_{W'}\|$, and double thick blue lines are $\|\E_{w, 1}\| = \|\St_W(w)\| \cap \|\E\|$. Red dotted lines represent the retraction $F_{W'}$, and the green ones are the extension of $F_{W'}$ to $I_w * F_{W'}$.} 
\label{fig-RETRACTION1}
\end{figure}
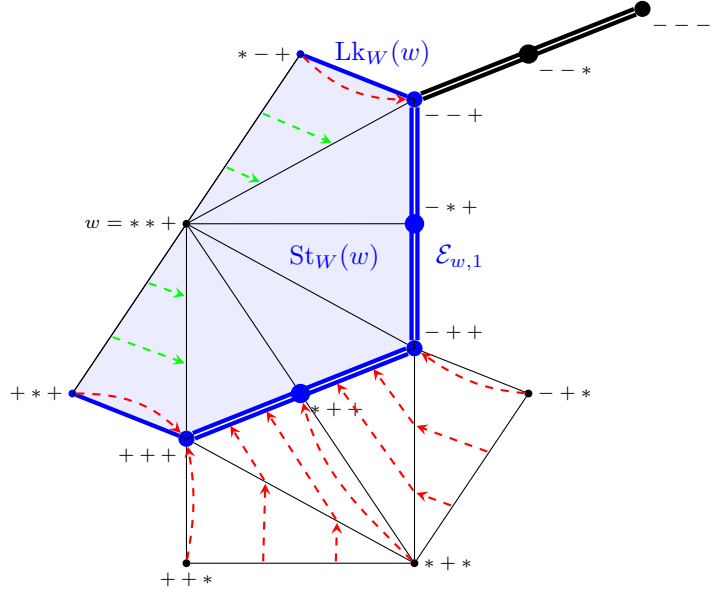

Let us, however, define $G$ explicitly:
\begin{itemize}
	\item For $x=w$, $G(t, w) \equiv w$;
	\item For $x\in \|\Q_{W'}\|$, $G(t, x) = F_{W'}(t, x)$;
	\item Finally, for $x \in \|\St_W(w)\| = w * \|\Lk_W(w)\|$, $x\neq w$, let $x = \alpha w + \beta x'$, for $x' \in \|\Lk_W(w)\|$, be the representation of $x$ in the barycentric coordinates of the join (see~\Cref{sec-joins}). Note that $x\neq w$ implies that $\beta >0$ and that $x'$ is uniquely defined. Then $G(t, x) = \alpha w + \beta F_{W'}(t, x') \in w * \|\Lk_W(w)\|$.
\end{itemize}

Finally, we need to check a few things about $G$. First, note that $G$ is defined separately on $\|\Q_{W'}\|$ as $F_{W'}$ and on $w * \|\Lk_W(w)\|$ as $(I_w * F_{W'})$. The fact that this definition is consistent on $\|\Lk_W(w)\| = \|\Q_{W'}\| \cap w * \|\Lk_W(w)\|$ follows from the construction of $(I_w * F_{W'})$. This also implies that $G$ is continuous, as it is defined in a continuous way on two closed subspaces and is consistent on their intersection. The fact that the range of the final map $G(1, \cdot)$ is $\|\E_1\| \cup w * \|\E_{w,1}\|$ follows from the inductive statement that $F_{W'}$ is a deformation retraction to $\|\E_1\|$ and that $F_{W'}(t, \|\Delta^{w}_{\E, 1}\|) \subseteq \|\Delta^{w}_{\E, 1}\|$, together with the fact that $\|\E_{w,1}\| = \|\E_1\| \cap \|\Delta^{w}_{\E, 1}\|$. And the fact that $G$ is a retraction, that is, an identity on $\|\E_1\| \cup w * \|\E_{w,1}\|$, follows from $F_{W'}$ being identity on $\|\E_1\|$ and, consecutively, on $\|\E_{w,1}\|$.

Let us now define a deformation retraction $H$ from $\|\E_1\| \cup w * \|\E_{w,1}\|$ to $\|\E_1\|$ as follows. Let $C_w$ be the contraction map from $\|\E_{w,1}\|$ to some $x_w\in \|\E_{w,1}\|$ from Proposition~\ref{prop-ext-contractible}. Recall that this means that $C_w(0, \cdot)$ is the identity on $\|\E_{w,1}\|$ and $C_w(1, \cdot) \equiv x_w \in \|\E_{w,1}\|$. Now, we define 
$H\colon [0,1] \times \left(\|\E_1\| \cup w * \|\E_{w,1}\|\right) \rightarrow \left(\|\E_1\| \cup w * \|\E_{w,1}\|\right)$ as identity on $\|\E_1\|$, and on  $w * \|\E_{w,1}\|$ we define it as
$$H(t, x) = \alpha(1-t) w + \left(\beta + \alpha t\right) C_w\bigl(\alpha t / (\beta + \alpha t), x'\bigr),$$
where $x = \alpha w + \beta x'$, for $x'\in \|\E_{w,1}\|$. We note that the term $\alpha t / (\beta + \alpha t)$ is undefined for $\alpha = 1$ (alternatively, $\beta=0$) and $t = 0$, in which case we put it to be $1$. Note, however,  that thus extended, this term becomes discontinuous at $\alpha=1$, $t=0$, which is something we will need to deal with later. This is illustrated in \Cref{fig-RETRACTION2} below.

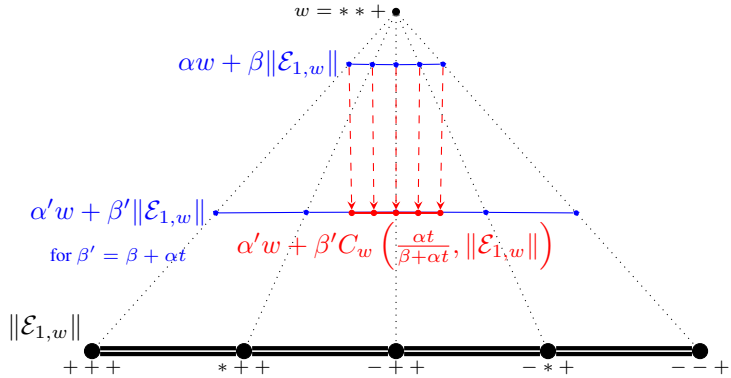
\begin{figure}[hbt]
\centering
\begin{tikzpicture}
[
xpt/.style={inner sep = 0.0pt},
pt/.style={inner sep = 0.9pt, circle, draw, fill=black},
ptb/.style={inner sep = 0.7pt, circle, draw=blue, fill=blue},
ptr/.style={inner sep = 0.7pt, circle, draw=red, fill=red},
ptl/.style={inner sep = 2pt, circle, draw, fill=black},
ptbl/.style={inner sep = 2pt, circle, draw=blue, fill=blue}
]

\begin{scope}[yscale=1.5, xscale = 2]  
	\node[pt] (**+) at (2,3) {};

	\node[ptl](+++) at (0, 0) {};
	\node[ptl](*++) at (1, 0) {};
	\node[ptl](-++) at (2, 0) {};
	\node[ptl](-*+) at (3, 0) {};
	\node[ptl](--+) at (4, 0) {};


    \draw[ultra thick, double] (+++) to (*++) to (-++) to (-*+) to (--+);

    \node [left] at (**+) {\scriptsize$w = **+$};
    \node [below] at (+++) {\scriptsize$+++$};
    \node [below] at (-++) {\scriptsize$-++$};
    \node [below] at (--+) {\scriptsize$--+$};
    \node [below] at (-*+) {\scriptsize$-*+$};
    \node [below] at (*++) {\scriptsize$*++$};

    \draw[dotted] (**+) to node[ptb, pos=0.15](+++a){} node[ptb, pos=0.6](+++b){} (+++);
    \draw[dotted] (**+) to node[ptb, pos=0.15](*++a){} node[ptb, pos=0.6](*++b){} (*++);
    \draw[dotted] (**+) to node[ptb, pos=0.15](-++a){} node[ptb, pos=0.6](-++b){} (-++);
    \draw[dotted] (**+) to node[ptb, pos=0.15](-*+a){} node[ptb, pos=0.6](-*+b){} (-*+);
    \draw[dotted] (**+) to node[ptb, pos=0.15](--+a){} node[ptb, pos=0.6](--+b){} (--+);
    
    \draw[blue] (+++a)--(*++a)--(-++a)--(-*+a)--(--+a);
    \draw[blue] (+++b)--(*++b)--(-++b)--(-*+b)--(--+b);

    \path (+++b) to node [ptr, pos=(1 - 0.15/0.6)/2](+++c) {} 
     			node [ptr, pos=(1 - 0.15/0.6)/2 + 1*0.15/0.6/4](*++c) {} 
     			node [ptr, pos=(1 - 0.15/0.6)/2 + 2*0.15/0.6/4](-++c) {} 
     			node [ptr, pos=(1 - 0.15/0.6)/2 + 3*0.15/0.6/4](-*+c) {} 
     			node [ptr, pos=(1 - 0.15/0.6)/2 + 4*0.15/0.6/4](--+c) {} 
     		(--+b);
    
    \node[left] at (+++a) {\textcolor{blue}{$\alpha  w + \beta \|\E_{1,w}\|$}};
    \node[left](x) at (+++b) {\textcolor{blue}{$\alpha'  w + \beta' \|\E_{1,w}\|$}};
    \node[below] at (x.south) {\scriptsize\textcolor{blue}{for $\beta' = \beta + \alpha t$}};
    \node[above left] at (+++) {$\|\E_{1,w}\|$};
    
    \draw[dashed, red, -stealth] (+++a)--(+++c);
    \draw[dashed, red, -stealth] (*++a)--(*++c);
    \draw[dashed, red, -stealth] (-++a)--(-++c);
    \draw[dashed, red, -stealth] (-*+a)--(-*+c);
    \draw[dashed, red, -stealth] (--+a)--(--+c);
    
    \draw[thick, red] (+++c)--(--+c);
    \node[below] at (-++c) {\textcolor{red}{$\alpha' w +  \beta'C_w\left(\frac{\alpha t}{\beta + \alpha t}, \|\E_{1, w}\| \right)$}};
\end{scope}

\end{tikzpicture}
\caption{Construction of the deformation retraction $H$ on $w * \|\E_{w,1}\|$ for the same parameters as in~\Cref{fig-RETRACTION1}, compare also to \Cref{fig-SUBDIVISION}.
The set $L_\beta = \alpha w + \beta \|\E_{1, w}\|$ is mapped by $H(t, \cdot)$ into $L_{\beta'}$ for $\beta' = \beta + \alpha t$. Note that $\diam(L_{\beta'}) = \beta' \diam(\|\E_{1, w}\|) = \beta/\beta' \diam(L_\beta)$ and so, for $\beta$ small and $t$ not close to $0$, $\diam(L_{\beta'}) \gg \diam(L_{\beta})$. However, we need to keep the diameter of $H(t, L_{\beta})$ comparable to $\diam(L_\beta)$. This motivates the somewhat peculiar choice of parameter $\alpha t/(\beta + \alpha t)$; in particular, taking it, more naively, to be $\alpha t$ would break condition i) below.} \label{fig-RETRACTION2}
\end{figure}

Let us now check the properties of $H$ on $w * \|\E_{w,1}\|$. First of all, note that the barycentric coordinates of the image are $\alpha(1-t)$ and $\beta + \alpha t$, and their sum is $\beta + \alpha t + \alpha (1-t) = \beta + \alpha = 1$, as needed. We now need to check four things about $H$: i) Recall that for $\alpha=1$, $x=w$, which does not depend on the choice of $x'$ in the decomposition $x=1 w + 0 x'$. Thus, for the definition of $H$ to be consistent, we need to check that $H(t, x)$ in this case also does not depend on the particular choice of $x'$; ii) For $\alpha = 0$, by the definition of retraction, $H(t, x) =x$. In particular, this is consistent with the definition of $H$ as the identity on $\|\E_1\|$, and hence makes $H$ to be consistently defined on the whole domain; iii) For $t=0$, $H(0, x) = x$; and iv) For $t=1$, $H(1, x)\in \|\E_{w, 1}\|$. Those conditions are summarized in \Cref{fig-RETRACTION-H} below. Apart from that, we need to separately argue that $H$ is continuous. 

\begin{figure}[hbt]
\centering
\begin{tikzpicture}
\begin{scope}[scale=1.5, xshift=1.5cm]  
    \draw(-1, -1)--(-1, 1)--(1,1)--(1, -1)--(-1, -1);
	\node (t0) at (-1.65, 0) {iii) $\mathbf{t=0}$};
	\node (t1) at (1.7, 0) {iv) $\mathbf{t=1}$};
	\node (a0) at (0, -1.2) {ii) $\mathbf{\alpha=0, \beta=1}$};
	\node (a1) at (0, 1.4) {i) $\mathbf{\alpha=1, \beta=0}$};

	\node at (-1.6,-0.3) {\centering  \scriptsize{$H(0, x) = x$}};
	\node at (1.75,-0.3) {\centering  \scriptsize{$H(1, x)\in \|\E_{w, 1}\|$}};

	\node at (0,1.15) {\centering  \scriptsize{$x \equiv w$, does not depend on $x'$}};

	\node at (0,-1.45) {\centering  \scriptsize{$x \equiv x'\in \|\E_{w, 1}\|$}};
	\node at (0,-1.65) {\centering  \scriptsize{$H(t,x)\equiv x$}};
	
	\draw[dashed, ->](-0.9, -0.9) to node[midway, right] {$\alpha$} (-0.9, 0.9);
	\draw[dashed, ->](-0.9, -0.9) to node[midway, above] {$t$} (0.9, -0.9);

\end{scope}

\end{tikzpicture}
\caption{Diagram summarizing the conditions on the intermediate deformation retraction $H$.}
\label{fig-RETRACTION-H}
\end{figure}
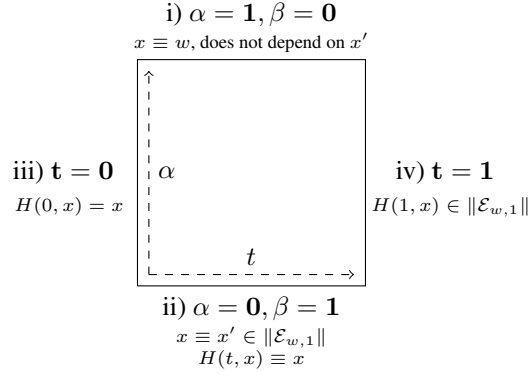

Let us proceed with checking i)--iv):

\emph{i) For $\alpha=1$, $H(t, x)$ does not depend on the choice of $x'$.}
$$H(t, x) = (1-t) w + t C_w(t/t, x') = (1-t) w + t C_w(1, x') = (1-t) w + t c_w,$$
as needed. Note also that changing $t/t$ into $1$ is consistent with our extension of the term $\alpha t / (\beta + \alpha t)$;

\emph{ii) For $\alpha = 0$, $H(t, x) =x$.}
$$H(t, x) = C_w(0, x') = x' = x,$$
as needed.

\emph{iii) For $t = 0$, $H(0, x)=x$.} If $\alpha\neq 1$ then
$$H(0, x) = \alpha w + \beta C_w(0, x') = \alpha w + \beta x' = x,$$
as needed. Otherwise, if $\alpha = 1$, then $x = w$ and 
$$H(0, x) = 1 w + 0 C_w(1, x') = 1 w + 0 c_w = w,$$
as needed. 

\emph{iv) For $t=1$, $H(1, x) \in \|\E_{w, 1}\|$.}
$$H(1, x) = \left(\beta + \alpha \right) C_w(\alpha, x') = C_w\bigl(\alpha/ (\beta + \alpha), x'\bigr) = C_w(\alpha, x') \in \|\E_{w,1}\|,$$
as needed.

The only thing left to be checked is that $H$ is continuous and, trivially, the continuity can only break at the discontinuity of the term $\alpha t / (\beta + \alpha t)$, that is, at $\alpha=1$ and $t=0$. So let $x_i \in w *  \|\E_{w,1}\|$ and $t_i \in [0,1]$ be the sequences converging to $w$ and $0$ respectively. Without losing generality, we can assume that $x_i \neq w$ for all $i$, which means that each $x_i$ has a unique representation in barycentric coordinates of the join, $x_i = \alpha_i w + \beta i x_i'$, $\alpha_i \neq 1$, with $\alpha_i \rightarrow 1$ and $\beta_i = 1 - \alpha_i \rightarrow 0$, but with no control over $x_i'$. Out goal is to show that $H(t_i, x_i)\rightarrow H(0, w) = w$. By the definition of $H$,
\begin{align*}
	H(t_i, x_i) &= \alpha_i(1-t_i) w + \left(\beta_i + \alpha_i t_i\right) C_w\bigl(\alpha_i t_i / (\beta_i + \alpha_i t_i), x'_i\bigr) \\
		&=  \alpha_i(1-t_i) w + \left(\beta_i + \alpha_i t_i\right) x''_i,
\end{align*}
for some $x''_i\in \|\E_{w,1}\|$. Again, we do not have much control over $x''_i$, but, as $\alpha_i(1-t_i) \rightarrow 1$ and $\left(\beta_i + \alpha_i t_i\right) \rightarrow 0$, this obviously converges to $w$, as needed.

Finally, the required deformation retraction $F_W$ is obtained by chaining $H$ after $G$. Foramlly, $F_W(t, x) = G(2t, x)$ for $t\in [0, 1/2]$ and $F_W(t, x) = H\bigl(2(t-1/2), G(1,x)\bigr)$ for $t\in [1/2, 1]$. As $G$ is the deformation retraction from $\|\Q_W\|$ to $\|\E_1\| \cup w * \|\E_{w,1}\|$ and $H$ is from $\|\E_1\| \cup w * \|\E_{w,1}\|$ to $\|\E_1\|$, $F_W$ is a deformation retraction from $\|\Q_W\|$ to $\|\E_1\|$. The only thing left to check is that for any $x\in \|\Q_W\|$ and $t\in [0,1]$, $F_W(t, x) \in \|\Delta^{h(x)}_{\E, 1}\|$. This, however, is rather easy: if $x\in \|\Q_W\|$ is such that $h(x)\neq w$, then $x\in \|\Q_{W'}\|$, and so, by the induction hypothesis, $G$, which coincides with $F_{W'}$ on $\|\Q_{W'}\|$, would keep the image of $x$ in $\|\Delta^{h(x)}_{\E, 1}\|$ and map it there into $\|\E_1\|$, where it will stay unaffected by further application of $H$. And if $h(x) =  w$ then $x\in \|\St_W(w)\| = \|\Delta^{w}_{\E, 1}\| = w * \|\Lk_W(w)\|$, where it will first be moved by $G$ to $w * \|\E_{w, 1}\|$, and then by $H$ to $\|\E_{w, 1}\|$, all without leaving $\|\Delta^{w}_{\E, 1}\|$.
\end{proof}

\begin{customthm}{\ref{th-sd-vc-ub}}
	If $\VC(\H)\leq 1$ then $\sd(\H) \leq 1$. Alternatively, if $\sd(\H) \geq 2$ then $\VC(\H)\geq 2$.
\end{customthm}
\begin{proof}
Let $\H$ be a class with $\VC(\H)\leq 1$. It is known (\cite{ben15}) that in this case $\H$ can be embedded into an extremal class of $\VC \leq 1$, hence $\VC^\extr(\H) \leq 1$. But then, by~\Cref{t-sd-for-extremal}, $\sd(\H) \leq  2\VC^\extr(\H) - 1 \leq 1$, as needed.
\end{proof}

\bibliographystyle{plainnat}
\bibliography{bib-disamb}

\end{document}